\documentclass[aps,rmp,twocolumn,amsmath,amssymb,superscriptaddress]{revtex4-2}
\usepackage[utf8]{inputenc}

\usepackage{url}
\usepackage{tikz}
\usepackage{graphicx}
\usepackage{amsfonts}
\usepackage{amsmath}
\usepackage{amssymb,bbold}
\usepackage{xcolor}
\usepackage{bm}
\usepackage{bbm}
\usepackage{dsfont}
\usepackage{enumerate}
\usepackage[abs]{overpic}
\usepackage{amsfonts, amsmath, amsthm, amssymb} 
\usepackage{mathtools}
\usepackage{array}
\usepackage[colorlinks,bookmarks=false,citecolor=blue,linkcolor=red,urlcolor=blue]{hyperref}

\makeatletter
\def\NAT@sort{\z@}   
\makeatother

\begin{document}

\title{Ion Coulomb crystals: an exotic form of condensed matter}

\author{Giovanna Morigi}
\affiliation{Theoretische Physik, Universit\"at des Saarlandes, D-66123 Saarbr\"ucken, Germany}
\author{John Bollinger}
\affiliation{Time and Frequency Div., National Institute of Standards and Technology, Boulder, Colorado, USA}
\author{Michael Drewsen}
\affiliation{Department of Physics and Astronomy, University of Aarhus, Denmark}
\author{Daniel Podolsky}
\affiliation{Department of Physics, Technion, Haifa 32000, Israel}
\author{Efrat Shimshoni}
\affiliation{Department of Physics, Bar-Ilan University, Ramat-Gan 52900, Israel}
\date{\today}

\begin{abstract}Ion Coulomb crystals are ordered structures formed by laser-cooled ions in traps that are characterized by interparticle distances of several micrometers and energy scales on the order of $\mu$eV. Their crystalline structure emerges from the interplay between Coulomb repulsion and the external confining potential, which can be readily tuned. Moreover, individual ions can be precisely manipulated with lasers and imaged via resonance fluorescence. These unusual and unique properties make ion crystals a powerful platform for studying phases of matter in the strongly correlated regime and at low temperatures where their dynamics is manifestly quantum mechanical.
 This review examines the theoretical framework and experimental characterization of ion Coulomb crystals from a condensed-matter perspective. We discuss their dynamical and thermodynamic properties in one, two, and three dimensions, and review recent investigations into their out-of-equilibrium behavior. We provide outlooks on future directions for exploring novel condensed matter phenomena with trapped ion crystals, as well as for exploiting these features for scientific and technical applications.
\end{abstract}

\maketitle

\tableofcontents

\section{Introduction} 
\label{Sec:1}

In a seminal article from 1934, Eugene Wigner predicted that conduction electrons in a metal at low densities can crystallize  \cite{Wigner:1934}. This is a peculiar situation in condensed matter, where otherwise crystallization is expected at high densities, and it results from the decay of the Coulomb interactions as the inverse of the interparticle distance $a$. For this reason, in an unscreened environment the Coulomb interactions scale with the charge density $n\sim a^{-3}$ as $E_{\rm pot}\sim n^{1/3}$.  By comparison, the typical scale of the kinetic energy in a degenerate free electron gas is $E_{\rm kin}\sim\frac{h^2}{2ma^2}\sim n^{2/3}$~\cite{Kittel:1991}.  As a result, at low densities, the Coulomb interactions are dominant over the kinetic energy, leading to crystallization.  
More generally, at low densities the Wigner crystal phase is favored regardless of the particle statistics, since the overlap of single-particle wave functions is small.

Wigner crystallization, here broadly defined as the formation of ordered arrays of particles that interact through Coulomb repulsion, can occur in a variety of systems.
It is expected to occur in the interior of white dwarfs and in the outer crust of neutron stars~\cite{VanHorn:1991}.  It has been experimentally observed with electrons trapped on the surface of liquid helium~\cite{Grimes:1980}, and in low-dimensional solid state devices including GaAs/GaAlAs heterojunctions \cite{Andrei:1988} and carbon nanotubes \cite{Shapir:2019}. The stability of Wigner crystalline islands formed in quantum dots was analyzed in  \cite{Shklovskii:1998}; Wigner molecular crystals have been reported in semiconductor moiré superlattices \cite{Yannouleas:2007,Li:2024}. Ordered arrays of charged, solid particles have been experimentally studied with dusty plasmas \cite{Thomas:1994,Piacente:2004b,Liu:2005} and colloidal suspensions of polystyrene spheres in water~\cite{VanWinkle:1986,Bechinger:2001}.  Wigner crystallization is being pursued through the formation of ultracold neutral plasmas of atomic ions immersed in a background of photo-ionized electrons~\cite{Bergeson:2019,Langin:2019}.  In many of these examples, the Coulomb repulsion of the charged objects is screened by a polarizing medium in which the objects are suspended.

Wigner crystallization has also been observed in the laboratory through laser cooling of atomic and molecular ions stored in radiofrequency (rf) Paul traps \cite{Diedrich:1987,Wineland:1987,Drewsen:1998}, in Penning traps \cite{Gilbert:1988}, and in ion storage rings \cite{Birkl:1992}.  From the point of view of many-body physics, Wigner crystals of trapped ions are unusual examples of solid phases with interparticle distances of $\sim5-10$ micrometers and energy scales below $\sim1 \,\mu$eV ($\sim$10 mK).  Crystallization is the result of the interplay between the Coulomb repulsion and the trap potential, and their intrinsic properties are significantly influenced by the confining environment. The ions interact via the unscreened long-range Coulomb repulsion, giving rise to dynamical and thermodynamical properties different from the ones of typical crystals in condensed matter. This unconventional condensed phase is the focus of the present review.

Wigner crystals of trapped ions, often dubbed ion Coulomb crystals, have been acquiring increasing importance for quantum technological applications \cite{Haffner:2008,Blatt:2012,Bruzewicz:2019,Monroe:2021} thanks to advances in cooling and trapping \cite{Eschner:2003,Johanning:2009,Schneider:2012}. These advances rely on the detailed knowledge and experimental control of the microscopic dynamics of these strongly-correlated phases of matter \cite{Johanning:2009,Schneider:2012}. This review reports the progress on the characterization of the equilibrium and out-of-equilibrium dynamics of ion Coulomb crystals, which has enabled this progress and has, in turn, unveiled the properties of an exotic form of condensed matter.

Our starting point is the knowledge summarized in the review by Dubin and O’Neil \cite{Dubin:1999} on the plasma phase and its transition to crystallization. We set our focus on the crystallized phase and discuss the insights gained over two decades of theoretical and experimental investigations. 
The content is complementary to recent overviews on ion Coulomb crystals~\cite{Drewsen:2015,Thompson:2015} and a wide-ranging review on strongly coupled Coulomb systems~\cite{Mihalcea:2023}.

\subsection{Historical framework and conditions for crystallization}
In 1977 Malmberg and O'Neil explained how collections of a single species of trapped charges (ions or electrons)  are ideal one-component plasmas ~\cite{Malmberg:1977}. One-component plasmas (OCP) have been the subject of detailed theoretical investigations for understanding  correlations in a diverse range of physical systems~\cite{Brush:1966,Baus:1980,Ichimaru:1982,Ichimaru:1987}. They consist of a collection of identical point charges that interact through Coulomb repulsion and are immersed in a uniform neutralizing background of given density $n_0$, which is assumed to be continuous and non-polarizable. 
In equilibrium, the point charges match their density to that of the neutralizing background. Malmberg and O'Neil showed that the forces that confine ions in a Penning trap are mathematically equivalent to a uniform, neutralizing background.  More specifically, in a frame rotating with the collection of ions (or electrons) confined in a Penning trap (see Sec.~\ref{sec:Penning trap}), the static thermodynamic properties of the trapped ions are the same as those for an OCP. The argument can be straightforwardly extended to ions confined in radio frequency traps in the so-called pseudo-potential approximation (see Sec.~\ref{sec:rf trap}).  

The thermodynamic properties of the classical OCP are determined by a single dimensionless parameter, typically called the Coulomb coupling constant $\Gamma$, defined by 
\begin{equation}
\label{eq:Gamma}
    \Gamma \equiv \frac{1}{4\pi\epsilon_0}\frac{Q^2}{a_\mathrm{WS}k_\mathrm{B}T} \,.
\end{equation}
Here, $\epsilon_0$ is the permittivity of the vacuum, $Q$ is the charge and $T$ is the temperature of the ion, $k_\mathrm{B}$ is Boltzmann's constant, and $a_\mathrm{WS}$ is the Wigner-Seitz radius, which is determined by the neutralising background density $n_0$ through the relation
\begin{equation}
    \frac{4}{3}\pi a^3_{\rm WS} = 1/n_0\,.
    \label{eq:WS_radius}
\end{equation}
The coupling constant $\Gamma$ is a measure of the ratio of the Coulomb potential energy between nearest neighbor ions (or more generally point charges) to the average kinetic energy of an ion and is sufficient to determine the phase of a OCP in three dimensions.
Plasmas with $\Gamma >1$ are called strongly coupled.  For a bulk (that is, infinite) OCP, the onset of fluid-like behavior, characterized by short-range oscillations in the pair correlation function, is predicted at $\Gamma \sim 2$.  These oscillations become more pronounced with increasing $\Gamma$ and the system is predicted to undergo a first-order, liquid-solid phase transition to a body-centered cubic lattice at a critical $\Gamma$~\cite{Pollock:1973,Slattery:1982,Ogata:1987,Dubin:1990}. The best estimates of $\Gamma_\text{crit}$ lie in the range $\Gamma_\text{crit} \approx 173\text{--} 175$~\cite{Farouki:1993,DeWitt:1999,Dubin:1999,Potekhin:2000}. 

The classical OCP in two dimensions consists of systems of point charges that interact with the $1/r$ Coulomb repulsion but are restricted to move on a plane in a neutralizing background. The coupling constant $\Gamma$ is now determined using the Wigner-Seitz radius in two dimensions, given by $a_{\mathrm{WS,2D}}=(\pi n)^{-1/2}$  where $n$ is now the number of point charges per unit area.
Here, a liquid-solid phase transition at $\Gamma \approx
131(7)$ has been measured~\cite{Grimes:1980} in agreement with theory~\cite{Gann:1979}. 
We note that the Mermin-Wagner theorem \cite{Mermin:1966,Hohenberg:1967}, that prohibits full-fledged crystallization in two and lower dimensions, does not apply for the unscreened Coulomb repulsion, where the energy is non-additive and the interactions stabilize the ordered structure against fluctuations \cite{Campa:2009}.

In ion Coulomb crystals the kinetic energy can be reduced through laser cooling, thus increasing the coupling constant $\Gamma$ to achieve crystallization. The dimensionality of the crystal is determined by the relative strength of the confinement of the trap in different directions.  In this way ordered structures can be realized that are one-, two-, or three-dimensional. Moreover, the transition between structures of different dimensions can be experimentally studied by adjusting the relative strength of the confining potential in different directions.  See Secs.~\ref{sec:linear-zigzag} and ~\ref{sec:1_3_instability} for a detailed discussion.

The temperature required for crystallization can be estimated from Eq.\ \eqref{eq:Gamma}.  We take the three-dimensional case and use typical trapped ion densities of $10^{15}\:\text{m}^{-3}$ and $T=1\:\text{K}$ as a reference, 
\begin{equation}
    \Gamma \approx 2.69 \: Z^2 \left( \frac{n_0}{10^{15} \:\text{m}^{-3}} \right)^{\frac{1}{3}} \left( \frac{T}{1\,\text{K}} \right)^{-1}\,,
\end{equation}
where $Z$ is the charge (in units of elementary charges) of the trapped ion. For singly charged ions ($Z=1$), the critical coupling parameter $\Gamma_\text{crit} \approx 174$ for the predicted liquid-solid phase transition for a bulk OCP is obtained with temperatures of $\sim15\:\text{mK}$, which is more than one order of magnitude higher than the limit of conventional laser cooling techniques (so-called Doppler laser cooling, see Sec.~\ref{Sec:laser-cooling}). 

We emphasize that, due to the long-range nature of the Coulomb interactions, boundary effects can affect significantly the resulting spatial correlations. For example, theoretical calculations indicate that observing a body-centered cubic structure predicted for the bulk OCP may require $N>10^5$ ions~\cite{Dubin:1989,Hasse:1991,Tan:1995}; see 
Sec.~\ref{Sec:ground state config}.

Temperatures close to the quantum zero-point motion can be reached with more refined laser cooling techniques, see Sec.\ \ref{Sec:laser-cooling}. In this regime the crystals' vibrations are quantum mechanical, but the ground state configuration is classical to a very good approximation since the typical ion densities are far from the conditions at which quantum fluctuations can melt the crystal. Below we provide an estimate for the density required to melt an ion Coulomb crystal and note that this estimate is closely related to the problem of tunneling in a two-ion crystal chain~\cite{Yin:1995}. Denoting by $\omega_\mathrm{p} = \left(\frac{Q^2 n_0}{\epsilon_0 m}\right)^{\frac{1}{2}}$ the ion plasma frequency, an approximate estimate of the zero-point energy of an ion localized in the crystalline phase is $\hbar\omega_\mathrm{p}/2$. Quantum melting is expected to occur when
\begin{equation}
    \frac{\hbar\omega_\mathrm{p}}{2} \gtrsim \frac{1}{4\pi \epsilon_0} \frac{Q^2}{a_\mathrm{WS}},
    \label{eq:ocp_melt}
\end{equation}
where the right side of inequality provides an estimate of the energy barrier impeding an ion from hopping to the neighboring sites of the Wigner crystal. Equation~(\ref{eq:ocp_melt}) sets an upper bound on the mean interparticle distance (here expressed in terms of $a_\mathrm{WS}$) for the ion crystal to melt:
\begin{equation}
    a_\mathrm{WS} \lesssim \frac{\hbar^2}{mQ^2/(4\pi\epsilon_0)}\,.
    \label{eq:aWS_melt}
\end{equation}
Maximum densities that can be achieved in current rf or Penning traps are less than $10^{16}\:\text{m}^{-3}$, resulting in Wigner-Seitz radii $a_\mathrm{WS}$ that are orders of magnitude larger than the criteria set by Eq.~\eqref{eq:aWS_melt}.  Recent proposals and experimental efforts for unveiling manifestations of the quantum statistics of trapped ions utilize highly symmetric ion crystals in a ring-trap geometry to attempt to realize an effective particle exchange operator, thereby measuring the symmetry of the wave function~\cite{Noguchi:2014,Li:2017,Roos:2017,Urban:2019}.

\subsection{Ion Coulomb crystals in a laboratory}
\label{sec:I.B}

It is not possible to disentangle any discussion on ion Coulomb crystals from their environment, the ion trap. Confining one-component plasmas in a laboratory requires one to either apply oscillating electric fields or to combine static electric and magnetic fields. Equilibrium, in the thermodynamic sense, can be challenging to achieve (see Sec.~\ref{sec:Equilibrium and temperature}). Below we review the basics of Paul (or rf) and Penning traps, which are the most frequently used experimental platforms. 

\subsubsection{rf trap \label{sec:rf trap}}
According to Earnshaw’s theorem \cite{Earnshaw:1842} it is impossible to confine charged particles in vacuum exclusively by static electric fields generated from voltages applied to external electrodes. However, by introducing time-varying electric fields, it is possible to realize spatially localized minima of the effective potential energy. For atomic and not very heavy molecular species, fields in the radio frequency (rf) regime are most often applied to achieve strong confining forces as well as deep effective trap potentials, as we detail below. Hence, such traps are often referred to as rf traps.

\begin{figure}
\includegraphics[width=0.45\textwidth]{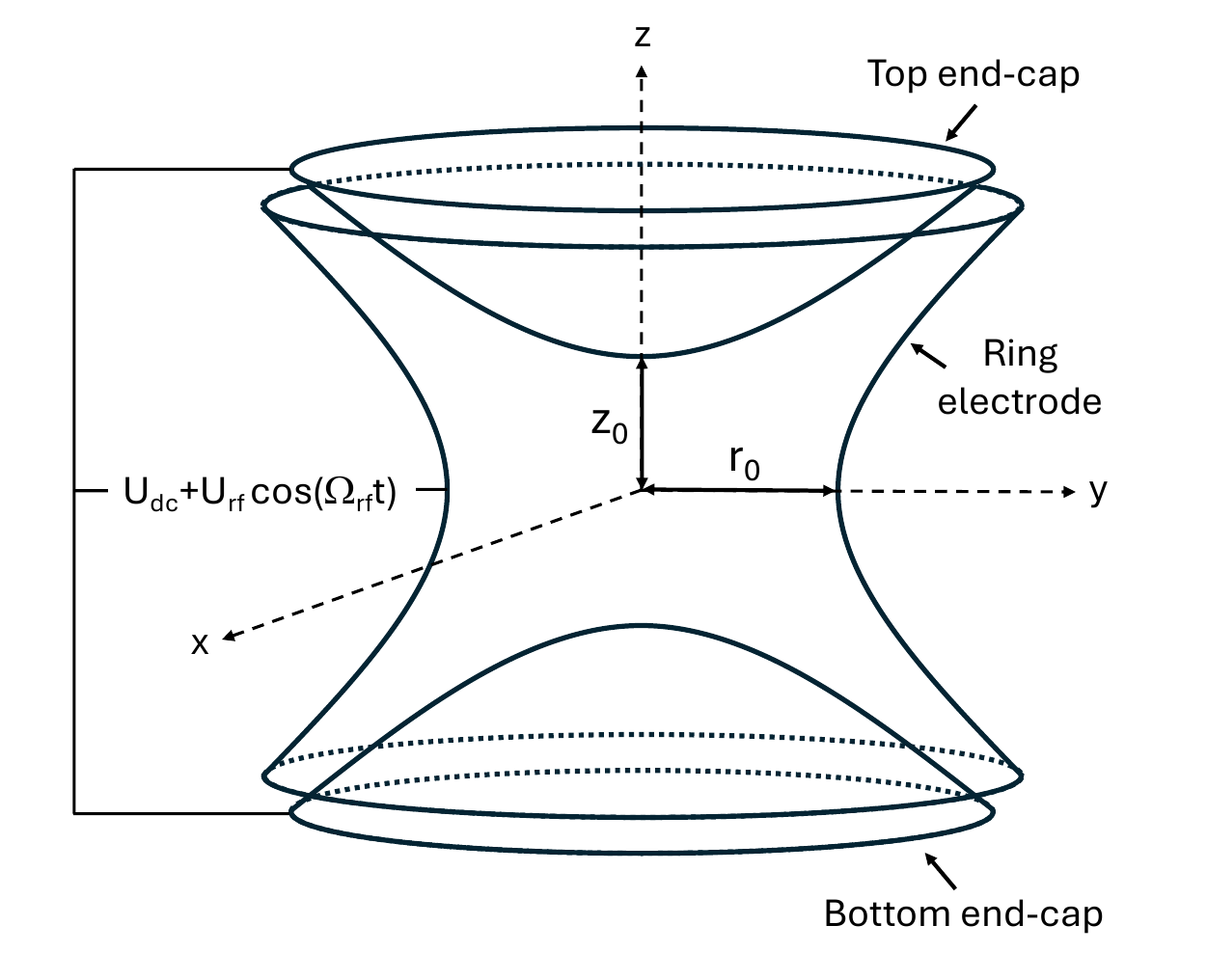} 
\caption{Schematics of the standard three-dimensional rf trap (Paul trap). The trap consists of a central ring electrode (named ``ring'' in the figure), and two so-called end cap electrodes (``Top/Bottom end-cap''). All three electrodes have hyperboloid-shaped surfaces surrounding the trap center. They produce an electrical quadrupole field when the voltage difference $\it{U_{\rm dc}+U_{\rm rf}}\cos(\Omega_{\rm rf}t)$ is applied between the ring and the end caps. The strength of the quadrupole field depends on the distances $\it{r_0}$ and $\it{z_0}$. A perfect quadrupole field is generated when $\it{r_0}$= $\sqrt{2}\it{z_0}$. \label{fig:Paul_trap_sketch}}
\end{figure}

\begin{figure}
\includegraphics[width=0.48\textwidth]{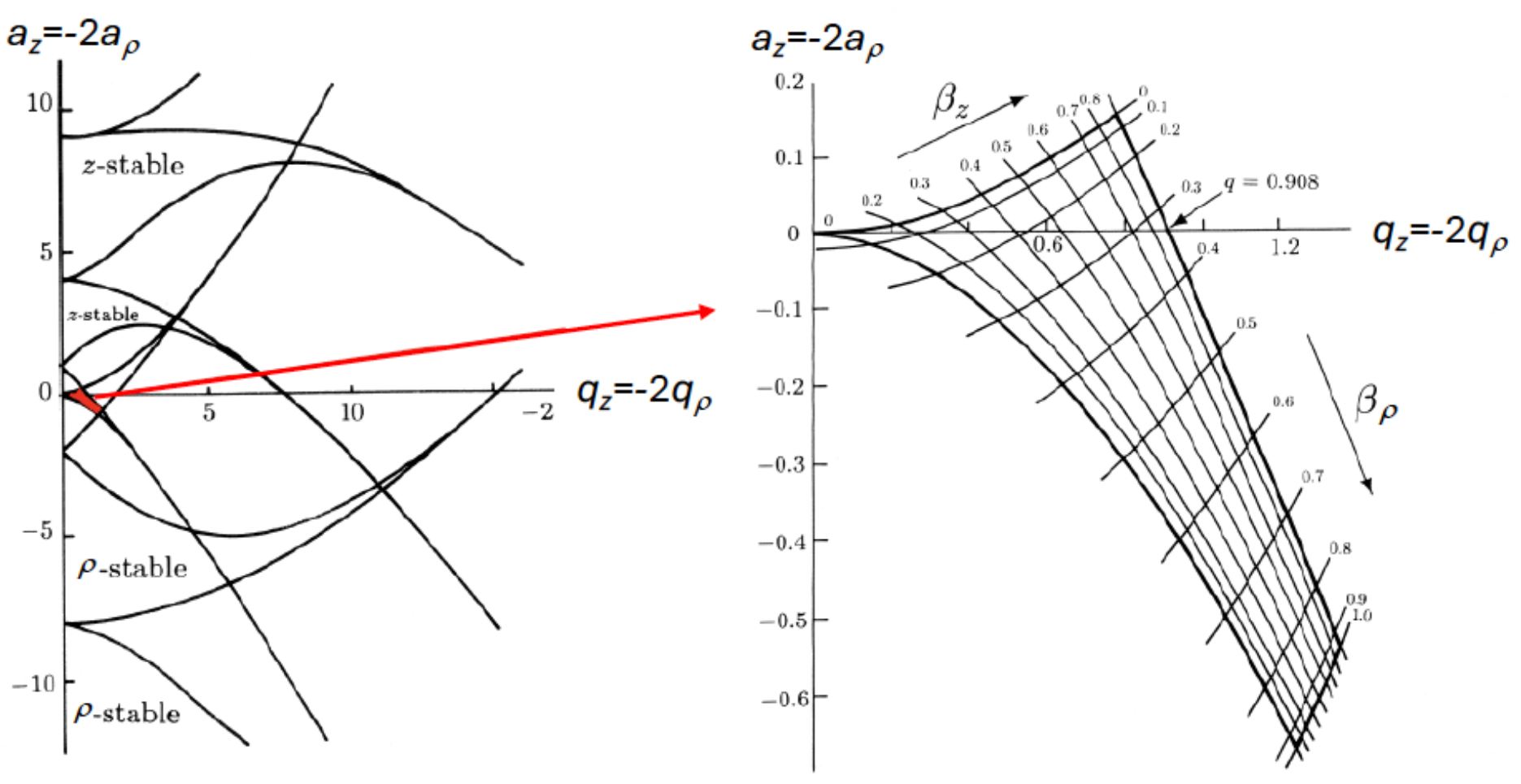} 
\caption{Stability diagram for standard rf trap. Left: Confined motion of a charged particle can only happen when its stability parameters $a_{z,\rho}$ and $q_{z,\rho}$ lie within common areas of stable motion ($z$- and $\rho$-stable). Right: Zoom on the largest area of common stability (marked in red in left panel). The parameters $\beta_{j}$ represent the ratio between the main motional frequencies of the trapped particle along the $j$ direction ($j=z,\rho)$ and the rf frequency $\Omega_{\rm rf}/2$. Reproduced from~\onlinecite{Gosh:1995} 
 \label{fig:Paul_trap_stability_diagram}}
\end{figure}

\begin{figure}
\includegraphics[width=0.45\textwidth]{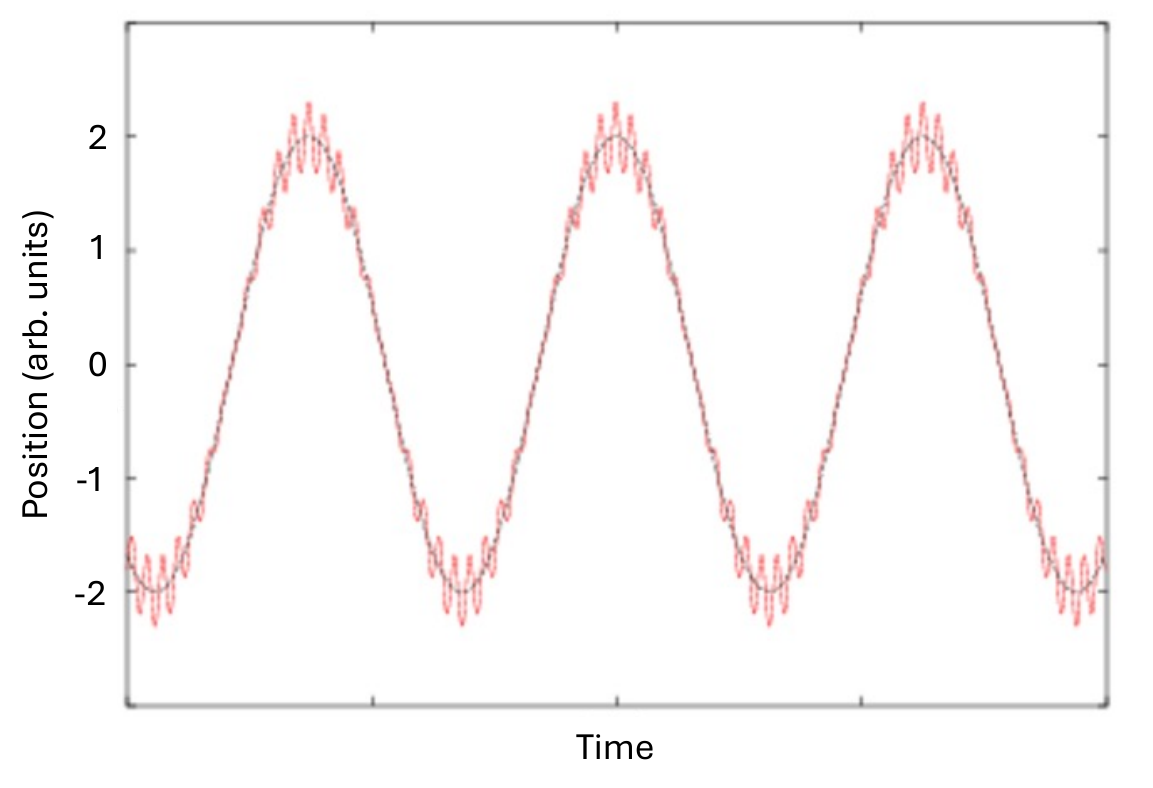} 
\caption{(Color online) Motion of a trapped particle in an rf trap. The plot shows the trajectory of a trapped ion along one of the principal directions ($x, y, x$) for $a_z=0.0$ and $q_z=0.3$. The oscillatory  red (solid) curve represents the full solution to Eq. (\ref{DiffEq_u1}), while the dark blue (dashed) curve is the solution obtained by using the harmonic pseudopotential of Eq. (\ref{eq:pseudopot}). The difference between the two curves is the so-called micromotion at the oscillation frequency $\Omega_{\rm rf}$.}      
\label{fig:ion_motion}
\end{figure}

The standard rf trap is named after Nobel Prize Laureate Wolfgang Paul, who invented the principle for two-dimensional confinement \cite{Paul:1953}. It is sketched in Fig.~\ref{fig:Paul_trap_sketch} and is based on three electrodes machined to have three-dimensional hyperbolic surfaces in order to create the perfect boundary conditions for an electrical quadrupole field with respect to the line of cylindrical symmetry of the electrodes ($z$-axis in Fig.~\ref{fig:Paul_trap_sketch}).
With reference to the parameters defined in Fig.~\ref{fig:Paul_trap_sketch}, the time-dependent electrical potential in cylindrical coordinates reads
\begin{equation}
  \phi\left(z,\rho,t\right) =  \frac{1}{2}\left(U_\mathrm{dc}+U_\mathrm{rf} \cos\left(\Omega_\mathrm{rf} t\right)\right) \,\frac{2z^2-\rho^2}{r_0^2}\,,
\end{equation}
where $\rho=\sqrt{x^2+y^2}$ is the distance from the symmetry axis. Newton’s equations of motion for a single charged particle read
\begin{equation}
	\frac{d^2u}{d\tau^2} + \left(a_u - 2 q_u\, \cos\left(2\tau \right)\right)\,u=0, \quad (u = z,\,\rho)\,,
    \label{DiffEq_u1}
\end{equation}
with the dimensionless parameters 
\begin{align}
&\tau = \frac{\Omega_\mathrm{rf}}{2}t \\
&a_z =-2a_\rho  = \frac{4 Q U_\mathrm{dc}}{m\,z_0^2\Omega_\mathrm{rf}^2} \\
&q_z=-2q_\rho  = \frac{2 Q U_\mathrm{rf}}{m\,r_0^2\Omega_\mathrm{rf}^2} \,.
\end{align}
Equations \eqref{DiffEq_u1} are examples of the so-called Mathieu Equations. They can lead to spatially confined motion whenever the parameters $a=|a_u|$ and $q=|q_u|$ lie within the areas of combined stability as illustrated in Fig.~\ref{fig:Paul_trap_stability_diagram}.
The motion within this so-called stability diagram is generally rather complex. However, for $a,q\ll 1$ the solution to the equation of motion can be well approximated by the trajectory
\begin{equation}
\label{eq:trap_ut}
	u(t) = u_0\left(1 - \frac{q_u}{2}\cos\left(\Omega_\mathrm{rf} t\right)\right) \cos\left(\omega_{u} t\right) \,,
\end{equation}
with $\omega_u$ given by
\begin{equation}
\label{eq:omegau}
  \omega_u = \frac{1}{2}\sqrt{a_u+\frac{q_u^2}{2}}\Omega_\mathrm{rf} \,.
\end{equation}
These solutions describe harmonic motion at frequencies $\omega_{z,\rho}$ superimposed with a much faster quiver motion at $\Omega_\mathrm{rf}$. The quiver motion is often referred to as $micromotion$ and has a small amplitude proportional to the distance of the harmonic (secular) motion from the origin (See Fig.~\ref{fig:ion_motion}). In many experimental situations, one can neglect the micromotion and is essentially left with pure harmonic motion. In such situations, the term proportional to $q_u$ in Eq.\eqref{eq:trap_ut} can be discarded and one assigns a potential energy to the trap, often referred to as the $\it{pseudopotential}$, which reads
\begin{equation}\label{eq:pseudopot}
	\phi_{\rm{pseu}}(z,\rho) = \frac{1}{2} m (\omega_{z}^2 z^2 + \omega_{\rho}^2 \rho^2)\,.
\end{equation} 

\begin{figure}
\includegraphics[width=0.45\textwidth]{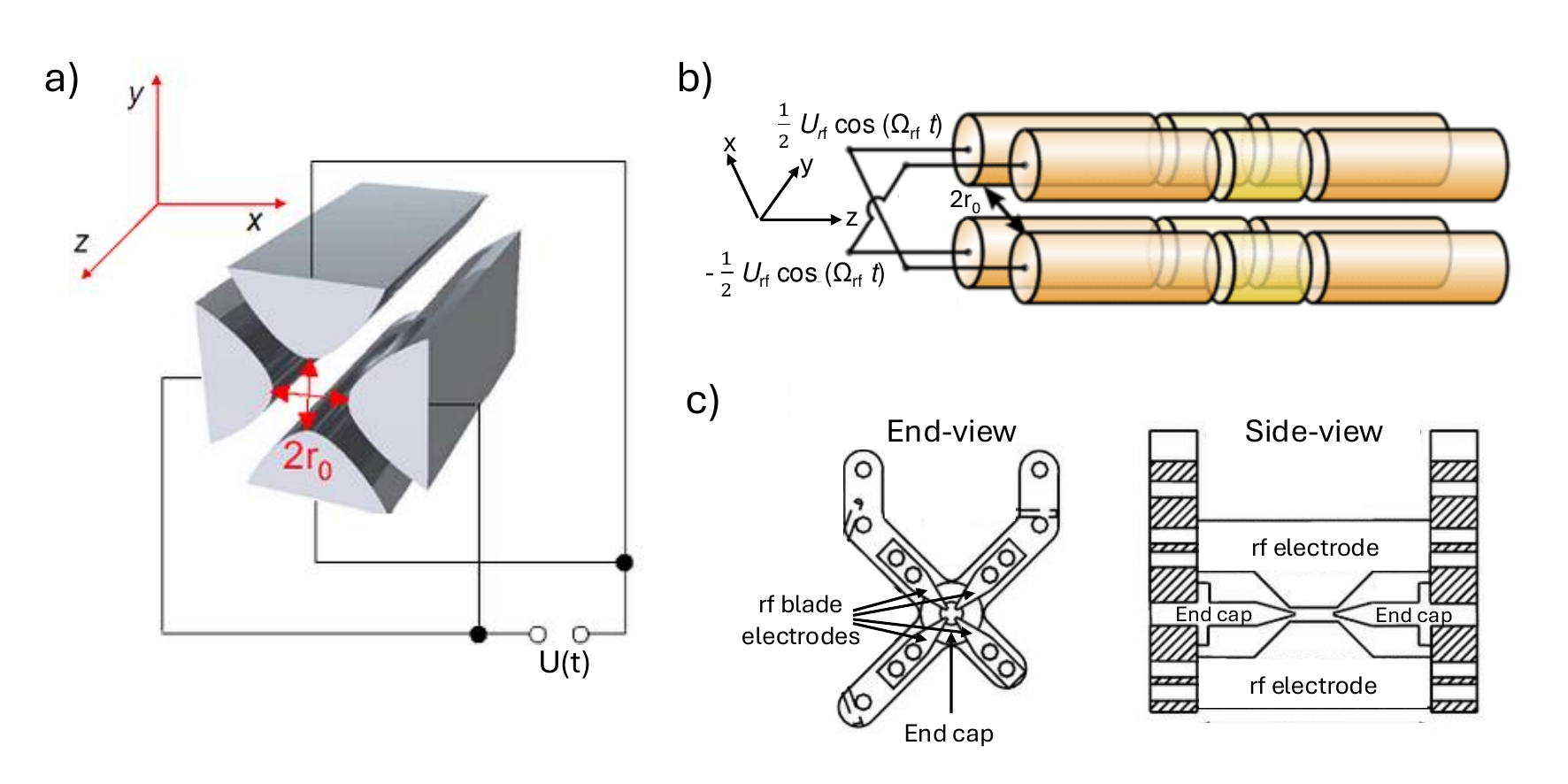} 
\caption{Linear trap geometries. a) Illustration of an ideal electrode geometry with hyperbolic-shaped electrodes to achieve two-dimensional trapping in the $x-y$ plane by a perfect electrical quadrupole field. This ideal geometry is often used in quadrupole mass spectrometers. b) An example of a three-dimensional linear rf trap with cylindrically-shaped electrodes. Here, one realizes a confinement in the $x-y$ plane like in $a)$ by applying the same rf voltage to each of the three sections of the electrodes. Axial confinement along the $z$-axis is achieved through application of the same dc voltage $U_{\rm end}$ to all the orange-colored electrode sections. c) Linear trap based on rf electrodes shaped like blades. This design creates a smaller effective value of $r_0$, and hence achieves higher trap frequencies in the $x-y$ plane at a given rf voltage (see Eqs. (\ref{eq:omegau}) and (\ref{eqs:stabaq1})-(\ref{eqs:stabaq4}). The built-in anharmonicity of this geometry eventually limits its use to one-dimensional ion Coulomb crystals or to small structures of two or three dimensions. Reproduced from \onlinecite{Splatt:2009}, \onlinecite{Poulsen:2012} and \onlinecite{Schindler:2013}.} 
 \label{fig:linear_rf_traps}
\end{figure}

The types of rf traps most used in experiments with ion Coulomb crystals are linear quadrupole rf traps \cite{Prestage:1989}. Ideally, a linear quadrupole rf trap consists of four rods with hyperbolic cross-sections arranged in a parallel configuration to create a perfect two-dimensional quadrupole field for confinement in the cross-section plane (See Fig.~\ref{fig:linear_rf_traps} a.). Along the rod axis, confinement is obtained by applying DC voltages either to two additional electrodes placed along the central axis between the rods or to the eight end-pieces after sectioning the rods into three pieces  (See Figs.~\ref{fig:linear_rf_traps}b and c).         

For the trap geometry of Fig.~\ref{fig:linear_rf_traps}, the equations of motion of a charged particle have the same form as Eq.\ (\ref{DiffEq_u1}), but now with $u = (x,\,y,\,z)$ and 
\begin{align}
&a_z = \frac{4 Q \kappa U_\mathrm{end}}{m\,r_0^2\Omega_\mathrm{rf}^2}\label{eqs:stabaq1} \\
&a_x   = \frac{4 Q (U_\mathrm{dc}-\kappa U_\mathrm{end}/2)}{m\,r_0^2\Omega_\mathrm{rf}^2}\label{eqs:stabaq2} \\
&a_y  = \frac{4 Q (-U_\mathrm{dc}-\kappa U_\mathrm{end}/2)}{m\,r_0^2\Omega_\mathrm{rf}^2}\label{eqs:stabaq3} \\
&q_x =-q_y  = 2\,\frac{Q U_\mathrm{rf}}{m\,r_0^2\Omega_\mathrm{rf}^2}\label{eqs:stabaq4} \\
&q_z =0.\label{eqs:stabaq5} 
\end{align}
When a voltage $U_{\rm end}$ is applied to all eight end-electrodes, $U_\mathrm{\rm dc}$ is applied to all sections of the rods in the $x-z$ plane and $-U_\mathrm{\rm dc}$ is applied to all sections of the rods in the $y-z$ plane. $\kappa$ is a geometrical factor, which depends on the length of the central section of the rods. 

\begin{figure}
\includegraphics[width=0.45\textwidth]{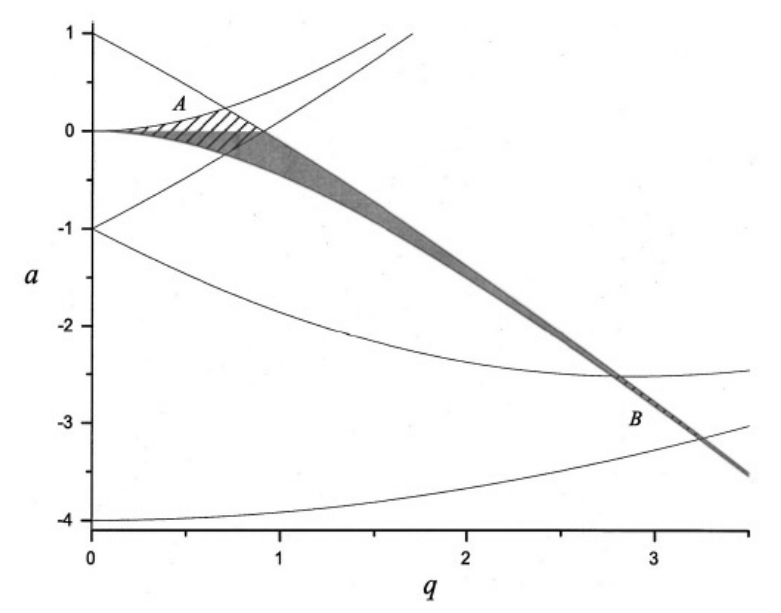} 
\caption{Stability diagram for a linear rf trap. The grey shaded area represents the stability region for the specific case of $U_\mathrm{dc}$=0, for which $a_x=a_y=a$ according to Eqs. (17) and (18), and $q_x=-q_y=q$. The area of the stability diagram is significantly 
larger than for the two-dimensional confinement of the quadrupole mass filter (the hatched areas A and B) and of the three-dimension stable trapping are of the standard rf trap (see Fig.~\ref{fig:Paul_trap_stability_diagram}). Reproduced from \onlinecite{Drewsen:1999}
 \label{fig:linear_trap_stability_diagram}}
\end{figure}

For $U_\mathrm{\rm dc}$=0, the stability criteria for the $x$ and $y$ motion become identical, and one can construct a stability diagram for $a=a_x=a_y=a_z/2$ and $q$ as presented in Fig.~\ref{fig:linear_trap_stability_diagram}. In this case when $a$, $q$ $\ll$1, one can establish the same approximate solutions to the equation of motion as Eqs. \eqref{eq:trap_ut} and \eqref{eq:omegau} for the standard rf trap, as well as define a pseudopotential of the form shown in Eq.\eqref{eq:pseudopot} with $\rho = \sqrt{x^2+y^2}$ and $\omega_{\rho}=\omega_{x}=\omega_{y}$.

The fact that all quadrupole traps close to the center give rise to an effective harmonic potential is a very important feature. Only such potentials will lead to a uniform particle density of three-dimensional Coulomb crystals, like traditional crystal structures known from solid state physics. In this case, for the linear rf trap presented above, the ion number density is given by
\begin{equation}
	n_0 = \frac{\epsilon_0 U_\mathrm{rf}^2}{m r_{0}^4\Omega_\mathrm{rf}^2}\,.
    \label{eq:n_0_rf_trap}
\end{equation}
For trap parameters $U_\mathrm{rf}\sim 100-1000\,\mathrm{V}$, $r_0\sim 1-5\,\mathrm{mm}$, and $\Omega_\mathrm{rf}\sim2\pi \times 10\,\mathrm{MHz}$, one typically reaches ion densities $n_0\sim10^{14}-10^{15} \,\mathrm{m}^{-3}$. 

The linear rf trap has two obvious advantages over the standard rf trap when it comes to experiments with ion Coulomb crystals. First, this trap configuration is more open for introducing laser beams for carrying out laser cooling needed for ion crystallization (see Sec.\ \ref{Sec:laser-cooling}). Second, the whole line defined by the $z$-axis is free of rf fields. 

Confinement of ions in higher order rf traps leads to new possibilities and challenges. A good review on trapping of charged particles in multipole rf traps is given in Ref.\,\cite{Gerlich:1992}. Rf traps of order 2$l$ leads in general to effective pseudo potential of order 2$l$-2, and hence to a much shallower potential than the harmonic potential, as well as to less micromotion at positions close to the trap center. This latter feature has been exploited for decades in experiments involving buffer gas cooling of dilute ensembles of molecular ions \cite{Gerlich:1995}. In the context of Coulomb crystallization multipole traps are expected to realize three-dimensional crystals with non-uniform ion densities, see the theoretical studies of \cite{Champenois:2009,Champenois:2010} and the experimental realization of \cite{Okada:2009}. The current lack of high quality experimental results is mainly due to the limitations in
the ability to operate perfect multipole traps. Lower order perturbations significantly distort the effective potential with the result of more local effective potential minima \cite{pedregosa:2018}. It is still unclear what is the effect of the anharmonicity in the effective potential on rf induced heating for dense ion ensembles.

\subsubsection{Penning trap\label{sec:Penning trap}}

\begin{figure}
\includegraphics[width=0.35\textwidth]{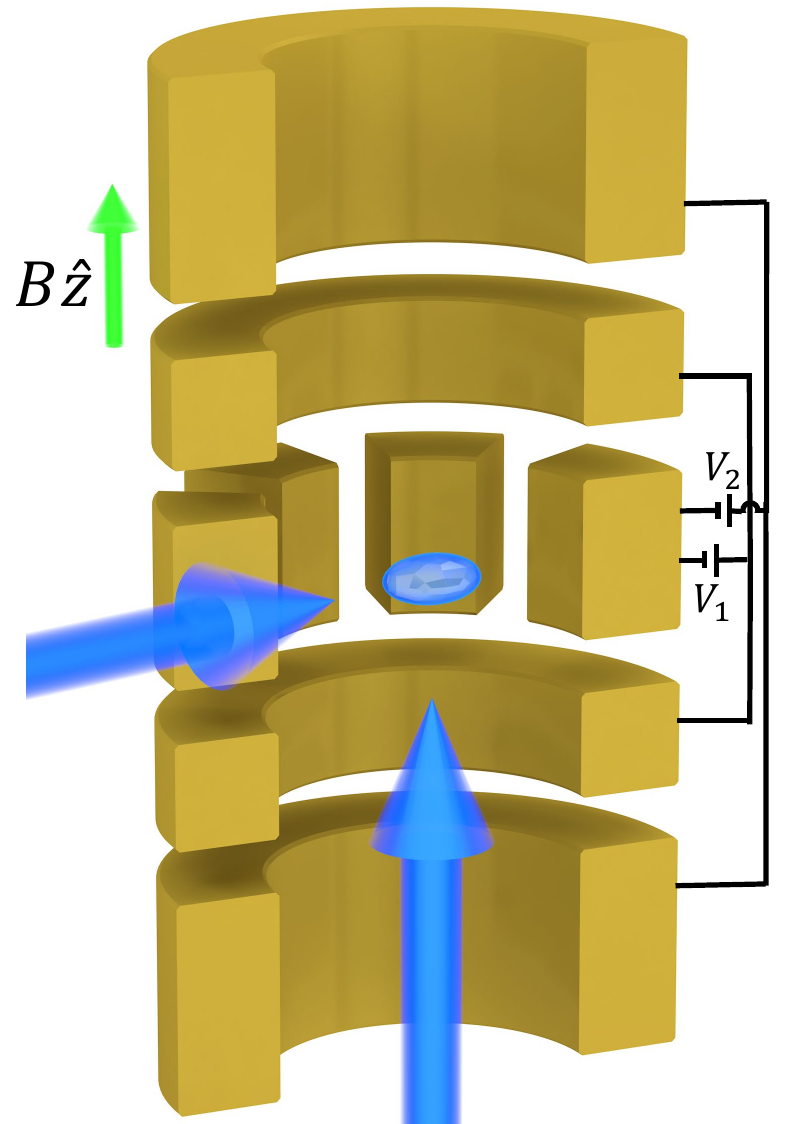} 
\caption{Cross-sectional sketch of a Penning trap that employs cylindrical electrodes, similar to that used in~\cite{Britton:2012}.  The thick arrows represent Doppler laser cooling beams that cool ion motion parallel and perpendicular to the magnetic field.  The spheroid they are pointing at is an ion crystal in the center of the trap. For trapping of positively charged ions, the outer electrodes are biased with a positive potential with respect to the central electrode, typically with $V_2 > V_1 >0$. The central electrode is segmented to enable the application of a rotating wall potential. See the text. Not shown are light collection optics that image ion fluorescence both parallel and perpendicular to the magnetic field. \label{fig:Pening_trap_sketch}}
\end{figure}

In contrast to rf traps, Penning traps (and the closely related Penning-Malmberg traps) employ both static electric and magnetic fields to confine charged particles \cite{Dubin:1999,Vogel:2018}.  A set of electrodes, frequently a co-axially aligned array of cylindrical electrodes as sketched in Fig.~\ref{fig:Pening_trap_sketch}, generates an electrostatic trap potential $\phi_T$ that has symmetry about the common axis of the cylinders, denoted by the $\hat{z}$-axis.  With a positive potential applied to the end cylinders relative to the central cylinder, $\phi_T$ provides axial confinement for positively charged ions.  However, the Laplace equation ${\nabla}^2 \phi_T = 0$ implies the electric field near the center of the Penning trap is anti-confining in the radial direction.  Radial confinement is then obtained by immersing the trap in a strong uniform magnetic field ${\mathbf B}=B\hat{z}$ directed along the symmetry axis of the trap.

In more detail, ions in the trap undergo an ${\mathbf E}\times{\mathbf B}$ drift, which, due to the radial electric field of the trapping potential, is a rotation of the ion crystal about the symmetry axis of the trap.  Rotation in the presence of the strong axial magnetic field produces a Lorentz force directed radially inward, providing radial confinement.  Theoretical and experimental work shows that for a trap with axial symmetry and in the absence of external forces, a collection of charged particles confined in a Penning trap evolves to a thermal equilibrium state characterized by rigid rotation \cite{Brewer:1988,Driscoll:1988,Dubin:1999,Davidson:2001}.  Throughout this manuscript we use $\omega_r$ to denote the rotation frequency of the ion crystal. Note that the energy of an ensemble of ions in a Penning trap is lowered when the ions in the crystal move radially outward. Radial confinement in a Penning trap is therefore not an energetic confinement.  It can be understood in terms of the conservation of 
the canonical angular momentum about the symmetry axis of the trap \cite{Dubin:1999}. This motivates the construction of Penning traps with good axial symmetry.

Penning traps have been employed for mass spectroscopy and precision measurements in atomic physics for many decades.  Typically the number of trapped charges is modest and the characteristic size of the collection of charged particles is small compared to the dimensional size of the trap electrode structure.  In this case the trap potential $\phi_T$ can be expanded in a Taylor series about the center of the trap,
\begin{equation}
    Q\phi_T(\rho,z) \approx \frac{1}{2} m{\omega_z}^2 \left( z^2-\frac{\rho^2}{2} \right) \:,
    \label{eq:quadratic}
\end{equation}
where $z$ and $\rho$ are axial and cylindrical coordinates, $m$ is the mass and $Q$ is the charge of the ion. The trap frequency $\omega_z$ characterizes the strength of the axial confinement: it scales as the square root of the potential difference between the end and central electrodes and inversely with the size of the trap.  For a trap with two end electrodes and one central electrode,
\begin{equation}
    \omega_z=\sqrt{\frac{QV_0}{m d_T^2}} \:,
\end{equation}
where $V_0$ is the voltage applied between the end and central cylinders and $d_T$ is a characteristic distance describing the trap. The volume over which Eq.~(\ref{eq:quadratic}) is a good approximation can be increased by employing trap electrodes with hyperboloid shapes~\cite{Brown:1986}.

It is frequently convenient to describe the dynamics of an ion crystal in a Penning trap in a frame rotating with rotation frequency $\omega_r$ of the ion crystal.  In this rotating frame, the radially deconfining trap potential of Eq.~(\ref{eq:quadratic}) turns into an effective, radially confining  potential
\begin{equation}
     Q\phi_R(x,y) = \frac{1}{2} m{\omega_z}^2 \left( z^2+\beta \rho^2 \right) \:,
     \label{eq:rotating_frame}
\end{equation}
where $\beta = \frac{\omega_r\left(\Omega_c-\omega_r \right) }{\omega_z^2} -\frac{1}{2}$.  Here, $\Omega_c = \frac{QB}{m}$ is the ion cyclotron frequency and the term proportional to $\omega_r \Omega_c$ is due to the Lorentz force generated by the rotation of the ion crystal in the presence of the strong magnetic field. The parameter $\beta$, sometimes called the trap anisotropy parameter, characterizes the relative strength of the radial confinement compared to the axial confinement.

So-called Penning-Malmberg traps \cite{deGrassie:1977,deGrassie:1980} are employed in non-neutral plasma physics for studying non-equilibrium dynamics and the transport of particles and energy across magnetic field lines, and for the long-term confinement of anti-matter \cite{Murphy:1992,Andresen:2011,Gabrielse:2012}. Here, the central cylindrical section of the trap (the segmented central electrode in Fig.~\ref{fig:Pening_trap_sketch}) is a cylinder whose length is large compared to its diameter. In Penning-Malmberg traps the number of trapped charged particles can be large ($> 10^8$), filling a volume where Eq.~(\ref{eq:quadratic}) is no longer a good approximation.  For a discussion of other configurations of trap electrodes, see \cite{Vogel:2018}.

The formation of ion Coulomb crystals in the laboratory requires low temperatures ($T\lesssim 10$~mK).  At these low temperatures the density $n_0$ of an ion crystal in a Penning trap is constant and determined by the rotation frequency (SI units),
\begin{equation}
    n_0 = \frac{2\epsilon_0}{Q^2}m\omega_r \left( \Omega_c -\omega_r \right) \:,
    \label{eq:3Ddensity}
\end{equation}
where $\epsilon_0$ is the permittivity of the vacuum.  In contrast to rf traps, this result is independent of the details of the trapping potential $\phi_T$.  Specifically, even if the trapping potential (Eq.~(\ref{eq:quadratic})) contains higher order terms ($z^4$, $\rho^4$, $z^2\rho^2$, $\ldots$), the three-dimensional density of the ion crystal is constant and given by Eq.~(\ref{eq:3Ddensity}). The density is maximized at $\omega_r = \Omega_c/2$, known as the Brillouin density limit~\cite{Brillouin:1945}. Typical densities $n_0$ demonstrated in the laboratory range between $10^{13}$~m$^{-3}$ to $10^{16}$~m$^{-3}$~\cite{Huang:1998b,Dubin:1999}.

To date, the formation of Coulomb ion crystals has been demonstrated with modest numbers of ions ($<10^6$) \cite{Tan:1995,Itano:1998,Mavadia:2013,Ball:2019} where the dimensional size of the ion crystal is small compared to the size of the trap electrodes.  In this case the quadratic trapping potential of Eq.~(\ref{eq:quadratic}) is a good approximation, resulting in some simplifications. In particular, bounded equilibrium crystals (corresponding to $\beta>0$ in Eq.~(\ref{eq:rotating_frame})) are obtained for rotation frequencies in the range
\begin{equation}
    \omega_m < \omega_r < \Omega_c -\omega_m \:,
    \label{eq:magnetron}
\end{equation}
where $\omega_m = \frac{\Omega_c}{2}-\sqrt{(\frac{\Omega_c}{2})^2 - \frac{\omega_z^2}{2}}$ is usually called the magnetron frequency. It is the frequency at which a single trapped ion undergoes a circular motion about the center of the trap due to ${\mathbf E}\times{\mathbf B}$ drift. In addition, in the limit that the uniform density ion crystal can be treated as a uniformly charged fluid (valid for ion crystals with dimensions more than a few interparticle spacings), the low temperature equilibrium shape of the fluid (or crystal) is a uniform density spheroid with an aspect ratio $\alpha = Z_p/R_p$ determined by $\omega_r$, $\Omega_c$ and $\omega_z$ through the parameter $\beta$. Here, $2Z_p$ is the axial extent and $2R_p$ the diameter of the spheroid. The detailed functional dependence is discussed in Refs.~\cite{Bollinger:1993,Dubin:1999} and is in good agreement with experimental measurements~\cite{Brewer:1988,Bollinger:1993}.

\begin{figure}
\includegraphics[width=0.5\textwidth]{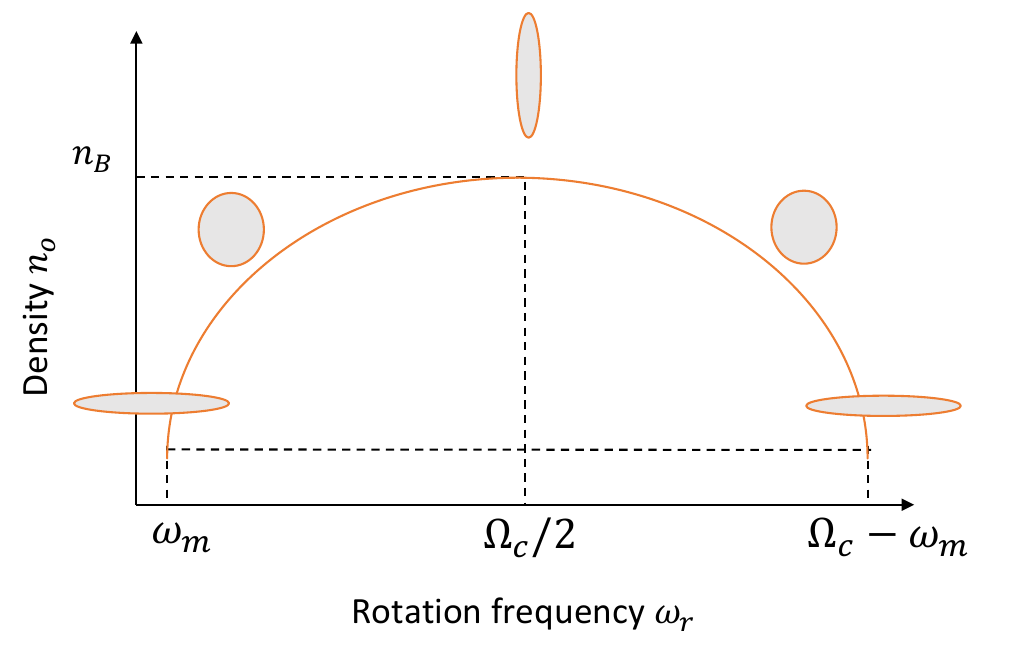} 
\caption{Simple qualitative sketch of the ion number density $n_0$ and different spheroidal crystal shapes obtained as a function of the ion crystal rotation frequency $\omega_r$ in a Penning trap. The maximum density $n_B$, called the Brillouin limit, occurs for $\omega_r = \Omega_c/2$. \label{fig:Pening_crystal_shapes}}
\end{figure}

Figure~\ref{fig:Pening_crystal_shapes} illustrates the qualitative dependence of the ion crystal density and shapes obtained as a function of the rotation frequency $\omega_r$ for fixed $\omega_z$ and $\Omega_c$. For $\omega_r$ slightly greater than $\omega_m$, the crystal consists of a single plane corresponding to $\alpha\approx 0$.  As $\omega_r$ is increased, the radius of the ion crystal decreases due to the increasing radial confinement and at some rotation frequency ions move out of the $z=0$ plane, forming a 3-dimensional crystal. Further increases in $\omega_r$ increases the aspect ratio $\alpha$ as well as the ion density $n_0$.  The largest $\alpha$ and highest density $n_0$, known as the Brillouin limit, is obtained for $\omega_r= \Omega_c/2$.  For $\omega_r >\Omega_c/2$, the centrifugal acceleration weakens the radial confinement, resulting in a decreasing ion crystal aspect ratio and ion density.  For ion crystals with small numbers of ions, the radial confinement with $\omega_r\sim\Omega_c/2$ can be made sufficiently strong compared to the axial confinement that the ions form a 1-dimensional crystal along the symmetry axis of the trap \cite{Mavadia:2013}.  These structural phase transitions (from two dimensions to three dimensions \cite{Mitchell:1998} and from three dimensions to one dimension \cite{Mavadia:2013}) have been observed in the lab.

Precise control of the ion crystal rotation frequency $\omega_r$ is important for controlling and maintaining a stable crystal whose shape and density are constant in time. Without this control $\omega_r$ will slowly decrease in time due to the influence of torques from small asymmetries in the confining fields of the trap. In the laboratory $\omega_r$ is precisely controlled through the use of a rotating electric field, frequently called a rotating wall. Rotating wall electric fields were first investigated by the non-neutral plasma physics community \cite{Huang:1997} and then borrowed by the atomic physics ion trap community where its use with low temperature ion crystals enables precise, phase-locked control of the ion crystal rotation \cite{Huang:1998a,Huang:1998b}. Rotating electric fields of different azimuthal symmetry can be employed. However, a frequently used geometry is the rotating quadrupole potential,
\begin{equation} \label{eq:quad_wall}
    Q\phi_W \left(\rho, \phi, t \right) = \frac{1}{2}m\omega_W^2 \rho^2 \cos{\left[ 2\left(\phi+\omega_r t\right)\right]} \:,
\end{equation}
where $\phi$ is the azimuthal angle in the laboratory frame. In a frame rotating clockwise about the $\hat{z}\text{-axis}$ at a frequency $\omega_r$, Eq.~(\ref{eq:quad_wall}) describes a static quadrupole potential 
\begin{equation} \label{eq:quad_wall2}
     Q\phi_W = \frac{1}{2}m\omega_W^2 \left( x_{\text{rot}}^2 - y_{\text{rot}}^2 \right) \:.
\end{equation}
Here, the frequency $\omega_W$ characterizes the strength of the quadrupole potential and $x_\text{rot}$ and $y_\text{rot}$ are rotating frame coordinates along the strong and weak axes of the rotating quadrupole potential. The rotating quadrupole potential breaks the symmetry of the trap potential and changes the shape of the ion crystal, providing a restoring force that controls the azimuthal orientation of the ion crystal in the frame of the rotating potential.

\subsection{Laser cooling}
\label{Sec:laser-cooling}

The low temperature required for the formation of the ion Coulomb crystal is typically reached by means of laser cooling \cite{Chu:1998,Cohen-Tannoudji:1998,Phillips:1998}. This mechanism is based on scattering of laser light and uses the mechanical effects of radiation to transfer energy from the ions' motion to the modes of the electromagnetic field. As such, the stationary state of the ions results from the dynamical equilibrium  of scattering processes that heat and that cool the motion. The width of the stationary energy distribution is determined
by the characteristic linewidth of the scattering processes and is often associated with a temperature. See Sec.\,\ref{sec:Equilibrium and temperature} for a detailed discussion. Laser cooling allows to prepare ions at temperatures ranging from hundreds of millikelvin down to fractions of a microkelvin, depending on the technique which is employed and on the ion species.

Laser cooling of ion crystals employs, typically, techniques appropriate for cooling individual ions.
In fact, multiple photon scattering events are very rare because 
the ion crystals are optically thin. In this limit, each ion of the crystal is effectively an independent scatterer \cite{Morigi:2001a,Morigi:2003b}. 
Nevertheless, there are some challenges to consider when applying laser cooling protocols to ion ensembles. One important challenge is that processes that heat in general become increasingly important with the number of ions in the crystal.  Another major challenge consists in the fact that the non-inertial forces of the trap cannot be neglected for the ions located away from the symmetry axes. This makes it difficult to ensure that the sample is uniformly cooled at the same temperature. We will discuss the implications on studies of thermalization in Sec.~\ref{Sec:3}. In what follows we provide a brief review of the state-of-the art in preparing ion ensembles at low temperatures.  For a comprehensive review of laser cooling with trapped ions, we refer the reader to~\cite{Eschner:2003}.

The most commonly used laser cooling technique is Doppler laser cooling, which can produce temperatures on the order of 1 mK or lower~\cite{Wineland:1979,Stenholm:1986,Eschner:2003}. Doppler laser cooling has been used to prepare one-dimensional (1D), linear chain crystals of up to $\sim100$ ions in linear rf traps (see Sec.~\ref{sec:the ion chain}) and two-dimensional (2D), single-plane crystals of ions of up to $\sim500$ ions in both Penning and rf ion traps (see Sec.~\ref{sec:planar_geometries}).  Much larger three-dimensional (3D) ion crystals have been formed through Doppler laser cooling in both Penning ($> 10^5$ ions) and rf ion traps ($> 10^4$ ions) (see Sec.~\ref{sec:three-dimensional crystals}).  

Once formed, 1D and 2D ion crystals are cooled to temperatures well below the mK regime by means of so-called sub-Doppler cooling techniques.  Examples of sub-Doppler cooling techniques include polarization-gradient cooling~\cite{Birkl:1994,Joshi:2020,Li:2022}, sideband cooling~\cite{Diedrich:1989,Monroe:1995} and Electromagnetically Induced Transparency (EIT) cooling~\cite{Morigi:2000,Roos:2000}.  For completeness we also mention protocols for simultaneously cooling the vibrational modes of an ion chain to the ground state that are based on designing the chain spectrum by means of an external field~\cite{Wunderlich:2005,Fogarty:2016}.  These sub-Doppler cooling techniques can prepare the crystal close to the zero-point motion. Sideband cooling is routinely used to sequentially cool the vibrational modes of an ion chain in a linear Paul trap to near the quantum mechanical ground state state~\cite{Eschner:2003}.  
Moreover, in Penning traps, small crystals with less than 10 ions have been successfully sideband-cooled~\cite{Stutter:2018}.

EIT cooling has been demonstrated to be particularly efficient for cooling crystals of ions to the ground state, thanks to the Fano-like profile of the scattering processes~\cite{Morigi:2003}, which permits to cool simultaneously several vibrational modes of the crystal~\cite{Schmidt-Kaler:2001}. In a linear Paul trap, ion chains ranging from a few ions~\cite{Lin:2013} to the transverse motion of a chain of 18 ions~\cite{Lechner:2016} and 40 ions~\cite{Feng:2020} have been EIT cooled to near the ground state. EIT cooling has also been employed to simultaneously cool the out-of-plane modes of single-plane crystals to near the ground state of the vibrational motion in Penning traps ~\cite{Jordan:2019,Shankar:2019} and rf traps~\cite{Kiesenhofer:2023,Guo2024}. 

While 1D and 2D crystal formation has been limited to a few hundred ions, Doppler laser cooling has been used to prepare much larger, mesoscopic, 3D crystals in both Penning and rf traps. Laser cooling may play a role in determining the size of ion crystals that can be produced in the lab.  Below we discuss experimental demonstrations of some of the largest trapped ion crystals that have been prepared in the lab. 

Penning traps employ static confining fields and the non-thermal motion of ions in the trap is a rigid body rotation that is not readily shared with the thermal degrees of freedom.  Therefore, the implementation of Doppler cooling of rotating ion ensembles in Penning traps is relatively straightforward, although Doppler shifts due the ion crystal rotation can produce temperatures that are slightly elevated from standard Doppler cooling limits~\cite{Torrisi:2016}. Three-dimensional crystals with $N \sim 5 \times 10^5$ ions have been Doppler cooled in Penning traps to milliKelvin temperatures ($\Gamma > 200$), where a body centered cubic (bcc) lattice structure in the crystal’s interior was observed~\cite{Tan:1995,Itano:1998}.  Doppler cooling has been demonstrated to be effective for larger ensembles: Nonneutral plasmas of $N \sim 10^8$ Mg$^+$ ions in Penning-Malmberg traps have been prepared at temperatures less than 50 mK~\cite{Anderegg:2009,Anderegg:2010}. Here, the ions are strongly correlated ($\Gamma \sim 20$) but above the temperature for crystal formation.  The lack of crystal formation is not understood. Its origin is probably due to the scaling with $N$ of heating processes. 
For example, the potential energy stored in the ion crystal increases with $N$ and in Penning traps small imperfections in the construction of the trap can drive a slow radial expansion of the ion crystal that converts potential energy to kinetic energy~\cite{Dubin:1999,Jensen:2004}.

Cooling large ion ensembles in rf traps is challenging due to the time-dependent potential. In fact, the rf-driven micromoton can be converted into thermal motion by the Coulomb interactions, heating the crystal. This heating rate increases as the temperature is lowered and the number of ions increases, giving rise to chaotic dynamics and limiting the cooling efficiency~\cite{Bluemel:1989,Prestage:1991,Nam:2014,Poindron:2023}.  Therefore, the formation of large ion crystals in rf traps requires large laser cooling powers.  Despite these difficulties, large crystals composed of $10^4$ up to $10^5$ ions have been prepared in rf traps~\cite{Drewsen:1998,Hornekaer:2002,Mortensen:2006}.  It has been observed that once the ion crystal forms, the rf heating can be significantly reduced. A lower limit on the Coulomb coupling parameter of $\Gamma \gtrsim 70$  has been estimated from the observed structure of these large ion crystals in rf traps. Direct measurements of the ion temperature are challenging with 3D crystals in rf traps because of the large driven micromotion.

We conclude this section by noting that, to date, sub-Doppler laser cooling techniques have not been demonstrated on large 3D ion crystals.

\subsection{Ion trap arrays and optical dipole traps \label{sec:arrays and ODT}}

In this review we focus on ion crystals that form at low temperatures in a configuration resulting from the interplay of a common ion trap potential and the Coulomb repulsion. In this section we provide a short outlook on other kinds of trapping potentials that have been discussed in the literature and are based on different mechanisms than the ones discussed in the rest of this review. One such alternative approach consists of using  an array of trapping potentials, where each cell stores a single ion~\cite{Schmied:2009}. This approach enables the generation of arbitrary lattice geometries.  Ion trap arrays are being pursued both with rf ion traps~\cite{Mielenz:2016,Lysne:2024,Niedermeyer:2025} as well as Penning traps~\cite{Jain:2020,Jain:2024}. To date the number of ions stored in an ion trap array is small (up to $N\sim 4$).

A further approach employs optical dipole traps for ions with optical dipole transitions. A major motivation is to avoid boundary effects that are significant in Penning and Paul traps, thereby ideally realizing strings of equidistant ions. The micromotion due to the coupling of the charge monopole to the fast laser oscillations at the optical frequency is negligible and the potential the ions experience is essentially the same as the potential experienced by neutral atom scattering in the visible \cite{Cormick:2011}.
Optical trapping of ion Coulomb crystals 
is challenging because of the strength of the Coulomb repulsion of closely spaced ions, requiring large laser powers. 
Experimental demonstrations include trapping linear strings of up to 6 ions by a single beam optical dipole trap~\cite{Schmidt:2018}, and confinement of a few ions in a one-dimensional optical lattice~\cite{Hoenig:2023}. An alternative approach is to use an array of optical tweezers  \cite{Mazzanti:2021}. Recent studies indicate that this architecture could provide a scalable approach for fault-tolerant digital quantum computation and for analog quantum simulations with ions arrays \cite{Arias-Espinoza:2021,Mazzanti:2021,Schwerdt:2024}.

Somewhat less demanding in laser power is the use of optical dipole forces to modify the normal mode spectrum of trapped ion crystal. We refer the reader to Sec.\ \ref{Sec:Frustration} for an extensive discussion.

\section{ Ion Coulomb crystals: from one to three dimensions}
\label{Sec:2}

Trapped ions can crystallize in different structures, determined by the geometry of the confining potential and by their number. Linear, planar, up to cubic-centered structures have been realized in the laboratory. In this section we review the equilibrium properties of the crystals from one- to three-dimensional spatial arrangements of the ions. We start with the linear chain, on which an extensive number of studies have been performed, review then planar structures, and finally report on the crystalline properties of three-dimensional structures. Differing from previous reviews, we set the focus on the dynamical and thermodynamical properties, for the purpose of establishing a direct link with condensed-matter systems.

\subsection{Linear geometries: the ion chain}
\label{sec:the ion chain}

\begin{figure}
\includegraphics[width=0.48\textwidth]{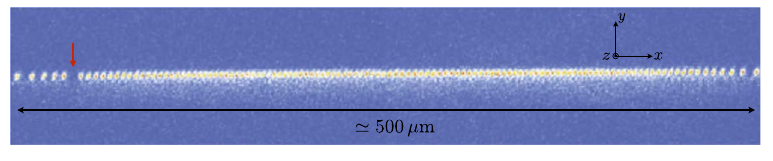}
\caption{ \label{fig:Pagano}
(Color online)  Linear chain of 121 $^{171}$Yb$^+$ ions in a cryogenic linear Paul trap with $\omega_y/2\pi = 1.5$ MHz and $\omega_x/2\pi = 35$ kHz. The red arrow marks a lone $^{171}$Yb$^+$ ion which is in a state that does not fluoresce. This dark ion retained its position after more than 4 hours, suggesting that the chain kept its ordered structure and never melted despite collisions with the background gas. From~\onlinecite{Pagano:2018}.
}
\end{figure}

Figure \ref{fig:Pagano} is the image of an experimentally realized ion chain composed of more than 100 ions \cite{Pagano:2018}. The interparticle distances are of the order of several micrometers, and the individual ions can be imaged on a CCD camera via fluorescence light emitted during laser excitation.  Ion chains typically form at Doppler cooling temperatures. First experimental demonstrations were in quadrupole traps \cite{Birkl:1992,Raizen:1992,Waki:1992}. In the meantime linear chains have also been realized in Penning traps \cite{Mavadia:2013} and ring geometries \cite{Li:2017}. The ion chain has been the workhorse of the ion-trap quantum computer for the versatility in controlling the individual components, such as the interactions between the electronic and the vibrational degrees of freedom at the quantum level \cite{Cirac:1995}.

Chains of ions are a laboratory realization of a one-dimensional Debye crystal \cite{Ashcroft:1976}. Due to the unscreened Coulomb interactions, it constitutes a peculiar one-dimensional structure. Below we review salient dynamic and thermodynamic properties, emphasizing analogies with and fundamental differences from textbook examples of condensed matter systems. 

\subsubsection{The ground state configuration}

Ion chains are the stationary configurations of anisotropic, elongated potentials. In the typical setups the ions are at distances of several micrometers and the ground state is fully characterized by the configuration that minimizes the potential energy $V$, consisting of the Coulomb repulsion and the external trap potential. For $N$ ions, the equilibrium positions are the solution of the coupled nonlinear equations $\bm{\nabla}_j V=0 $ with $\bm{\nabla}_j$ the gradient for the particle at position  ${\mathbf r}_j=(x_j,y_j,z_j)$ and $j=1,\ldots,N$.

For a linear-Paul trap with cylindrical symmetry about the chain axis, the potential takes the form
\begin{equation}
\label{Eq:potential}
    V=\frac{1}{2}\sum_{j=1}^Nm\left(\nu^2x_j^2+\nu_t^2(y_j^2+z_j^2)\right)+\frac{1}{2}\sum_{\substack{i\ne j}}^N\frac{Q^2}{|{\mathbf r}_i-{\mathbf r}_j|}
\end{equation}
where $m$ is the mass, $Q$ the charge, and $\nu$ and $\nu_t$ are respectively the trap axial frequency and the transverse trap frequencies in the pseudopotential approximation, such that $\nu<\nu_t$. The linear chain is stable for $\nu_t/\nu$ exceeding a critical value, as we will detail below.

\begin{figure}
\includegraphics[width=0.45\textwidth]{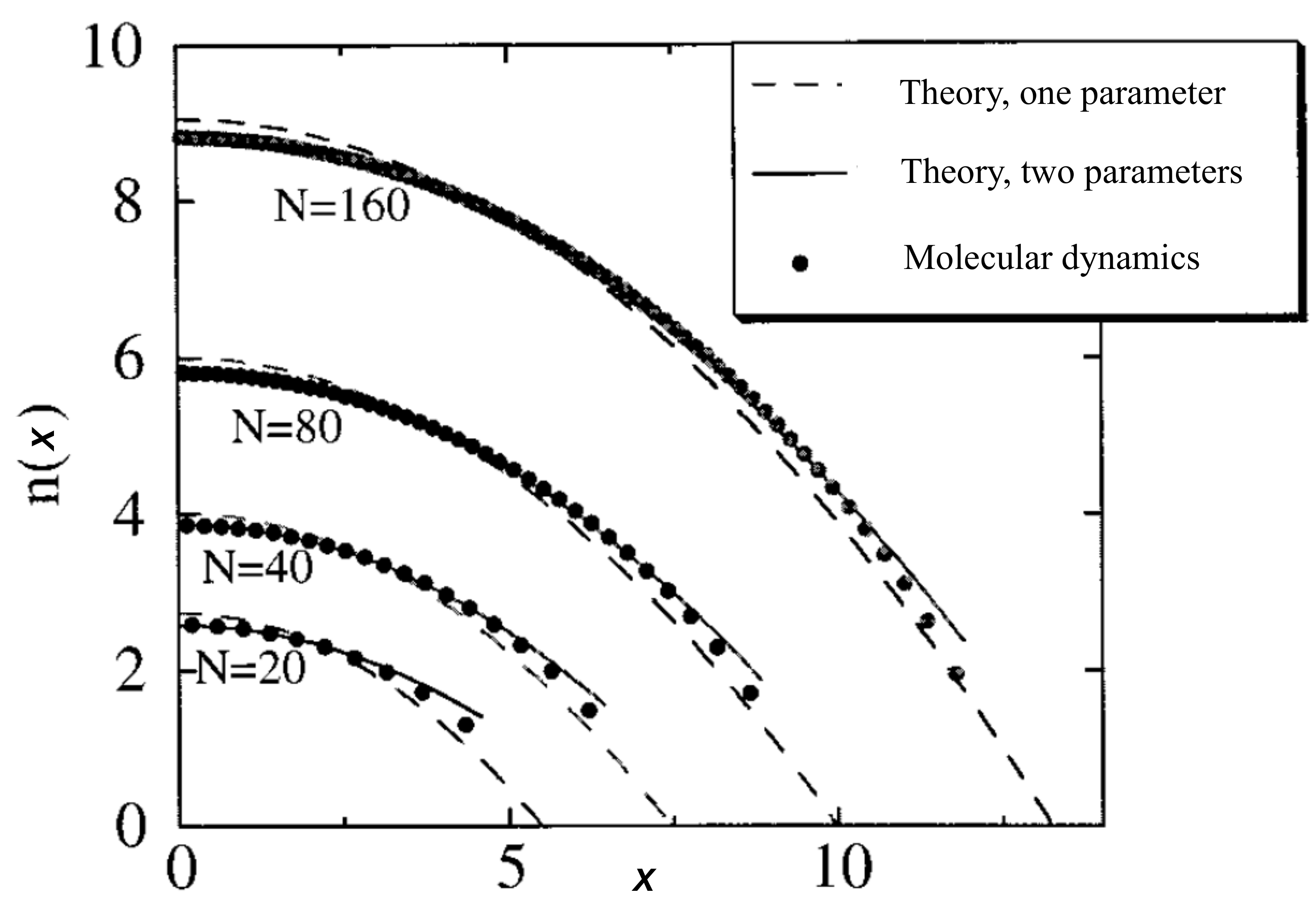}
\caption{ \label{Fig:Dubin:1997}
Density of charges per unit length, $n(x)$, as a function of position $x$ from the center of the chain for different numbers $N$ of ions in a linear Paul trap. The dots are the equilibrium positions determined by molecular dynamics simulation for a finite number of ions $N$. The lines are obtained using the method of trial variational functions with two parameters. The dashed lines are the one-parameter function of Eq.\ \eqref{Gauss}. Both $x$ and $n(x)$ are scaled to $(Q^2/m\nu^2)^{1/3}$. From~\onlinecite{Dubin:1997}.
}
\end{figure}

The equilibrium positions of the ions in a linear Paul trap are typically determined numerically \cite{Schiffer:1993,James:1998}. An approximate analytical expression can be found in the limit of large $N$ by means of the local density approximation. In this limit $n(x)$ is a smooth function of the position $x$ along the axis and takes the shape of the Thomas-Fermi distribution \cite{Dubin:1997}:
\begin{equation}
\label{Gauss}
n(x)=\frac{3}{4}\frac{N}{L}\left(1-\frac{x^2}{L^2}\right)\,,
\end{equation}
with $|x|\le L$, and vanishes otherwise. The parameter $L$ is the chain half-length at equilibrium and is determined by means of a variational ansatz, such that to leading order in an expansion in $\log N$ it reads \cite{Dubin:1997}
\begin{equation}
    L(N)=\left[N\log N(3Q^2/(m\nu^2)\right]^{1/3}\,.
\label{eq:L}
\end{equation} 
At the chain center the density varies very slowly about the value $n(0)\sim N/L$, where the interparticle distance is minimal, $a(0)=1/n(0)$. Figure \ref{Fig:Dubin:1997} shows that expression \eqref{Gauss} gives a good estimate of the charge distribution sufficiently far away from the chain edges. This expression has been experimentally validated for chains composed of more than 100 ions and has been used to determine the number of trapped ions without counting the individual components \cite{Kamsap:2017}.

Similar to solid-state systems, a thermodynamic limit is identified by scaling $N$ and $L$ leaving $n(0)$ constant. Due to the dependence of $L$ on $\nu$ and $N$, this imposes that the axial trap frequency shall scale with $N$ as \cite{Morigi:2004a} 
\begin{eqnarray}
\label{Thermodynamic:Limit}
\nu(N)\sim\omega_0\sqrt{\log N}/N\,,
\end{eqnarray}
with the reference frequency
\begin{equation}\omega_0^2=Q^2/(4\pi\epsilon_0 ma_0^3)\,,
\end{equation}
and $a_0=a(0)$ the interparticle distance at the chain center.

The above analysis ignores thermal and quantum fluctuations, which tend to suppress long-range order in one dimension. Strictly speaking, a perfectly ordered linear chain is unstable in the thermodynamic limit ($L\rightarrow\infty$). In practice, however, under suitable conditions the crystalline order can persist over sufficiently large length scales so that the chain effectively appears as fully ordered.
Stability against quantum fluctuations has been studied in \cite{Schulz:1993} by means of a treament based on bosonization, showing that at zero temperature the density-density correlations of a one-dimensional Wigner crystal decay with the distance $x$ as $\exp(-c\sqrt{\log x})$ with $c$ a constant \cite{Schulz:1993}. This is an extremely slow decay: any large (but finite) chain can be considered crystallized. Stability against thermal fluctuations requires $Q^2\log N/(4\pi\epsilon_0 a_0)\gg \kappa_BT$, namely, the potential energy scale is larger than the thermal fluctuations where $T$ is the temperature and the classical equipartition theorem has been used \cite{Morigi:2004b}. This shows that, for sufficiently large chains, the long-range Coulomb correlations dominate over thermal disorder. 

\subsubsection{Normal modes of the ion chain}
\label{Sec:NormalModes}

\begin{figure}
\includegraphics[width=0.48\textwidth]{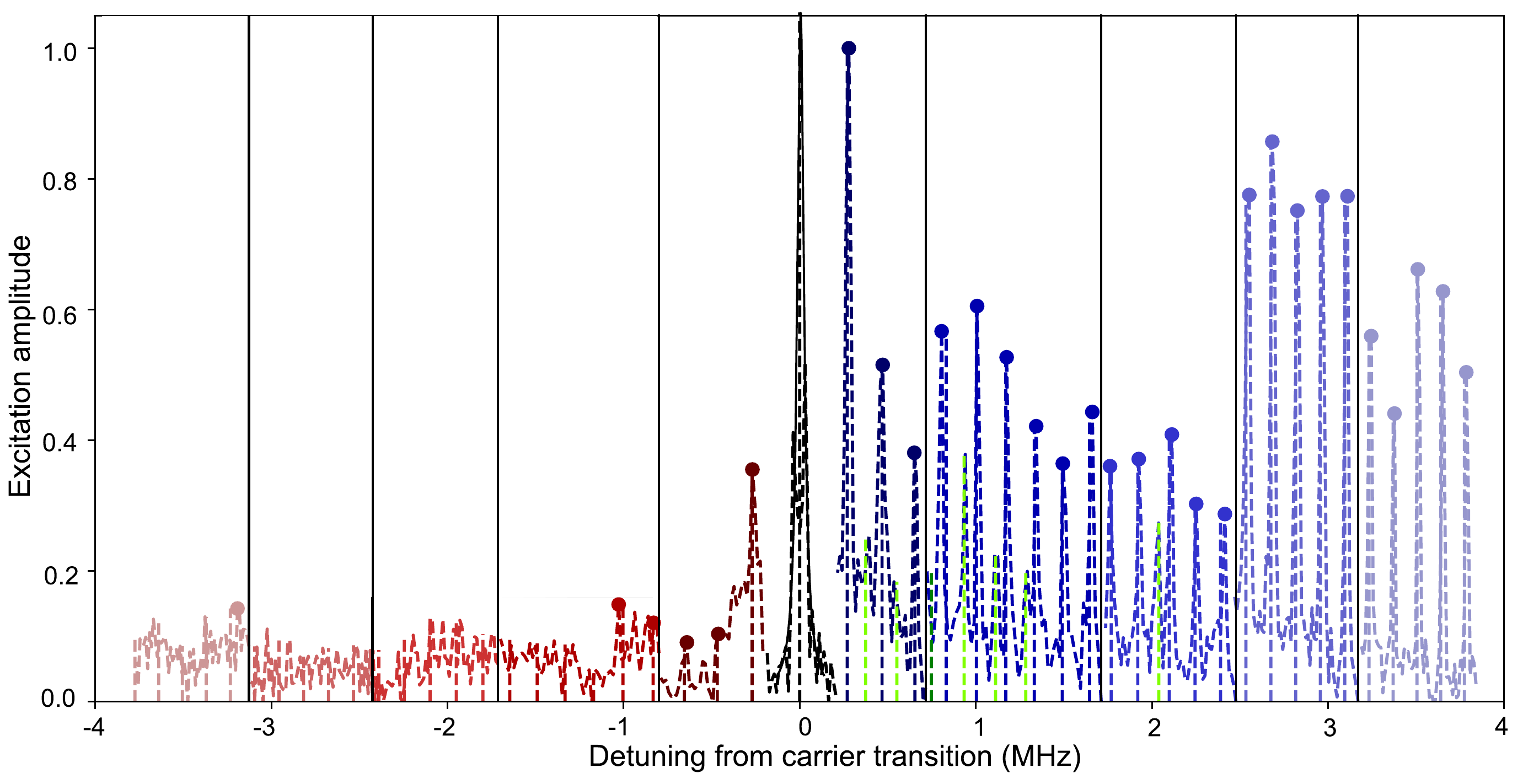}
\caption{ \label{Fig:Wu:2023}
(Color online) The frequency sideband spectrum of the axial modes of a laser cooled chain composed of 24 $^9$Be$^+$ ions. Dashed lines are the experimental
results of normalized excitation level versus frequency, where red and blue denote the scan over red-shifted sideband (anti-Stokes), carrier, and blue-shifted sideband (Stokes) transitions,
respectively. The brightness of the dashed line shows the variable probe times throughout the spectrum to ensure that all modes are excited with high visibilities. Dashed vertical blue and red (dark) lines show all 24 calculated
first-order red and blue sideband frequencies. Dashed vertical green (light) lines show the mode mixings between center-of-mass and a few higher frequency modes,
coinciding with the smaller peaks in the experimental data. Details on the figures can be found in Ref.~\onlinecite{Wu:2023}.
}
\end{figure}

At laser cooling temperatures, the ions perform vibrations about the equilibrium positions with displacements of the order of tens of nanometers, and thus about three orders of magnitude smaller than the equilibrium interparticle distance. In this regime the restoring forces are harmonic and the ion chain is a one-dimensional harmonic crystal. Figure \ref{Fig:Wu:2023} displays the Stokes sidebands, measured in the resonance fluorescence of a chain of 24 ions. The measured spectra are reproduced by theoretical models that approximate the potential of Eq.\ \eqref{Eq:potential} by only keeping the quadratic terms in the displacements about the equilibrium position. 

Below we focus on ion chains in a linear Paul trap and in the pseudopotential approximation. We note that the analysis will also apply to linear chains in Penning traps \cite{Mavadia:2013} at the so-called Brillouin limit, where the effective magnetic field is zero. We refer to Refs.\ \cite{Lin:2009,Home:2011,Cartarius:2013} for studies of the normal modes of ion chains in anharmonic traps.

In a harmonic trap axial and transverse motion are decoupled and the harmonic spectrum can be numerically determined even for relatively large chains. The set of linear differential equations are obtained by expanding the potential Eq.\ \eqref{Eq:potential}  to second order in the axial displacements from the axial equilibrium positions, $q_i=x_i-x_i^{(0)}$, and in the transverse displacements from the chain, $\zeta_i=y_i,z_i$. They read
\begin{eqnarray}
\label{Eq:ax}
&&\ddot{q}_i=-\nu^2q_i-K_0\sum_j \mathcal I_{ij}\, q_j\,,\\
&&\ddot{\zeta}_i=-\nu_t^2 \zeta_i+\frac{K_0}{2}\sum_j\mathcal I_{ij}\, \zeta_j\,,
\label{Eq:y-z}
\end{eqnarray}
where $K_0=2Q^2/(mL^3)$ scales the long-range Coulomb interaction, and $\mathcal I_{ij}$ is a $N\times N$ real symmetric matrix:
\begin{equation}
\label{Eq:I}
\mathcal I_{ij}=-(1-\delta_{ij})\frac{1}{|\xi_i-\xi_j|^3}+\delta_{ij} \sum_{l\neq i}\frac{1}{|\xi_i-\xi_l|^3}\,,
\end{equation}
with $\delta_{ij}$ Kronecker's delta. Here, $\xi_i=x_i^{(0)}/L$ is the dimensionless equilibrium position rescaled by the half-length $L$ of the chain at equilibrium. The set of differential equations \eqref{Eq:ax} and \eqref{Eq:y-z} is solved by diagonalizing the matrix $I_{ij}$, thus solving the linear set of equations 
\begin{equation}\sum_j \mathcal I_{ij} w^{(n)}_j=\lambda_n w^{(n)}_i\,,
\label{eq:I}
\end{equation}
whose eigenvalues $\lambda_n$ are real and positive. Correspondingly, the axial and transverse dispersion relations read respectively 
\begin{eqnarray}
\label{Eq:w:ax}
&&\omega^{\|\,2}_n=\nu^2+K_0\lambda_n\,,\\
&&\omega^{\perp\,2}_n=\nu_t^2-K_0\lambda_n/2\,,\label{Eq:w:tr}
\end{eqnarray}
with $n\in [1,N]$ and $\omega^{\perp}_n$ doubly degenerate in a cigar-shaped trapping potential. 
While $\omega^{\|}_n$ is always real, $\omega^{\perp}_n$ becomes imaginary when $\lambda_n\ge 2\nu_t^2/K_0$. This identifies the critical value of the transverse trap frequency $\nu_t^{(c)}$ at which the chain becomes mechanically unstable.
 
In the rest of this section we make general considerations on the structure of the eigenvalues for $\nu_t>\nu_t^{(c)}$, which are based on the treatment of Ref.\ \cite{Morigi:2004a,Morigi:2004b}. For convenience, we adopt the convention $\lambda_n<\lambda_m$ for $n<m$. With this convention the label $n$ directly gives the number of nodes $\ell=n-1$ of the corresponding eigenmode $w^{(n)}$. Since the position of the nodes is generally constrained to the position of the ions within the chain, the value of $\lambda_n$ generally depends on the number of ions $N$. There are two exceptions: (i) The eigenvalue $\lambda_1=0$, corresponding to the bulk eigenmodes where the chain oscillates as a rigid body at the frequencies of the trap, and (ii) the eigenvalue $\lambda_2=2\nu^2/K_0$, corresponding to the breathing mode, which has a single node at the chain center. These two eigenvalues are related to two symmetries of the Hamiltonian: point (i) is due to the fact that the dynamics is separable into center-of-mass and relative coordinates even in the presence of the harmonic trap, point (ii) is a consequence of the symmetry by reflection about the trap center. The breathing oscillations $\omega_2^{\|}=\sqrt{3}\nu$ and $\omega_2^\perp=\sqrt{\nu_t^2-\nu^2}$, in fact, are also characteristic of a one-component plasma in an ellipsoidal trap \cite{Dubin:1996}. 

\begin{figure}
\includegraphics[width=0.4\textwidth]{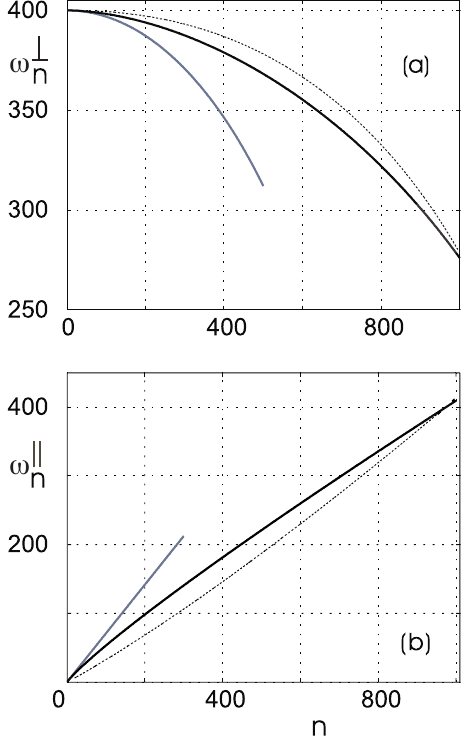}
\caption{ \label{Fig:Morigi2004}
(a) Transverse and (b) axial spectrum of eigenfrequencies
(in units of $\nu$) for a chain of $N=1000$ ions and with $\nu_t=400\nu$.
The solid line is obtained from the numerical solution of Eq.\ \eqref{eq:I},
the grey line from the spectrum of Eq.\ \eqref{Jacobi} (the grey curves
have been truncated, as they do not correctly reproduce the shortwavelength
eigenmodes). The dotted line gives the spectra evaluated
extending a method first developed for determining the density of states of a disordered chain. Adapted from Ref.~\onlinecite{Morigi:2004b}.
}
\end{figure}

Some information on the analytical behavior of the other modes can be extracted for large numbers of ions $N$. For $n\ll N$, thus for "long-wavelength" eigenmodes, the displacements $w_i^{(n)}$ can be approximated by continuous functions, $w_i\to w(\xi)$, and away from the edges the secular equation \eqref{eq:I} reduces to a differential equation to leading order in an expansion in $1/\log N$ \cite{Morigi:2004a,Morigi:2004b}:
\begin{equation}
\label{Jacobi}
\log N\left[(1-\xi^2)w^{\prime\prime}(\xi)-4\xi w^{\prime}(\xi)\right]=\lambda_nw(\xi)\,.
\end{equation}
The eigenmodes are Jacobi polynomials $w^{(n)}=P_\ell^{(1,1)}$ with $n=\ell +1$ and $\ell=0,1,\ldots$ The eigenvalues read ${\lambda}_n=\ell(\ell+3)$ and exactly reproduce the eigenvalues $\lambda_1$ and $\lambda_2$ for $\ell=0,1$, as also visible in Fig.\ \ref{Fig:Morigi2004}. Interestingly, 
the model of a plasma in a cigar-shaped trap potential gives the same expression for the axial spectrum $\omega_n^{\|}=\nu\sqrt{n(n+1)/2}$ in the limit of vanishing aspect ratio, see Ref.\ \cite{Dubin:1991}. The same plasma model, however, fails to predict the transverse oscillations of the chain since it becomes singular.

The short-wavelength spectrum is also amenable to some analytical treatment. In Ref.\ \cite{Morigi:2004b} the density of states at short wavelengths was determined by extending analytical methods, originally developed for calculating the dispersion relation of disordered chains \cite{Dyson:1953}. The same treatment allows one to determine the largest eigenvalue $\lambda_N$. At leading order in an expansion in $1/\log N$ and using 
the thermodynamic limit of Eq.\ \eqref{Thermodynamic:Limit}, one obtains $K_0\lambda_N=(9/8)\nu^2 N^2/\log N=8Q^2/(ma_0^3)=8\omega_0^2$. This shows that the frequency $\omega_0$ scales the bandwidth of the axial mode spectrum. It also scales 
the critical value of the transverse trap frequency, $\nu_t^{(c)}$, at which the chain becomes unstable. The latter is found setting $\omega^\perp_N=0$, giving $\nu_t^{(c)}=\sqrt{K_0\lambda_N/2}=2\omega_0$.
The corresponding critical aspect ratio reads $\alpha^{(c)}=(\nu/\nu_t^{(c)})=\sqrt{(16/9)\log N}/N$ \cite{Dubin:1991,Schiffer:1993,Morigi:2004b}. In Sec.\ \ref{sec:linear-zigzag} we review the structural transition at $\nu_t=\nu_t^{(c)}$ and its statistical properties. 

In the regime of stability of the linear chain, but for $\nu_t\sim 8\omega_0$, axial and transverse eigenfrequencies can be degenerate and the coupling due to anharmonicities becomes relevant. The effects of this coupling have been measured even at the level of vibrational quanta \cite{Roos:2008}. 

When instead the trap frequency $\nu_t\gg 8\omega_0$, there is a gap between axial and transverse modes. Moreover, the transverse spectrum becomes almost flat, so that the transverse dynamics is well approximated by local oscillators weakly coupled by the Coulomb interactions. This regime is at the basis of the proposals for realizing effective Hubbard models \cite{Porras:2004} and for implementing quantum information processors \cite{Serafini:2009} using ion chains.

\subsubsection{Thermodynamics of the ion chain} 

\begin{figure}
\includegraphics[width=0.43\textwidth]{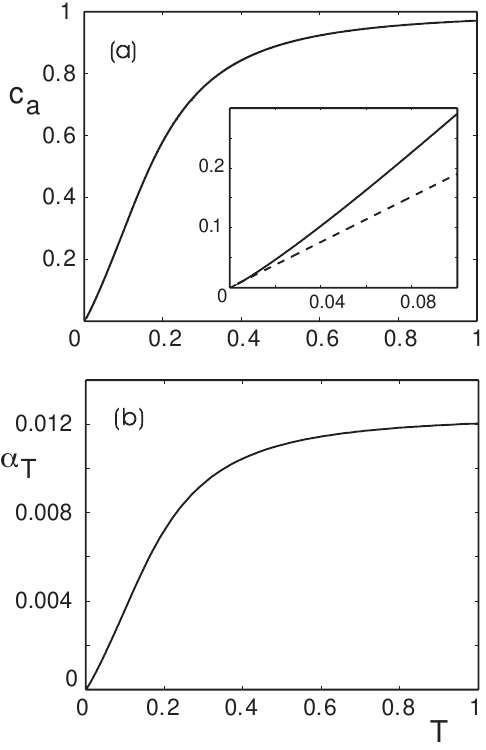}
\caption{ \label{Fig:Morigi:2004:spec-heat}
(a) Specific heat $c_a$, in units of $k_\mathrm{B}$, and (b) coefficient
of thermal expansion $\alpha_T$, in units of $\hbar\omega_N^{\|}$, as a function of the temperature $T$. The temperature is in units of the Debye temperature $\Theta_D=\hbar \omega_N^\|/k_\mathrm{B}$. 
The thermodynamic functions are calculated for $N=1000$ ions in the limit in which the transverse excitations are frozen out.  This
would correspond to an experiment with $\nu=2\pi\times 31$ kHz and $\nu_t
=2\pi \times 34$ MHz. For these parameters and Beryllium atoms, $\Theta_D\approx 20~\mu$K. The inset of (a) shows the low-temperature behavior of the specific heat. The dashed line is the
 slope extracted from a theoretical analysis which uses the density of states for $\omega_n^{\|}=\nu\sqrt{n(n+1)/2}$, see Sec.\ \ref{Sec:NormalModes}. Adapted from Ref.~\onlinecite{Morigi:2004b}.
}
\end{figure}

The thermodynamic properties of Debye crystals are textbook examples, and a linear chain provides a laboratory to verify hypotheses and predictions down to the quantum regime and even close to the absolute zero, thanks to laser cooling. Quantities such as the specific heat could be measured by a calorimetric experiment, where the temperature of the crystal is determined as a function of the energy transferred to the chain. The temperature can be extracted by resonance fluorescence, as in the protocols discussed in \cite{Rossnagel:2015,Gajewski:2022}, or by means of the so-called sideband thermometry \cite{Eschner:2003,Vybornyi:2023}. 

From the theoretical point of view, there is an essential difference between the textbook solid state crystals, where the interaction is typically nearest-neighbor, and the long-range Coulomb repulsion of ion Coulomb chains. In Ref.\ \cite{Morigi:2004a,Morigi:2004b} the influence of the Coulomb repulsion on the thermodynamic behavior has been studied by determining  the specific heat per particle and the coefficient of thermal expansion.  This work considered the case in which the motion is effectively one-dimensional, which is realized for an ion chain in a steep transverse trap potential, assuming temperatures for which the transverse vibrations are frozen out. Using the notation of the previous section, this requires that $\nu_t\gg 8\omega_0,k_\mathrm{B} T/\hbar$. 

The free energy is evaluated assuming quantized vibrations and a canonical ensemble with temperature $T$, using the scaling of Eq.\ \eqref{Thermodynamic:Limit}, which warrants a well defined thermodynamic limit. 
The free energy is a function of the number of ions $N$, of the temperature $T$, and of the volume. The latter is determined by the length $L$ from end to end, which, for fixed $N$, is controlled by the axial trap frequency $\nu$, see Eq.\ \eqref{eq:L}. The specific heat is $c_a=(1/N)\partial U_{\rm th}/\partial T|_{\nu,N}$, where $U_{\rm th}$ is the thermal energy of a quantum gas of phonons, $U_{\rm th}=\sum_{n}\hbar\omega_n^{\|}/({\rm e}^{\beta\hbar\omega_n^{\|}}-1)$ with $\beta=(k_BT)^{-1}$ the inverse temperature. 
Figure \ref{Fig:Morigi:2004:spec-heat}(a) shows that the behavior of the 
specific heat with the temperature is very similar to the one of a one-dimensional Debye chain: it saturates at large temperatures, characteristic of Dulong-Petit law, while at low temperatures $c_a$ scales linearly with $T$. Interestingly, the limit $T\to 0$ and $N\to\infty$ do not commute. By keeping $N$ constant, the proportionality coefficient in the relation $c_a=\gamma T$ is a function of $N$, namely, $\gamma\sim N/\sqrt{\log N}$~\cite{Morigi:2004b}. This deviation from extensivity is due to the long-range interactions and becomes evident in the quantum regime. Despite the non-intensive behavior of the specific heat (namely, the fact that it depends on $N$), the usual ensemble equivalence holds, because the relative energy fluctuations at low temperatures scale as $(\sqrt{\log N}/N)^{1/2}$, and hence vanish in the thermodynamic limit $N\to\infty$.

The coefficient of thermal expansion takes the form $\alpha_T =(1/L)\partial L/\partial T|_{P,N}$ and is calculated at constant pressure $P$ and number of ions $N$. Using thermodynamic relations, one can show that its temperature dependence is determined by $c_a$:
\begin{equation}
\alpha_T=\frac{3}{2}\frac{N}{L}\kappa_T c_a\,,
\end{equation}
where $\kappa_T$ is the isothermal compressibility and is independent of $T$ at leading order \cite{Morigi:2004a}. Figure \ref{Fig:Morigi:2004:spec-heat}(b) shows that the functional dependence on the temperature is similar to the specific heat. Due to the non-intensive behavior of the isothermal compressibility, the coefficient of thermal expansion is also a non-intensive function of $N$ over the whole range of temperatures. 

The non-intensive behavior of specific heat and of $\alpha_T$ is a manifestation of the long-range, Coulomb interactions in one dimension. It is typically cured by theoretical procedures, so-called Kac's scaling \cite{Campa:2009}, which in this case would consist of artificially assuming that the ions electric charge depends on $N$ in such a way to re-establish extensivity of the energy \cite{Landa:2020}. In experiments the charge is fixed and measurements at low temperatures would detect a relatively strong dependence of $c_a$ on the size of the chain, while they shall converge to a well defined, size-independent limit at relatively large temperatures.

\subsubsection{The linear-zigzag transition}
\label{sec:linear-zigzag}

We devote the rest of this section to the structural instability of the linear chain. The demonstration that this is a second order phase transition, of the Ising universality class of magnetic systems, has permitted to directly connect the physics of ion Coulomb crystals with statistical mechanics and condensed matter concepts and models. This has paved the way to a series of theoretical and experimental investigations on critical properties which we review in what follows. It has  further inspired and set the basis for subsequent works on the formation and stability of topological defects, see Secs.\ \ref{Sec:Defects}-\ref{Sec:quantum-Kink}. 

\begin{figure*}
\includegraphics[width=0.88\textwidth]{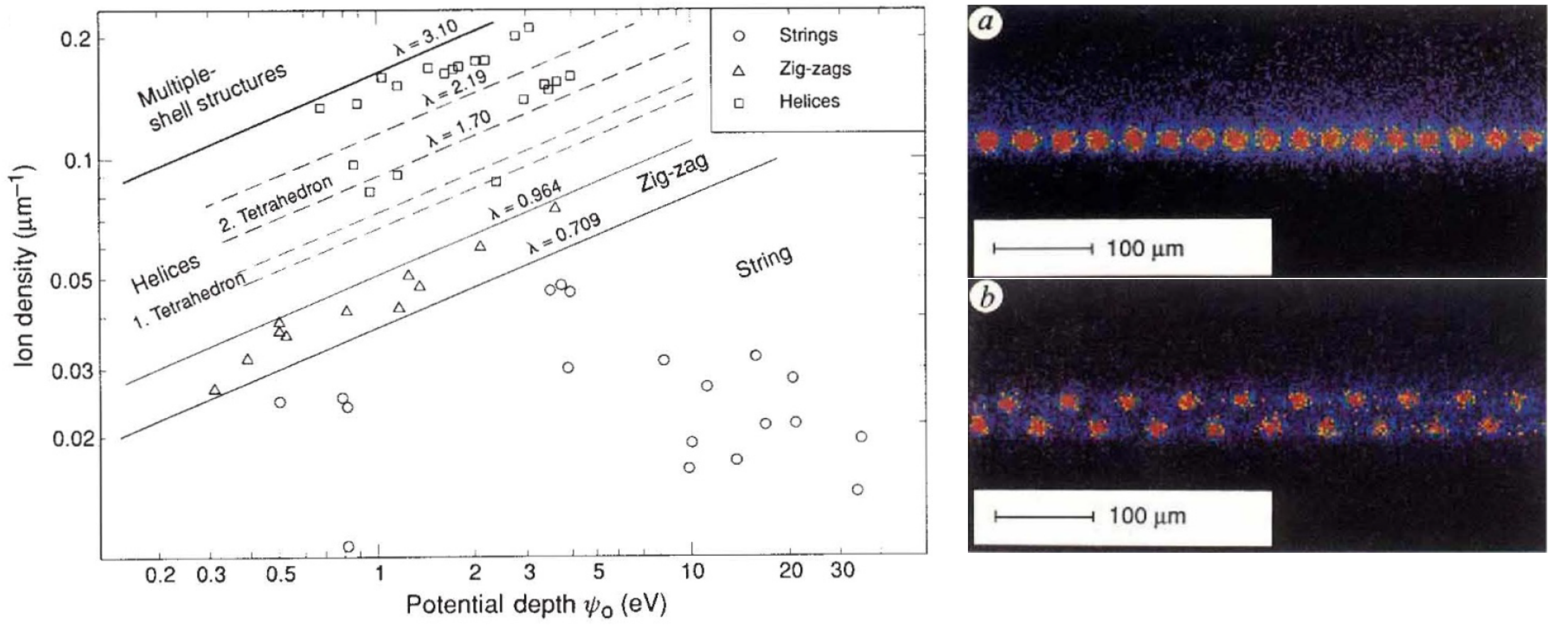}
\caption{\label{Fig:LinZigzag}
Left panel: Summary of the experimentally observed configuration as a function of the potential depth and of the linear ion density. These two parameters can be combined to give the normalized linear particle density $\lambda$, which fully determines the ion configuration (see text).  The observed configurations are labeled by different symbols for each structure. The straight lines show critical $\lambda$ values separating the regions of different theoretically expected structures, see \cite{Hasse:1990}. Right panel: images of crystalline structures of laser-cooled $^{24}$Mg$^+$ ions in a quadrupole trap. 
Adapted from \protect\cite{Birkl:1992}.}
\end{figure*}

Structural order in ion crystals is stabilized by the external trap, which compensates the repulsive Coulomb interactions. The resulting geometries are determined by the number of ions and by the trap aspect ratio. Starting from a linear chain in a cigar shaped potential, a rich phase diagram of structures was observed by decreasing the trap aspect ratio \cite{Birkl:1992,Waki:1992}, see the left panel of Fig.\ \ref{Fig:LinZigzag}. The experimental observations are in agreement with the theoretical results of Ref. \cite{Hasse:1990,Dubin:1993}.
The mechanical instability of the linear chain separates it from a planar structure: the ions order in a zigzag array, like the one shown in the right panel of Fig.\ \ref{Fig:LinZigzag}. Evidence of the linear-zigzag  transition is reported by numerous studies, starting from the first experimental \cite{Birkl:1992,Raizen:1992,Waki:1992} and numerical \cite{Dubin:1993,Schiffer:1993} investigations. Further studies seeked evidence that this transition is of second order \cite{Schiffer:1993,Enzer:2000,Piacente:2004} starting from the observation that the zigzag structure breaks the cylindrical symmetry of the linear chain. This conjecture is supported by numerical results on the derivatives of the energy of finite chains at the linear-zigzag transition \cite{Dubin:1993,Piacente:2004}.

These conjectures are put on a solid ground by a statistical mechanics approach, that derives the Landau theory of the linear-zigzag instability starting from the classical potential energy of an ion chain \cite{Fishman:2008}. This theory shows that the linear-zigzag transition in ion Coulomb chains is of the same universality class of the mean-field phase transition from paramagnet to ferromagnet. Its predictions were so far confirmed by experimental studies, as we detail in Sec.\ \ref{sec:KZ}. In what follows we review the main arguments and steps of the derivation in Ref.\ \cite{Fishman:2008}.

\begin{figure*}
\includegraphics[width=0.88\textwidth]{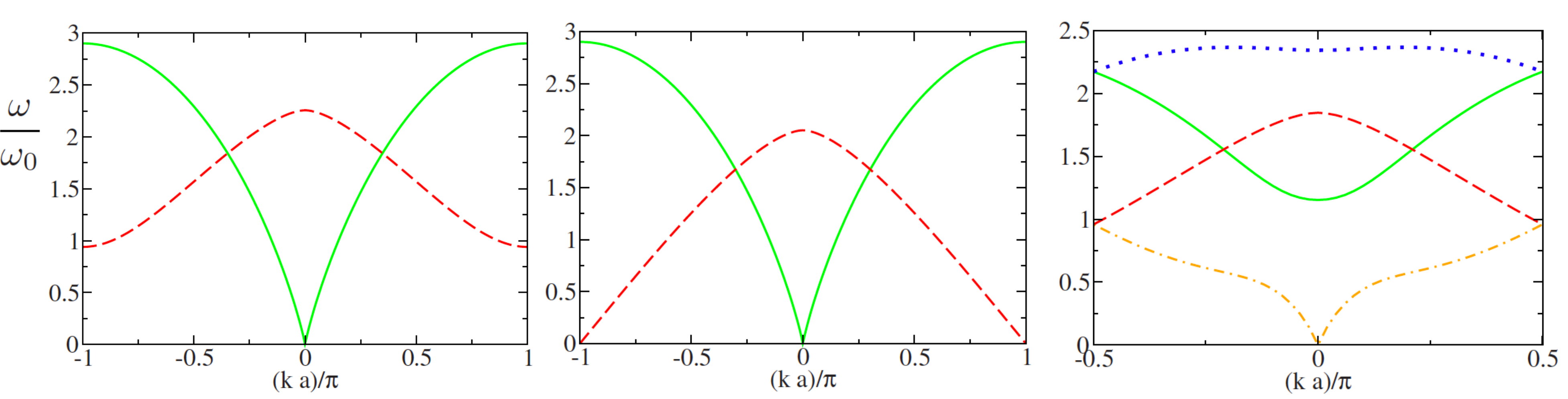}
\caption{\label{Fig:LinZigzag:Modes}
(Color online) Normal modes of a uniform chain as a function of the quasimomentum $k$ for (left) the linear array, $\nu_t>\nu_t^{(c)}$, (middle) at the transition,  $\nu_t=\nu_t^{(c)}$, and (right) a zigzag array,  $\nu_t<\nu_t^{(c)}$. Left and central panels: The red dashed line denotes the the transverse vibrations, Eq.\ \eqref{eq:omega:perp}. Central panel: At the critical frequency, the mode at vanishing frequency is the zigzag mode. Right panel: The Brillouin zone of the zigzag array is half the size of the linear chain, due to the period doubling of the zigzag structure. Adapted from Ref.~\onlinecite{Fishman:2008}.}
\end{figure*}

Assume now the thermodynamic limit, consisting of an infinite chain with particles at the uniform distance $a\equiv a(0)$. Away from the critical aspect ratio, the equilibrium positions are $x_n^{(0)}=na$, $y_n^{(0)}=z_n^{(0)}=0$ and the potential \eqref{Eq:potential} is approximated by its quadratic expansion $V\simeq V^{(0)}+V^{(2)}$. The chain is invariant under discrete displacements, hence the eigenmodes are standing waves with quasi momentum $k=\pi \ell/Na$\,. The quadratic term $V^{(2)}$ reads
\begin{equation} \label{V:2} V^{(2)} =
\frac{m}{2}\sum_{k>0,s=\pm}
\left(\omega_{\|}(k)^2\Theta_k^{(s)2}+\beta(k)(\Psi^{y(s)2}_k+\Psi^{z(s)2}_k)\right)\,,
\end{equation} 
where $\Theta_k^{(s)}$ are the Fourier transform of the axial displacements $q_j$ and $\Psi^{y(s)}_k,\Psi^{z(s)}$ of the transverse displacements $y_j$ and $z_j$, respectively, while $s=\pm$ denotes the $\cos$- and $\sin$- standing waves. The left panel of Fig.\ \ref{Fig:LinZigzag:Modes} displays the normal mode spectrum. For the linear-zigzag instability, the dispersion relation of the transverse excitation is crucial. Its explicit form \cite{Hasse:1992,Fishman:2008},
\begin{equation}
\label{eq:omega:perp}
   \beta(k)^2=\nu_t^2-K_0\left(\frac{L}{a}\right)^3\sum_{j=1}^{N/2}\frac{1}{j^3}\sin^2\frac{jka}{2}\,, 
\end{equation}
allows one to extract several key observations: (i) the minimal transverse frequency $\omega^{\perp}(k)$ is at the quasimomentum $k_0=\pi/a$. These are the eigenmodes with the shortest wavelength, $2a$, also known as zigzag eigenmodes because neighboring ions oscillate with opposite phase, $y_i=(-1)^i\Phi_{k_0}^{(y)}$ and $z_i=(-1)^i\Phi_{k_0}^{(z)}$. (ii) The linear chain becomes unstable when the eigenfrequency of the zigzag mode vanishes, $\beta(k_0)=0$ (see central panel of Fig. \ref{Fig:LinZigzag:Modes}). This determines the critical transverse trap frequency
\begin{equation}
\label{Eq:nu:t}
\nu_t^{(c)}=\omega_0\sqrt{7\zeta(3)/2}\,,
\end{equation}
with $\zeta(j)$ Riemann's zeta function and $\omega_0=\sqrt{{Q^2}/(4\pi \epsilon_0ma^3)}$. In terms of this notation, the normalized particle density of Fig.\ \ref{Fig:LinZigzag} becomes $\lambda=(3\omega_0^2/\nu_t^2)^{1/3}$, so that its critical value for linear-zigzag transition is $\lambda_c=(3/7\zeta(3))^{1/3}\approx 0.709$.

For $\nu_t<\nu_t^{(c)}$, when the linear chain is unstable, the quadratic term of Eq.\ \eqref{V:2} gives imaginary eigenfrequencies. The new ground state is then determined by including higher order terms in the expansion. Nevertheless, for $\nu_t$ yet sufficiently close to $\nu_t^{(c)}$, the transverse displacement is finite but smaller than $a$, and one expects that only the modes with frequency close to zero 
contribute to determine the new structure. These are the transverse eigenmodes with wave number $k=k_0-\delta k$ and $k=-k_0+\delta k$, such that $|\delta k\, a|\ll 1$. 
By taking the Taylor expansion of the potential $V$ about the equilibrium position of the linear chain, but truncating now at fourth order, the lowest (zero'th) order in an expansion in the small parameter $|\delta k\, a|$ (gradient expansion) is the Landau potential for the transverse displacement $\varrho=\sqrt{\Phi_0^{z2}+\Phi_0^{y2}}$ in a cigar-shaped trap:
\begin{eqnarray} \label{eq:V0} V_{\rm 0}(\varrho) =\frac{1}{2}m \left(\nu_t^2-\nu_t^{(c)2}\right)\varrho^2+ {\mathcal A}\varrho^4\,,
\end{eqnarray}
where ${\mathcal A}=\frac{3}{2}\frac{31}{32}\zeta(5)\frac{Q^2}{
a^5}$. 

\begin{figure}
\includegraphics[width=0.42\textwidth]{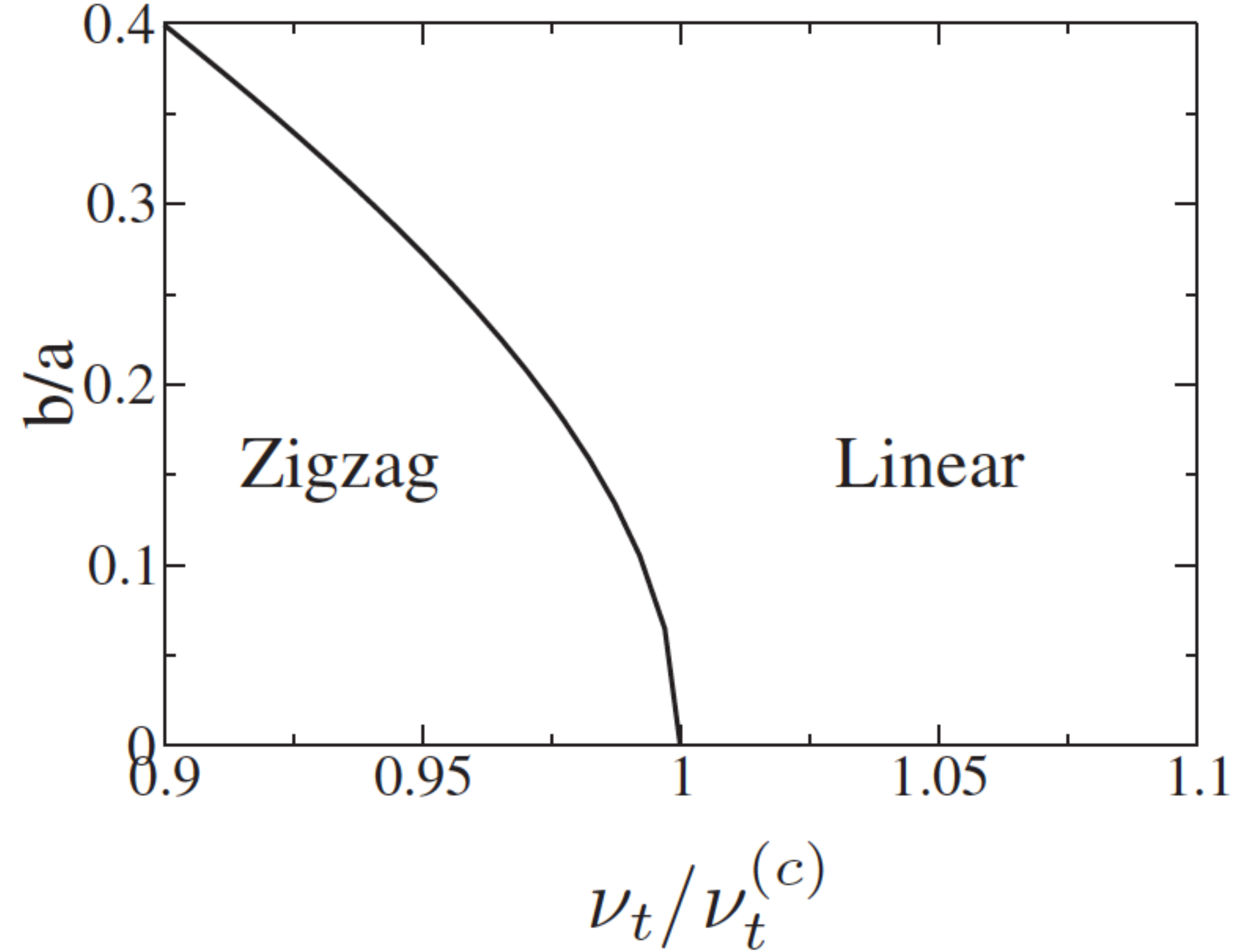}
\caption{\label{Fig:LinZigzag:Displacement}
Transverse equilibrium displacement $b=\varrho$, in units of the
interparticle spacing $a$, as a function of the transverse frequency $\nu_t$,
in units of $\nu_t^{(c)}$. On the right of the curve the ion crystal is a linear
chain. In the region on the left of the curve it exhibits a zigzag
structure. Adapted from Ref.~\onlinecite{Fishman:2008}.}
\end{figure}

For $\nu_t>\nu_t^{(c)}$ the potential $V_{\rm 0}$ has a single minimum at $\varrho=0$, and the linear chain is the ground state structure. By contrast, for $\nu_t<\nu_t^{(c)}$ the potential landscape has the characteristic form of a ``Mexican hat" with degenerate zigzag ground states at different angles around the symmetry axis. The magnitude $\varrho$ of the transverse displacement is fixed by minimizing $V_0(\varrho)$: 
\begin{eqnarray}\label{eq:rhobar}
\bar{\varrho}=\sqrt{\frac{m}{4\mathcal A}\left(\nu_t^{(c)2}-\nu_t^2\right)}\,,
\end{eqnarray}
and is in full agreement with the exact solution close to the critical value, see Fig.\ \ref{Fig:LinZigzag:Displacement}. The plane of the zigzag structure is given by the angle $\varphi=\arctan(\Phi^y_0/\Phi^z_0)$ and can take any value between 0 and $2\pi$. The system hence possesses a gapless ``Goldstone mode", which is a consequence of the spontaneously broken symmetry to rotations around the trap axis.

Using Eq.\eqref{eq:rhobar} one can verify that the transition is continuous by evaluating the difference between the ground state energy of the linear and of the zigzag structure.  This yields $\Delta E=-\frac 12 m \mathcal C a^2 (\nu_t-\nu_t^{(c)})^2$, 
with $\mathcal C = 112\;\zeta(3)/[93\;\zeta(5)]$. The second derivative with respect to $\nu_t$ is discontinuous at the critical point, in agreement with the behavior of a second-order phase transition. 

This derivation is straighforwardly extended to geometries where the transverse confining potential is anisotropic, such that the confining frequency $\nu_t$ appearing in Eq.~\eqref{Eq:potential} is replaced by two different values, $\nu_{t,y}$ and $\nu_{t,z}$. In this case the anisotropy pins the zigzag orientation and aligns it in the direction of the weakest transverse confinement. For, e.g., $\nu_{t,y}<\nu_{t,z}$, Eq.\ \eqref{eq:rhobar} gives the displacement $\varrho=\Psi_{k_0}^{(y)}$ and the angle, determining the plane, takes the two possible values $\varphi=0,\pi$: the Landau potential $V_0$, Eq. \eqref{eq:V0}, now possesses two degenerate minima at $\phi_j=\pm\bar{\varrho}$ with $\bar{\varrho}$ given by Eq. \eqref{eq:rhobar}. Thus, the anisotropy favors a zigzag structure confined to a particular plane. This zigzag configuration spontaneously breaks the reflection symmetry about the axis. As a consequence, the continuous XY symmetry of the transition is reduced to a discrete Ising symmetry. The potential close to the transition, including the first non-vanishing order in the small parameter of the gradient expansion $|\delta k|^2 a^2$, takes the form \cite{Shimshoni:2011a}:
\begin{eqnarray}\label{eq:GinzburgLandau:0}
V[\{\phi_j\}]=\sum_{j=1}^N V_0(\phi_j)+\frac{1}{2}K\sum_{j=1}^N \left(\phi_j-\phi_{j+1}\right)^2\,,
\end{eqnarray}
where $V_0$ is the local potential Eq. \eqref{eq:V0} and
$y_i=(-1)^i\phi_i(t)$ describe the transverse excitations within the interval $\delta k$, such that $\phi_i(t)$ is a slowly varying function of position $i$ and time $t$ and the position-averaged expectation value in the zigzag phase is $\Phi_0^y$. The non-local part of the interaction is accounted for via the nearest neighbors coupling term, with $K=m\omega_0^2\log 2=\frac{Q^2}{a^3}\log 2$ (now chosen to match the leading gradients in space of the order parameter). Notably, the nearest-neighbor interaction results from the systematic gradient expansion about the soft mode \cite{DeChiara:2008,Fishman:2008,Silvi:2014}. 

The Landau model is based on several conditions: critical behavior is strictly found for $N\to \infty$ and temperatures $T=0$. In a linear Paul trap, where the density is inhomogeneous, the zigzag structure forms first at the chain center, where the density is maximal \cite{Schiffer:1993}. For small chains the zigzag structure displaces the ions not only in the transverse but also in the axial direction. This effect becomes negligible close to the transition in chains of dozen of ions \cite{Fishman:2008}, where the zigzag eigenmodes are localized at the chain center and are well approximated by phononic waves \footnote{At the trap center Eq.\ \eqref{Jacobi} reduces to a wave equation in the thermodynamic limit, 
$I[w(\xi)]\sim \log N\frac{\partial^2}{\partial \xi^2}w(\xi)$ and the eigenmodes at the center are phononic waves \cite{Morigi:2006}. This expression also reproduces the behavior of the axial dispersion relation of a periodic chain for $k\sim \pi/(Na)$, and  $\omega^\|\sim k\sqrt{|\log k|}$ \cite{Ashcroft:1976}.}\cite{Morigi:2004b}. In \cite{Kaufmann:2012} it was experimentally shown that the micromotion modifies the eigenfrequencies  close to the transition. The measurements are in agreement with a theoretical treatment, which includes the periodic drive of the Paul trap \cite{Landa:2012} (see also Sec.\ \ref{sec:Equilibrium and temperature}).

\begin{figure*}
\includegraphics[width=0.95\textwidth]{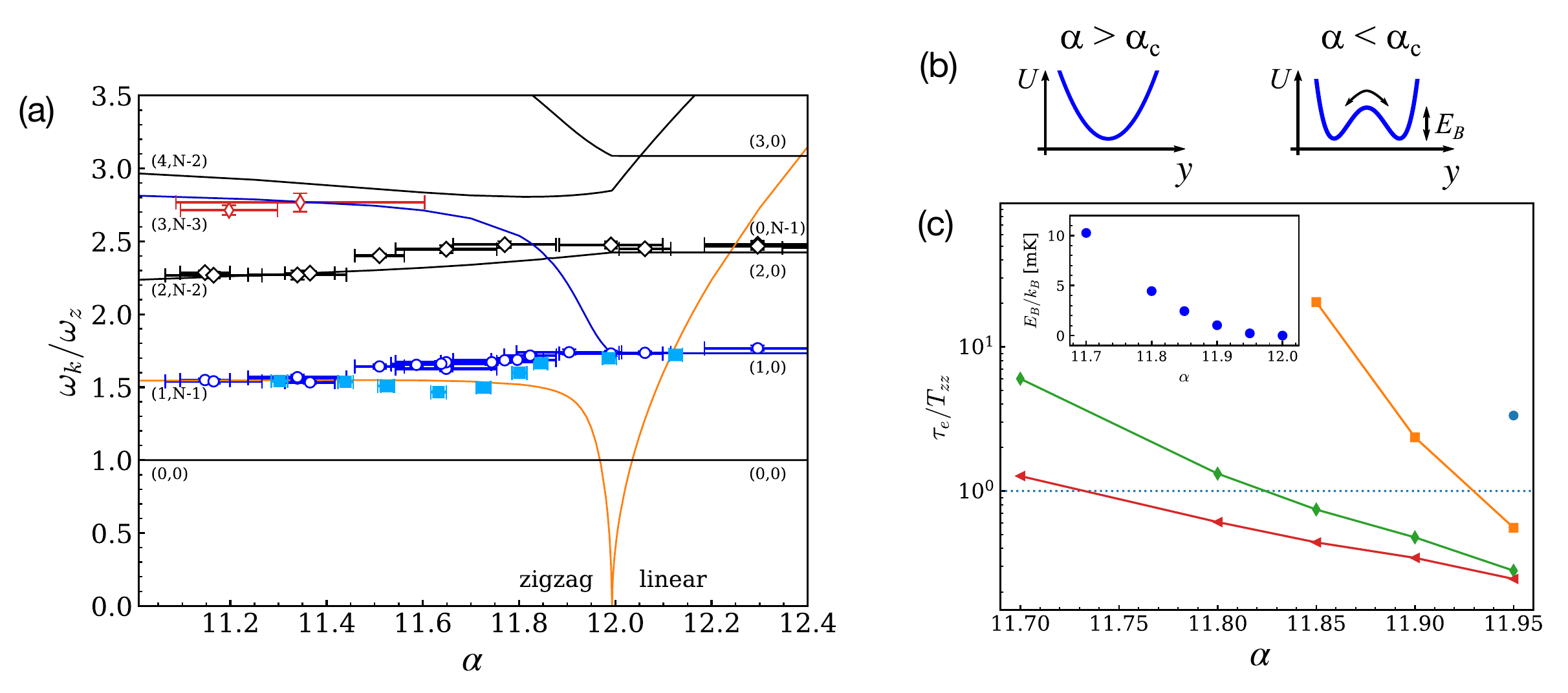}
\caption{\label{Fig:Kiethe_1} (Color online) (a) Low-frequency spectrum as a function of the aspect ratio $\alpha$. The solid lines are the modes of the harmonic crystal at T = 0, the symbols refer to the experimental measurements obtained with different techniques along with estimated uncertainties. The ions are laser cooled close to the Doppler limit. Comparison with molecular dynamics simulations gives $T \approx 3.5$ mK. (b) Schematic illustration of the Landau potential in the linear ($\alpha > \alpha_c$) and in the zigzag symmetry-broken phase ($\alpha < \alpha_c$). The coordinate $y$ indicates the transverse displacement of the central ion from the chain axis. Thermal excitations can overcome the barrier between the two degenerate zigzag configuration by inducing
collective jumps of the crystal configuration. (c) Average dwelling
time $\tau_e$, calculated using Kramers formula, in units of the zigzag mode period $T_{zz}=1/\omega_{k_0}$. In the simulations, temperatures $T$ are 0.1 mK (blue circles), 0.5 mK
(orange squares), 2.0 mK (green diamonds), 3.5 mK (red triangles).
The lines are a guide for the eye. For $T = $0.1 mK and 0.5 mK, missing
points indicate no switches were observed during the simulation
time. The dotted horizontal line indicates $\tau_e = T_{zz}$ . Inset shows the
potential barrier $E_B$ for different trapping ratios $\alpha$. Adapted from Ref.~\onlinecite{Kiethe:2021}.}
\end{figure*}

At finite, non-vanishing temperatures, the theory predicts that the linear-zigzag transition is a smooth crossover. For finite chains, nevertheless, a relatively sharp transition can be measured when the chain is prepared at temperatures below a critical value $T_c$ that depends on the number of ions $N$. In Ref.\ \cite{Kiethe:2021} the normal-mode spectrum of a trapped ion chain was measured at the symmetry-breaking linear to zigzag
transition and at finite temperatures. Spectroscopy was performed by modulating the amplitude of the Doppler cooling laser to
excite and measure mode oscillations of a chain composed of 30 $^{172}$Yb ions. The expected mode softening at the critical point, a signature of the second-order transition, was not observed, see the left panel of Fig.\ \ref{Fig:Kiethe_1}. The spectroscopic signal at the transition is reproduced by molecular dynamics simulations, suggesting that the disappearance of the mode softening is due to the finite temperature of the chain. The experimental low-frequency spectrum about the transition point is also reproduced by an effective theory based on Landau model, where thermal effects determine the weight of anharmonicities coupling the soft mode with high-frequency modes \cite{Kiethe:2021}. Inspection of the numerical trajectories shows that the ions collectively jump between the two ground-state configurations of the symmetry-broken phase, as illustrated in the right panel of Fig.\ \ref{Fig:Kiethe_1}. This jump rate is qualitatively reproduced using Kramers' formalism \cite{Delfau:2013}. In this regime, where the ion dynamics consists of thermally activated jumps between the two zigzag configurations, a time-averaged measurement of the mean transverse displacement detects a linear chain where one would instead expect a zigzag structure. The transition to a zigzag is measured at trap aspect ratios below the critical one, resembling a temperature-induced shift of the transition point \cite{Gong:2010,Delfau:2013,Li:2019}. 

In Ref.\ \cite{Cosco:2017} a quantum thermodynamic study of the linear-zigzag transition was performed. Here, it is argued that the average work distribution diverges at the linear-zigzag critical point. It is proposed to reveal this prediction by performing small quenches of the ion trap frequency across the transition and measuring the ion transverse displacements distribution. Interferometric protocols, based on spin-motion entanglement across the linear-zigzag instability, could access the autocorrelation function of the crystal \cite{DeChiara:2008} and measure the normal mode spectrum at criticality \cite{Baltrusch:2011,Baltrusch:2012}. However, 
the interferometric signals are relatively sensitive to thermal effects, which tend to suppress entanglement \cite{Baltrusch:2013}.  So far, the universal critical exponents of Landau theory have been indirectly verified by experiments analyzing the scaling of defects by slowly quenching the trap aspect ratio across the linear-zigzag phase transition, see Sec.\ \ref{sec:KZ}. 
In Ref.\ \cite{Zhang:2023} sideband spectroscopy was performed at the linear-zigzag transition, revealing quantum effects. This suggests a potential access to a structural transition driven by quantum fluctuations, that is discussed in the next subsection.  

\begin{figure*}
\includegraphics[width=0.9\textwidth]{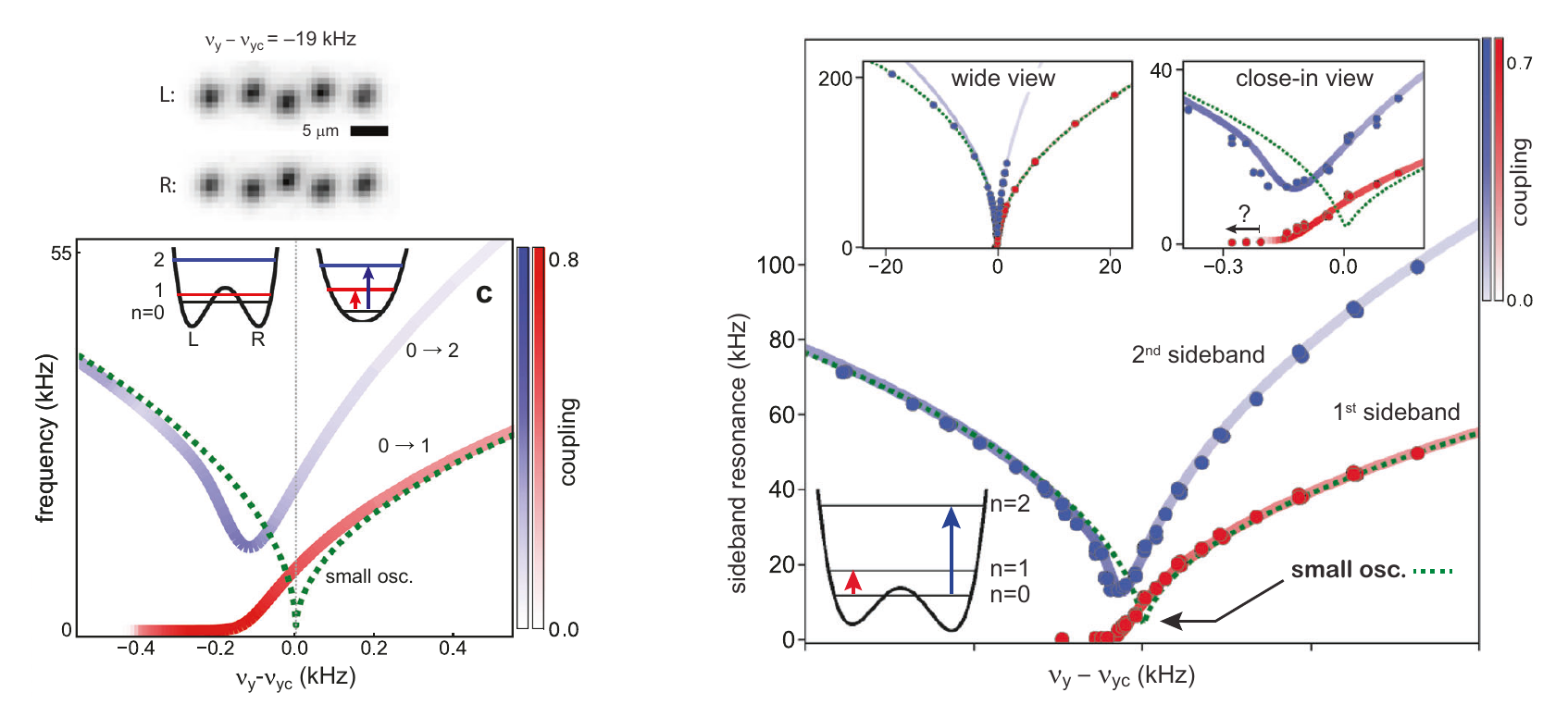} 
\caption{(Color online) Left panel. Top: Sample images of the two symmetry broken equilibrium structures of the five  $^{171}$Yb ion crystal far from the critical point. Bottom: Simulated quantum energy-level spectrum for the first two excited states of the transverse-$y$ zigzag mode of a four-ion crystal, calculated relative to the
ground state. Color shading of the levels indicates the strength of Raman sideband coupling to the
ground state for a central ion (see sidebar scales: right-handed bar for $0\rightarrow 1$ and left-handed bar for $0\rightarrow 2$). Also shown is the classical small-oscillation prediction and the form of the zigzag potential
on either side of the transition. Right panel: measured frequency of first and second upper Raman sidebands
for the transverse-y zigzag mode as a function of secular trap frequency $\nu_y$ relative to critical value $\nu_{y,c} = 760$ kHz. Insets show the full range of data acquisition and a close-in view near the critical point. Solid lines are quantum energy-level
differences of $n = 1$ and $n = 2$ number states with respect to the $n = 0$ ground state for a quartic potential.
Line shading corresponds to Raman coupling strength (see side-bar scale). Green dotted line shows the classical small-oscillation prediction
for the full pseudopotential. From~\onlinecite{Zhang:2023}.}
\label{fig:Zhang:2023}
\end{figure*}

\subsubsection{The quantum linear-zigzag transition}
\label{Qlinear_zigzag}

Structural phase transitions in ion Coulomb crystals are typically modelled by the configurations that minimize the potential energy. Although this approach discards the kinetic energy, it proves to be sufficiently accurate to reproduce the experimental results for Doppler-cooled chains. It works even down to low dimensions, thanks to the long-range nature of the Coulomb interactions. For this reason, the predictions of Landau theory for the linear-zigzag transition are confirmed by several experimental results within the level of accuracy permitted by the temperature and by the measurement apparatus. The possibility to cool relatively long chains close to the zero-point motion, on the other hand, paves the way for verifying the effects of quantum fluctuations at the linear-zigzag transition. In the near vicinity of the transition, in fact, the competing influences of the Coulomb interaction and the confining potential balance each other out, and fluctuations are expected to become significant. In Ref.\ \cite{Retzker:2008} tunnelling across the linear-zigzag transition was discussed for an anisotropic trap, which confines the motion of the ions on a plane. By means of a field-theoretical model, which was postulated on the basis of plausible assumptions, the tunneling rate between the two symmetry-broken zigzag ground states were estimated for chains of few ions. In Ref.\ \cite{Zhang:2023} it was possible to experimentally reveal quantum tunnelling between zigzag configurations in a chain composed of three to five $^{171}$Yb ions and confined by a linear Paul trap. Figure \ref{fig:Zhang:2023} displays the 
spectroscopic signal of a Raman transition resolving the anharmonic spectrum of excitations of the zigzag mode, demonstrating the effect of quantum dynamics.

In Refs.\ \cite{Shimshoni:2011a,Shimshoni:2011b} a quantum field theoretical model was derived for an anisotropic potential  as in Ref.\ \cite{Zhang:2023}, which pins the zigzag orientation in the $x-y$ plane. The procedure permits to map the quantum linear-zigzag transition to an Ising model in transverse field. The derivation incorporates quantum fluctuations in the classical model, Eq.\ \eqref{eq:GinzburgLandau:0}, and
leads to a 1+1 dimensional quantum field theory described by the partition function $Z=\int \mathcal{D}\phi e^{-S[\phi]/\hbar}$ with the Euclidean action \cite{Shimshoni:2011a} 
\begin{eqnarray}
S[\phi]=\int_0^{\hbar\beta}d\tau \sum_{j=1}^N\left[\frac{1}{2}m(\partial_\tau\phi_j)^2+V_0(\phi_j)\right.\nonumber\\+\left.\frac{1}{2} K(\phi_j-\phi_{j+1})^2\right]\,,
\label{eq:Euclidean}
\end{eqnarray}
where $\beta=1/k_\mathrm{B}T$ is the inverse temperature. The action \eqref{eq:Euclidean} is mapped to a quantum Ising model by writing the continuous field $\phi_j=\bar{\varrho} \sigma_j^z+\delta\phi_j$, where $\sigma_j^z$ can take the values $\pm 1$, corresponding to approximating the double well (and equivalently, the displacement of the ions from the chain axis) with a spin, and $\delta\phi_j$ are the fluctuations. Integrating out the fluctuations $\delta\phi_j$, 
the model takes on the form of a quantum transverse field Ising model, compactly described by the Hamiltonian
\begin{eqnarray}\label{eq:HIsing}
H_{\mathrm{I}}=-\sum_{j=1}^N\left(J\hat{\sigma}_j^z \hat{\sigma}_{j+1}^z+h\hat{\sigma}_j^x\right)\,,
\end{eqnarray}
where the spin-flip operator $\hat{\sigma}_j^x$ describes the tunnelling between the two minima.  
The magnetic field is $h=\frac{\hbar\Delta \omega}{2}$ and is proportional to the tunnel splitting $\hbar\Delta \omega$ of a quantum particle in the double-well potential. The interaction strength $J$ takes the form 
\begin{equation}
\label{eq:J2epsilon}
J=K\bar{\varrho}^2=K\frac{m}{4\mathcal A}\left(\nu_t^{(c)2}-\nu_t^{2}\right)\,,
\end{equation}
with $\bar{\varrho}$ given in Eq.\ \eqref{eq:rhobar}. Following \eqref{eq:J2epsilon}, the ratio $h/J$ is controlled by the dimensionless parameter
\begin{equation}
\label{eq:epsilon_def}
\varepsilon=(\nu_t^{(c)}-\nu_t)/\nu_t^{(c)}\,,    
\end{equation} 
which encodes the shift of the trap frequency from its mean-field critical value of Eq.\ \eqref{Eq:nu:t}. 

\begin{figure}
\includegraphics[width=0.45\textwidth]{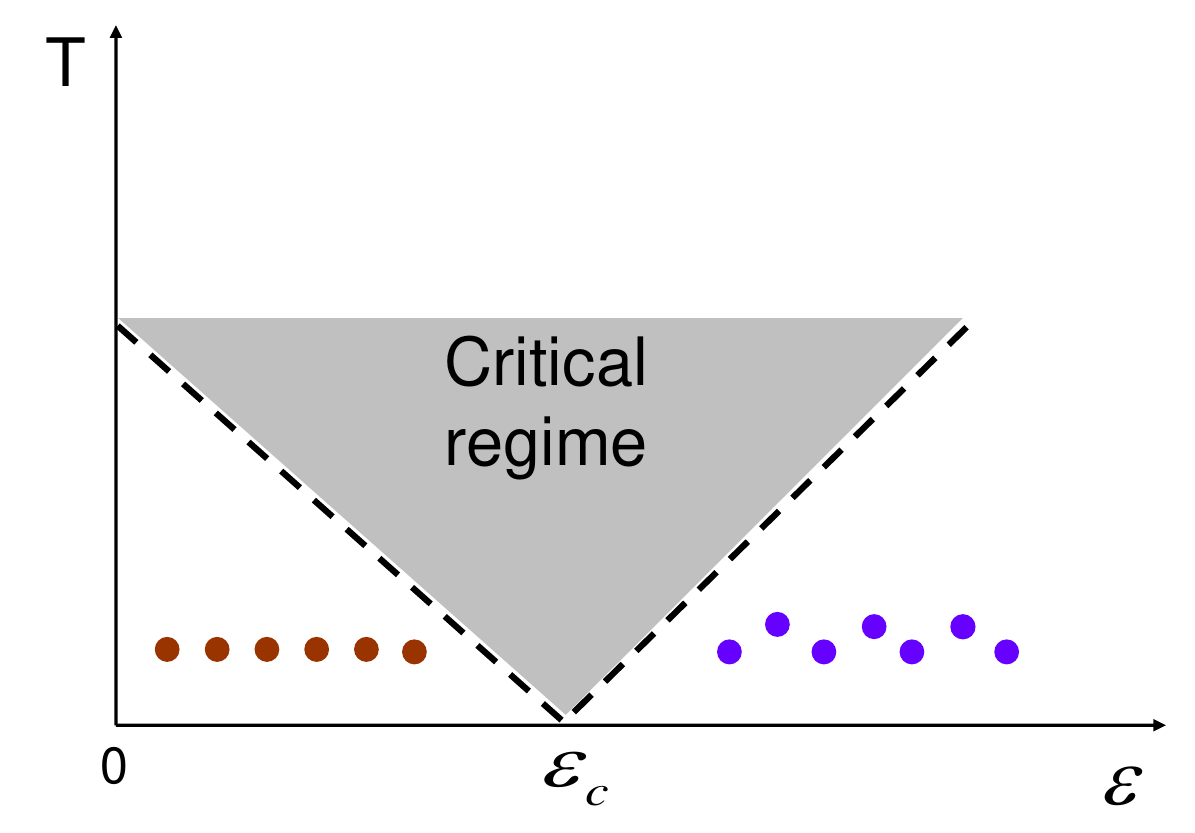} 
\caption{Phase diagram for a linear-zigzag transition, according to the
mapping to the quantum Ising chain, where $T$ is the
temperature and $\varepsilon$ is defined in \eqref{eq:epsilon_def}.
The quantum critical point, at
$\varepsilon_c>0$, separates the linear from the zigzag phase at
$T=0$. For $0<\varepsilon<\varepsilon_c$ quantum fluctuations
dominate, and the crystal is in the linear (disordered) phase. The
dashed lines indicate the boundaries of the quantum critical
region. From~\onlinecite{Shimshoni:2011a}.}\label{fig:IsingQCP}
\end{figure}

The model \eqref{eq:HIsing} is known to have a zero temperature quantum phase transition at $h_c=J$ \cite{Sachdev:2011} which corresponds to a finite positive value of the critical detuning,  $\varepsilon_c>0$; this implies a shift of the critical trap frequency below the putative mean-field value.  For $h<J$ (corresponding to $\varepsilon>\varepsilon_c$), the ground state has a non-zero expectation value of $\hat{\sigma}_z$, indicating zigzag order; whereas for $h>J$ (corresponding to $0<\varepsilon<\varepsilon_c$), the expectation value vanishes.  Furthermore,  on both sides of the transition the spectrum of excitations is separated by an energy gap $\Delta$ from the ground state.  However, the gap softens as the transition is approached from either side, as $\Delta=2|J-h|\propto |\varepsilon-\varepsilon_c|$. This yields the phase diagram shown in Fig. \ref{fig:IsingQCP}. Note that the phase transition is confined to a single critical point on the $T=0$ axis. However, on the ordered side of the quantum critical point ($\varepsilon>\varepsilon_c$), for $T$ below the gap $\Delta$ (dashed line in Fig. \ref{fig:IsingQCP}), zigzag correlations are maintained up to an exponentially-long correlation length $\xi\propto e^{\Delta/T}$.  Hence, for chains with a finite number of ions $N$, at sufficient low temperatures
the zigzag correlations can exceed the system size and extend throughout the full chain, giving the appearance of long-range order.  On the other hand, on the disordered side ($\varepsilon<\varepsilon_c$),  the correlation length $\xi$ remains finite for all $T\ge 0$.  For $T\to 0$, it diverges upon approaching the transition:  $\xi \sim|\varepsilon-\varepsilon_c|^{-1}$.  Finally, a critical regime emerges for $T>\Delta$ where the correlation length is controlled by the temperature, $\xi\propto 1/T$, and where correlations at distances smaller than $\xi$ decay as a power-law \cite{Sachdev:2011}. 

\begin{figure}
\includegraphics[width=0.5\textwidth]{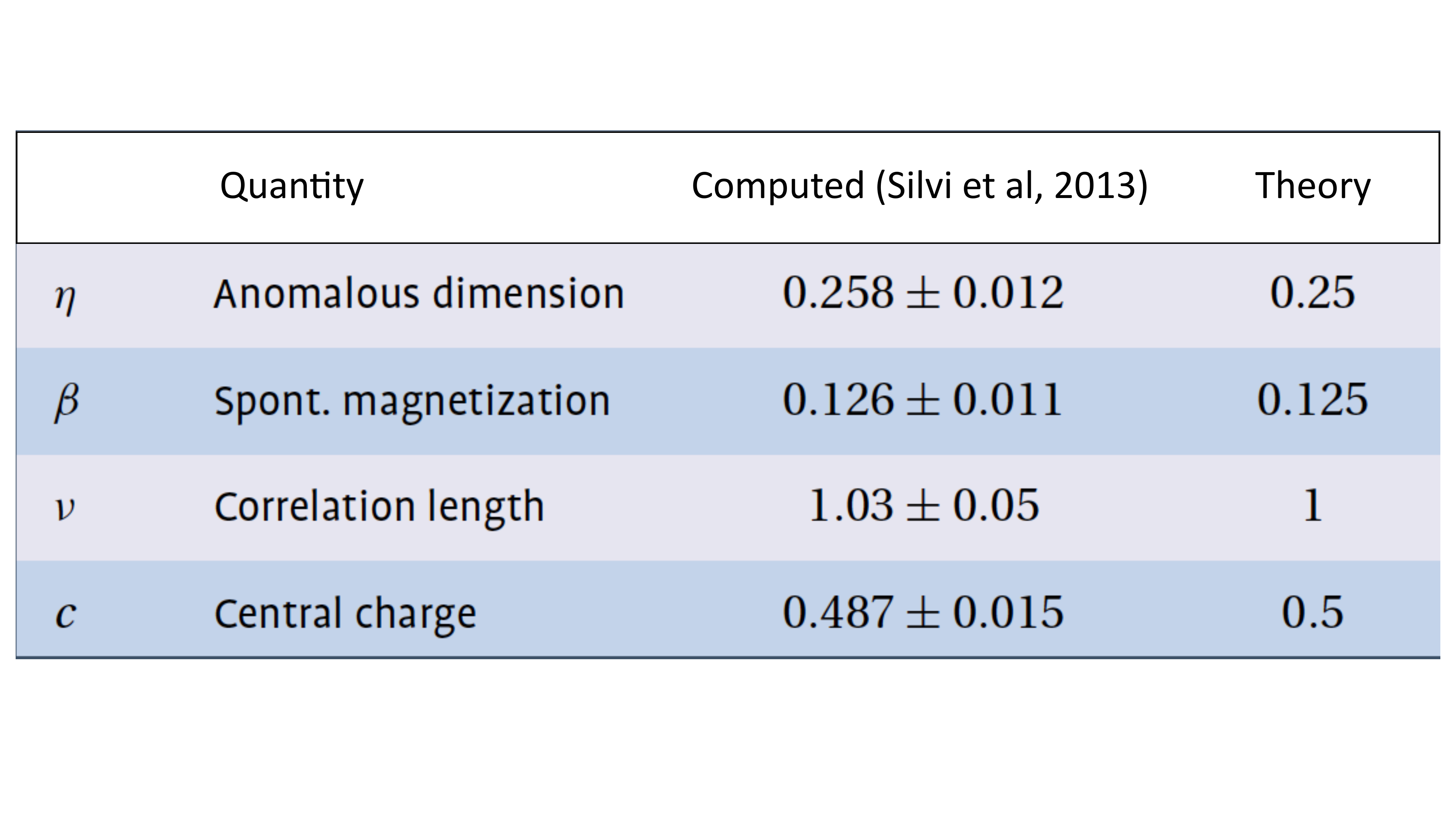} 
\caption{Computed critical exponents and central charge of the action in Eq.\ \eqref{eq:Euclidean}. The theory values are the ones of the Ising model in transverse field, see e.g.\ \onlinecite{Sachdev:2011}. From~\onlinecite{Silvi:2013}.}\label{Table:Ising}
\end{figure}

The mapping of Ref.\ \cite{Shimshoni:2011a} was numerically verified by means of a Density Matrix Renormalization Group (DMRG) analysis \cite{Silvi:2013,Silvi:2014}, a technique designed for studying quantum many-body systems in one-dimensional lattices. DMRG was used to simulate a lattice version of \eqref{eq:Euclidean}.
Figure \ref{Table:Ising} reports the critical exponents, the central charge and the corresponding error. The excellent agreement with the values of the Ising model in transverse field confirms that the quantum linear-zigzag transition belongs to the critical Ising model universality class. 

The quantum theory predicts a critical point that is shifted from the mean field value $\nu_t=\nu_t^{(c)}$ ($\varepsilon=0$). The shift scales with the ratio between kinetic and Coulomb energy, 
\begin{equation}
\label{eq:hbartilde}
\tilde{\hbar}= 
\frac{\hbar}{\sqrt{ma Q^2 \log 2}}\,, 
\end{equation} 
where the dimensionless variable $\tilde{\hbar}$ is the effective Planck constant. For the typical parameters of ion chains $\varepsilon$ takes values ranging between $10^{-5}$ and $10^{-4}$, see \cite{Silvi:2013}. 
The smallness of the effective Planck constant imposes demanding experimental conditions for observing the quantum critical behavior. The temperatures required are found by estimating the gap of the double well potential, giving $T[\textrm{mK}]\ll 1/(4n_A^{2/3} (a/a_0)^{5/3})$   with $n_A$ the atomic number and $a_0=1\mu$m a reference interparticle distance, corresponding to sub-Doppler cooling temperatures \cite{Shimshoni:2011a}. 

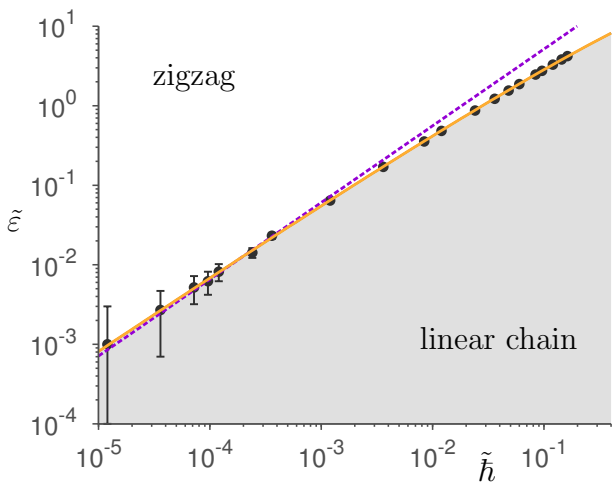
\begin{figure}
 \begin{center}
 \begin{overpic}[width = \columnwidth, unit=1pt]{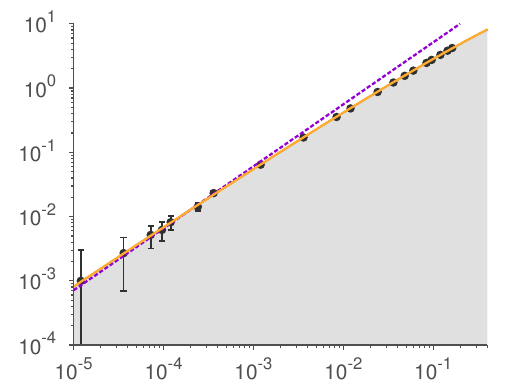}
  \put(0, 95){\large $\tilde{\varepsilon}$}
  \put(178, 0){\large $\tilde{\hbar}$}
  \put(55, 150){\large zigzag}
  \put(156, 50){\large linear chain}
 \end{overpic}
 \end{center}
\caption{ \label{fig:DMRG}
(Color online) Phase diagram in the
$(\tilde{\hbar}, \tilde{\varepsilon})$ parameter space for $g \simeq 6.027$.
The black dots are the critical points $\tilde{\varepsilon}_c(\tilde{\hbar})$ calculated via DMRG along with estimated uncertainties.
The solid orange curve is a fit to Eq.~\eqref{eq:RGepsilonShift} with $\log \tilde{\hbar}^* = 2.632 \pm 0.008$. 
By comparison, the dashed violet curve is a power-law fit to $\tilde{\varepsilon}_c\propto \tilde{\hbar}^\zeta$ restricted to data in the narrow
interval $[10^{-5},3\times 10^{-4}]$, with fitted exponent $\zeta=0.97 \pm 0.05$.  From~\onlinecite{Podolsky:2014}.
}
\end{figure}

In Ref. \cite{Podolsky:2014} the shift $\varepsilon_c$ was computed as a function of the small dimensionless parameter, $\tilde{\hbar}$. This was perfomed by rewriting the partition function $Z=\int{\mathcal{D}}\phi e^{-S[\phi]/\hbar}$ in the form 
\begin{eqnarray}
\label{eq:dimensionlessZ}
Z=\int{\mathcal{D}}\tilde{\phi} e^{-\tilde{S}[\tilde{\phi}]/\tilde{\hbar}}\,,
\end{eqnarray}
where $\tilde{S}$ is a dimensionless action written in terms of the rescaled field $\tilde{\phi}(x,\tau)=\phi(x,\tau)/a$:
\begin{equation}
\label{eq:Stilde}
\tilde{S}[\tilde{\phi}]= \frac{1}{2}\int d\tilde{\tau} d\tilde{x} \left[(\partial_{\tilde{\tau}}\tilde{\phi})^2+(\partial_{\tilde{x}}\tilde{\phi})^2-\tilde{\varepsilon} \tilde{\phi}^2+g\tilde{\phi}^4\right]\,,
\end{equation}
with $\tilde{\varepsilon}=\varepsilon 7\zeta(3)/\log 2 $ and  $g=93\zeta(5)/(32\log 2)$. \footnote{Equation \eqref{eq:Stilde} was obtained  by taking the continuum limit of Eq. \eqref{eq:Euclidean} after replacing the sum over sites by an integral in space and the finite difference term by spatial gradients.  Moreover, position and Euclidean time were rescaled as $\tilde{x}=x/a$ and $\tilde{\tau}=\sqrt{Q^2\log 2/ma^3}\tau$, respectively.  
All factors  $\log 2$ come from the value of $K$ in Eq.~(\ref{eq:GinzburgLandau:0}).  They do not appear in the expressions in \cite{Podolsky:2014}.} 
The parameter $\tilde{\hbar}$ controls the strength of quantum fluctuations in the theory: when $\tilde{\hbar}\to 0$, the partition function is given by a saddle-point minimization of the action. This corresponds to the mean-field equations of motion, with a phase transition at the mean-field value $\tilde{\varepsilon}_c=0$.  Adding small fluctuations, for small but non-zero $\tilde{\hbar}$, one can attempt to evaluate the partition function perturbatively in $\tilde{\hbar}$. However, the leading term in this expansion exhibits a logarithmic divergence in the long wavelength limit, implying that the shift in $\tilde{\varepsilon}_c$ is not analytic in $\tilde{\hbar}$. To sidestep the infrared divergences plaguing the perturbative expansion, a renormalization group (RG) analysis was performed in \cite{Podolsky:2014}. The resulting shift of the critical point is 
\begin{eqnarray}
\label{eq:RGepsilonShift}
\tilde{\varepsilon}_c\sim \frac{3 g \tilde{\hbar}}{\pi}\log \frac{\tilde{\hbar}^*}{\tilde{\hbar}}\,,
\end{eqnarray}
where $\tilde{\hbar}^*$ is a non-universal number. This result is predicted to be asymptotically exact for small $\tilde{\hbar}$ and is confirmed by numerical simulations using the DMRG program of Ref.\ \cite{Silvi:2013,Silvi:2014}. Figure~\ref{fig:DMRG} displays the numerical resuls and the prediction of \eqref{eq:RGepsilonShift}, showing agreement over the entire range of values of $\tilde{\hbar}$ considered, with a single fitting parameter, $\tilde{\hbar}^*$. 

These results indicate that the quantum critical region is very small, yet measurable using stable trapping potential. The shift could be revealed by means of interferometric protocols \cite{DeChiara:2008,Baltrusch:2012}. The critical exponents at the quantum linear zigzag transitions could be extracted by measuring the scaling of the excess heat produced by slow quenches across the phase transition \cite{Silvi:2016}. This is discussed in Sec.\ \ref{sec:KZ}.

\subsection{Planar geometries}\label{sec:planar_geometries}

Single-plane crystals of ions have been prepared in both Penning and rf ion traps by adjusting the confinement to be much stronger in one direction relative to the two orthogonal directions. Figure~\ref{fig:2D_crystals} shows some representative examples. In contrast with ion chains,
near ground state cooling of the transverse motion of single-plane crystals has only been demonstrated very recently~\cite{Jordan:2019,Kiesenhofer:2023,Guo2024,Pham:2024}. We note that two-dimensional ion crystals are at the basis of protocols for realizing robust quantum information processing \cite{Taylor:2008,Baltrusch:2010}. They have been employed in quantum simulation experiments of long-range Ising spin models, producing entanglement of the internal spin degrees of freedom~\cite{Bohnet:2016,Garttner:2017}, and in the preparation of geometrically frustrated ground states~\cite{Qiao:2022,Guo2024}. Two-dimensional ion crystals have also been used to simulate spin-boson models~\cite{Safavi-Naini:2018,Gilmore:2021}.

In this section, after reviewing the different trap configurations for preparing 2D ion crystals, we discuss their thermodynamic (ground state configuration) and dynamic (normal mode) properties, and the structural instability when a single-plane crystal transitions to multiple planes. Many of the results of this section can be applied to single-plane crystals in rf traps, in the limit in which the pseudopotential approximation holds and micromotion can be neglected.

\subsubsection{Trap configurations}
\label{sec:2D_trap_configuration}

\begin{figure}
\includegraphics[width=0.45\textwidth]{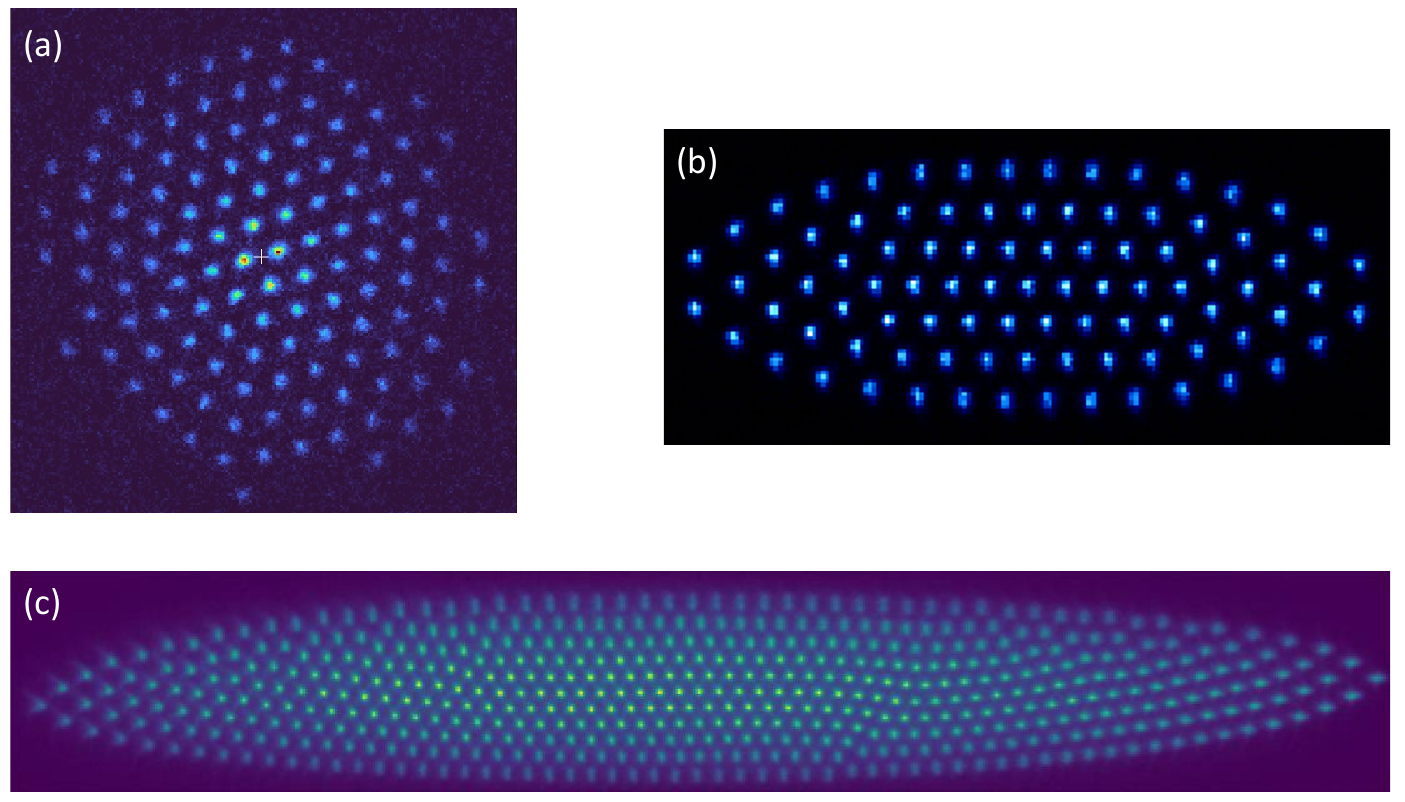} 
\caption{Examples of 2D trapped ion crystals formed in the lab.  In all cases the image is obtained in a direction normal to the plane. (a) 2D ion crystal with 121 $^9$Be$^{+}$ ions formed in a Penning trap.  The average ion-ion spacing is approximately 35 $\mu$m. From~\cite{Wolf:2024}. (b) 2D ion crystal with 91 $^{40}$Ca$^{+}$ ions formed in an rf trap operated in a lateral configuration (see Sec.~\ref{sec:2D_trap_configuration}).  The average ion-ion spacing is approximately 6.5 $\mu$m.
 From~\cite{Kiesenhofer:2023}. (c) 2D ion crystal with 512 $^{171}$Yb$^{+}$ ions formed in an rf trap operated in a lateral configuration. The average ion-ion spacing is approximately 4 $\mu$m. From~\cite{Guo2024}. \label{fig:2D_crystals}}
\end{figure}

The effective confining potential of an ion trap used to prepare trapped ion crystals can typically be written as
\begin{equation}
V(x,y,z) =\frac{1}{2} m\omega_z^2(z^2 + \beta_x x^2 + \beta_y y^2 )\:,
\label{eq:harmonic_confine}
\end{equation}
where $(x, y, z)$ denote coordinates along the principal axes of the confining potential and relative to the center of the trap.  Harmonic confinement is assumed, which is typically a good approximation. For a Penning trap (see Sec.~\ref{sec:Penning trap}), $(x, y, z)$ are coordinates in the rotating frame of the crystal with $z$ denoting distance along the magnetic field axis.  For a linear rf trap, $V$ denotes the confining pseudopotential and $z$ the distance along the nodal line in the center of the trap that is free from rf fields (see Sec.~\ref{sec:rf trap}). Single-plane crystals have been prepared in a variety of trap configurations, which we summarize below.

In Penning traps, single-plane ion crystals are formed in the plane perpendicular to the magnetic field of the trap by adjusting $\beta_x, \,\beta_y \ll 1$. This is achieved by controlling the crystal rotation frequency to produce a weak radial confinement compared to the axial confinement (see Fig.~\ref{fig:Pening_crystal_shapes}).  The relative strength of $\beta_x$ and $\beta_y$ is set by the strength of the rotating wall electric field. Typically $\beta_x \sim \beta_y$, giving rise to crystal shapes that are approximately circular.  Single-plane crystals of up to 400 to 500 ions have been produced and controlled in Penning traps \cite{Mitchell:1998,Mavadia:2013,Ball:2019}.
Readout of the spin state of an individual ion in a single experiment~\cite{McMahon:2024,Wolf:2024} and addressing of individual ions~\cite{McMahon:2024} in the rotating frame of the crystal have been demonstrated.

Similar to the Penning trap, single-plane crystals have been prepared in rf traps by adjusting the confinement in the axial $(\hat{z})$ direction to be strong compared to the confinement in the radial direction~\cite{Drewsen:2003,Richerme:2016,D'Onofrio:2021,Kato:2022}. With traps operated in this configuration (``radial configuration''), single-plane crystals of up to $\sim$50 ions have been prepared and well characterized.  Another approach for forming 2D ion crystals is to weaken the confinement in a direction transverse to the axis of a linear rf trap~\cite{Block:2000,Bermudez:2012,Kaufmann:2012,Szymanski:2012,Wang:2020,Qiao:2022,Kiesenhofer:2023}.  Starting with an ion chain along the $\hat{z}$-direction, weakening the lateral confinement in the $\hat{y}$-direction results in a 2D zig-zag structure as discussed in Sec.~\ref{sec:linear-zigzag}.  By further weakening the lateral confinement and careful controlling of the relative confinement strengths in the transverse and axial directions to ensure $1<\beta_y \ll \beta_x$, single-plane ion crystals are obtained that have an approximate elliptical shape, see Figs.~\ref{fig:2D_crystals}(b) and (c). With traps operated in this lateral configuration, single-plane crystals of up to 150 ions have been observed in a surface electrode trap~\cite{Szymanski:2012}. Stable crystal structures have been realized and studied with up to 105 ions in a three-layer rf trap operated at room temperature~\cite{Kiesenhofer:2023} and with up to 512 ions in a monolithic 3-layer trap operated at cryogenic temperatures~\cite{Guo2024}. 

In contrast to chains of ions in the linear rf trap, single-plane crystals in rf traps undergo substantial micromotion. In both the radial and lateral configurations discussed above, the driven micromotion of the ions at the rf drive frequency can be designed to lie in the plane of the ion crystal~\cite{Richerme:2016, Kiesenhofer:2023}.  A nice feature of the lateral configuration is that the trap can be designed so that the micromotion occurs in a single direction~\cite{Kiesenhofer:2023}, which enables micromotion-free interactions with laser beams from more than one direction.    

\subsubsection{Ground state configuration} 
\label{Sec:GroundState2D}

\begin{figure}
\includegraphics[width=0.35\textwidth]{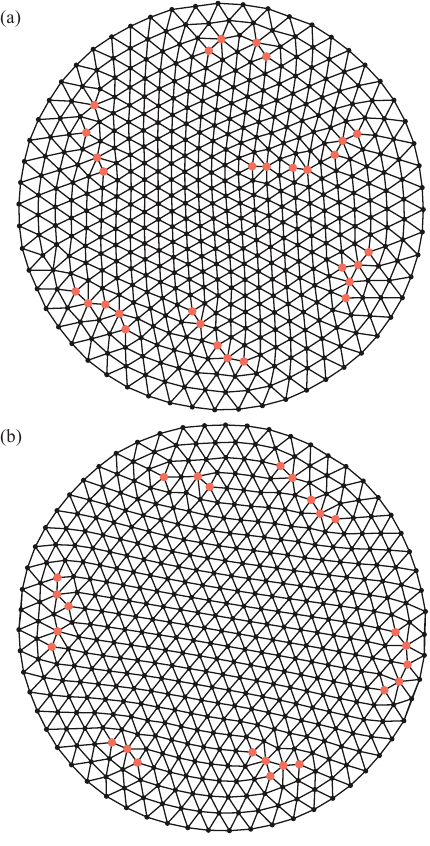} 
\caption{(Color online) Charge distribution in the ground state of a two-dimensional ion crystal. The dots correspond to the local minimum energy configuration evaluated numerically for $N=500$ identical charges in (a) a purely quadratic trap potential and (b) a quadratic trap potential plus a lowest-order anharmonic moment $(C_4\neq0)$ designed to generate a more uniform lattice in the interior of the crystal. The red (enlarged) dots indicate dislocations in the crystal lattice. From~\onlinecite{Dubin:2013}. \label{fig:triangular_lattice}}
\end{figure}

The ground state equilibrium configuration of a single-plane, trapped ion crystal is an approximate triangular lattice~\cite{Dubin:1989,Dubin:1993,Schiffer:1993}, like the one shown in Fig.~\ref{fig:triangular_lattice}.  In the interior of the ion crystal, the ions arrange themselves at the vertices of equilateral triangles, characteristic of an infinite 2D system.  
Such triangular lattices have frequently been called “hexagonal planes” in the literature as it emphasizes the 6-fold symmetry of the lattice. 

For a harmonic trap potential such as Eq.~(\ref{eq:harmonic_confine}), the 2D number density of a single-plane crystal is inhomogeneous, decreasing from the center to the  boundary of the crystal.  For simplicity, consider the azimuthally symmetric confinement potential (Eq.~(\ref{eq:rotating_frame}) in Sec.~\ref{sec:Penning trap}) 
\begin{equation}
V(\rho, z) = \frac{1}{2} m\omega_z^2( z^2+\beta \rho^2)\:.
\label{eq:azimuth_symmetric}
\end{equation}
obtained with $\beta_x = \beta_y \equiv \beta$.  single-plane ion crystals are formed in the $\hat{z} = 0$ plane of the trap when $\beta \ll 1$.  For a single-plane crystal of $N$ ions, $\beta$ must be less than a critical $\beta_c(N)\approx 0.665/\sqrt{N}$~\cite{Dubin:1993}.  We can estimate the radial dependence in the 2D number density of a single-plane crystal by considering three-dimensional crystals formed in the potential of Eq.~(\ref{eq:azimuth_symmetric}).  This potential produces a crystal in a spheroidal volume of axial extent $2Z_p$  and diameter $2R_p$ with constant, three-dimensional density $n_0$ given by
\begin{equation}
n_0 = \frac {\epsilon_0 m}{Q^2}\omega_z^2(1+2\beta)\,,
\label{eq:density_w_z}
\end{equation}
 see Eqs.~\eqref{eq:n_0_rf_trap}, \eqref{eq:3Ddensity}. 
For this spheroidal shaped crystal, the 2D number density per unit area, $\sigma(\rho)$, is 
\begin{equation}
\sigma(\rho) = \sigma \sqrt{1-\rho^2/R_p^2} \:,
\label{eq:2D inhomogeneity}
\end{equation}
where $\sigma=n_0 2Z_p$ is the 2D number density at the center.  Equation~(\ref{eq:2D inhomogeneity}) provides a good estimate for the 2D number density of a single-plane crystal, obtained in the limit of small $Z_p/R_p$.  It is the equivalent expression for the inhomogeneity in the 2D number density for a single-plane crystal as Eq.~(\ref{Gauss}) is for the inhomogeneity in the linear density for chains of ions.  

In the thermodynamic limit, $\beta\to 0$ and $R_p\to\infty$, the crystal is a triangular lattice with constant planar density $\sigma$. The single plane has lattice constant $a_{\mathrm{lat}} = (2/\sqrt{3}\sigma)^{1/2}$~\cite{Dubin:1989,Dubin:1993} and is stabilized by transverse potential $V(\rho, z)=m\omega_z^2 z^2/2$ for $\sigma {a_0}^2 \lesssim 1.11$, where  $a_0 \equiv (\frac{Q^2}{\epsilon_0}/m \omega_z^2)^{1/3}$ is a characteristic length derived from the ratio between the Coulomb interaction and the transverse confinement \cite{Dubin:1993}. Note that $a_0^{-3} = n_0$ in Eq.~(\ref{eq:density_w_z}) for $\beta=0$.

For finite planes, the inhomogeneous charge density of single-plane ion crystals is necessarily associated with dislocations. These are illustrated by the red dots in Fig.~\ref{fig:triangular_lattice}(a)~\cite{Dubin:2013}. Near the boundary of the planar crystal, shells are formed that conform to the overall symmetry of the confining potential~\cite{Dubin:1988}.  Shell structure is typical for systems with long-range interactions~\cite{Bubeck:1999}.  For azimutally symmetric potentials, the shells are circular and the locations of ions in neighboring shells is staggered, see Fig.~\ref{fig:triangular_lattice}. This shell structure has a different symmetry than the underlying triangular lattice and contributes to the formation of dislocations.

The number of dislocations in the interior of the crystal can be reduced by modifying the confining potential through the addition of anharmonic terms, thereby approaching a regular triangular lattice. The configuration of Fig.~\ref{fig:triangular_lattice}(a), for instance, can be made more regular by adding a fourth-order term to the harmonic confining potential. The resulting structure is presented in Fig.~\ref{fig:triangular_lattice}(b) for an optimized potential and in a configuration relevant for the Penning trap. The fourth-order term is frequently called a $C_4$ term. Structures even closer to uniform triangular lattices can be achieved by adding corrections to the confining potential with the same underlying symmetry as the target structure.
In Penning traps the symmetry of the boundary is determined by the symmetry of the rotating wall potential. Theoretical studies indicate that a nearly perfect triangular lattice can be obtained by fine tuning the strength of an $m=3$ rotating wall potential~\cite{Dubin:2013}, where $m$ refers to the azimuthal symmetry of the applied rotating wall potential ($\propto \cos[m(\phi+\omega_r t)]$, see Eq.~(\ref{eq:quad_wall})). Experimental work demonstrated that this strategy leads to the formation of more regular ion crystals in Penning traps~\cite{McMahon:2024}. Overall, these strategies are at early stages of experimental investigation.

\subsubsection{Normal modes}\label{sec:planar_geometry_mode}

In what follows we discuss the normal modes of single-plane ion crystals. We assume that the ions are confined to a single plane by a harmonic, conservative potential. The discussion applies to single-plane crystals in Penning traps and in linear rf traps; for the latter, micromotion is neglected and the pseudopotential approximation is assumed to be valid.

\begin{figure}
\includegraphics[width=0.48\textwidth]{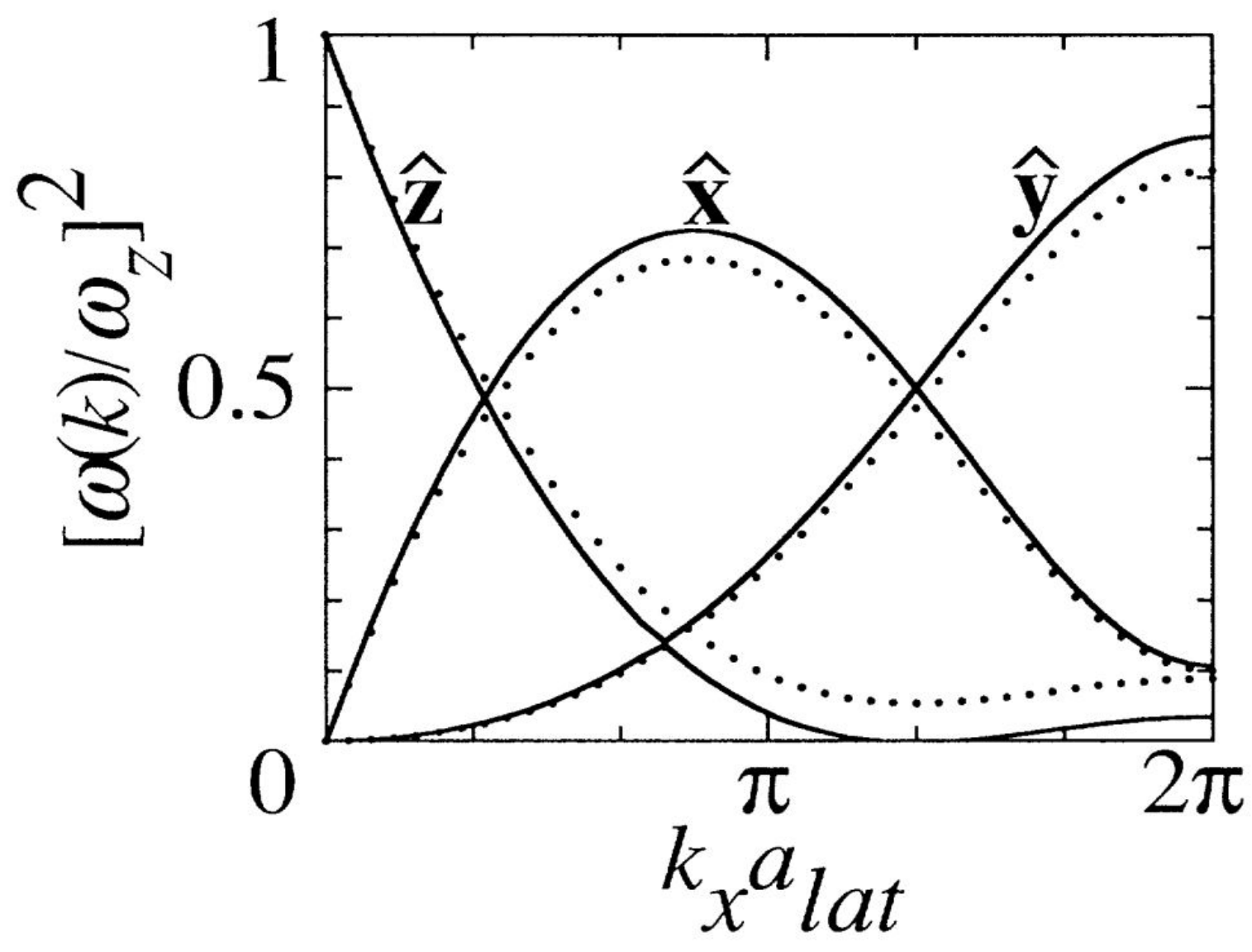} 
\caption{Normal mode frequencies of a single-plane triangular lattice as a function of $k_x$, in units of the inverse of the lattice constant $a_{\mathrm{lat}} = (2/\sqrt{3}\sigma)^{1/2}$. The frequencies $\omega({\mathbf k})$ are squared and in units of the frequency $\omega_z$ of the transverse harmonic trap, confining the ions on the $x-y$ plane. The modes shown are the in-plane (along $\hat{x}$ and $\hat{y}$) and the transverse (along $\hat{z}$) modes for $k_y=0$. Dotted lines: $\sigma a_0^2 = 1.07$; solid lines $\sigma a_0^2 = 1.11$, corresponding approximately to the density $\sigma$ where the single plane becomes mechanically unstable. The branch of the transverse modes approaches 0 for $k_x=4\pi/3$. The phase transition of this mechanical instability is discussed in Sec.~\ref{sec:1_3_instability}. From~\onlinecite{Dubin:1993}.} \label{fig:infinite_2D_dispersion}
\end{figure}

We start with the normal modes of an infinite, single-plane ion crystal. This is an idealization, that provides intuition for understanding the normal modes of finite-sized, two-dimensional structures. In the infinite triangular lattice, the equilibrium positions are $\bm{\rho} = [(n_1+n_2/2)\hat{x} +(\sqrt{3}/2)\,n_2\,\hat{y}] a_{\mathrm{lat}}$, with $n_1,n_2$ integers. The normal modes are waves with wave vector ${\mathbf k}$ in the $x-y$ plane and displacements proportional to $\exp\left( i{\mathbf k}\cdot \bm{\rho} \right)$. For every ${\mathbf k}$ there are three different polarizations: two are in the $x-y$ plane while the third is in the $\hat z$-direction, perpendicular to the plane. As an example consider the two in-plane modes when $k_y=0$. They have polarizations in the $\hat{x}$ and $\hat{y}$ directions, corresponding to compressional and transverse modes. Their dispersion relation, along with the dispersion relation for oscillations normal to the plane, is displayed in Fig.~\ref{fig:infinite_2D_dispersion} for $\sigma a_0^2 = 1.07$ and at the instability of the single plane  ($\sigma a_0^2 = 1.11$). Note that the bandwidth of the in-plane modes overlaps with the bandwidth of the mode for oscillations normal to the plane.

The dispersion relation  of the in-plane modes exhibits the characteristic features of acoustic modes in solids: the frequency increases as the wavelength decreases. The longitudinal (compressional) mode displays a sublinear dispersion at long wavelengths ($\hat{x}$-mode in Fig.~\ref{fig:infinite_2D_dispersion}), $\omega(k)\sim\sqrt{k}$, arising from the long-range nature of the Coulomb repulsion \cite{Stern:1967}. 
The dispersion relation of the out-of-plane oscillations in the $\hat{z}$ direction is similar to optical modes in solids: the mode frequency decreases as the wavelength increases. The largest frequency is the transverse confining frequency $\omega_z$, where the crystal oscillates as a bulk. The shortest wavelengths, comparable to the inter-ion spacing, are the smallest frequencies. As visible in Fig.~\ref{fig:infinite_2D_dispersion}, at the single-plane stability limit, $\left ( \sigma {a_0}^2 = 1.1 \right )$, the mode frequency becomes zero at $k_x a_{\mathrm{lat}} = 4\pi/3$, producing a second order phase transition discussed in Sec.~\ref{sec:1_3_instability}.

The normal modes of trapped, single-plane ion crystals take on the character of the infinite single-plane crystal modes, including a similar dispersion relation. The general theoretical approach consists of diagonalizing the stiffness matrix, obtained by performing the Taylor expansion of the Coulomb potential, with some salient differences between the rf- and the Penning trap due to the respective micromotion and the strong magnetic field~\cite{Wang:2013,Shankar:2020}. In rf traps, some of the ions in the crystal undergo significant micromotion: Including the effects of micromotion in describing normal modes is challenging and has only been incorporated in a few studies~\cite{Kaufmann:2012,Kiesenhofer:2023}.  In Ref.~\cite{Kaufmann:2012}, including micromotion through a full time-dependent Coulomb theory was necessary to obtain good agreement between the measured and predicted eigenfrequencies for the in-plane mode spectrum of a three-ion crystal.  In Ref.~\cite{Kiesenhofer:2023}, some discrepancies with the predictions of pseudopotential theory were obtained for some of the measured transverse mode frequencies of a single-plane crystal in a rf trap composed by 8 ions. Theoretical calculations including micromotion did not resolve the discrepancies.

In a Penning trap, the natural frame in which to analyze the normal modes is the frame that co-rotates with the ion crystal. Because the non-thermal ion motion is a rigid body rotation about the trap symmetry axis, it does not impact the calculation of the normal modes beyond adding a centrifugal potential. However, the strong magnetic field requires one to modify the theoretical procedure for determining the ion crystal normal modes.

In general, a single-plane crystal of $N$ ions supports $2N$ in-plane modes and $N$ transverse modes. The latter are frequently called drumhead modes because they resemble the modes of a drum with open boundary conditions~\cite{Sawyer:2012,Wang:2013,Shankar:2020}.  In a Penning trap, the drumhead modes for a single-plane crystal involve ion motion parallel to the magnetic field. Therefore, the magnetic field does not impact the drumhead mode frequencies or eigenvectors in the harmonic approximation, and the drumhead mode structure is the same for single-plane crystals in Penning traps and in rf traps, provided the pseudopotential approximation applies to the latter.  

\begin{figure}
\includegraphics[width=0.45\textwidth]{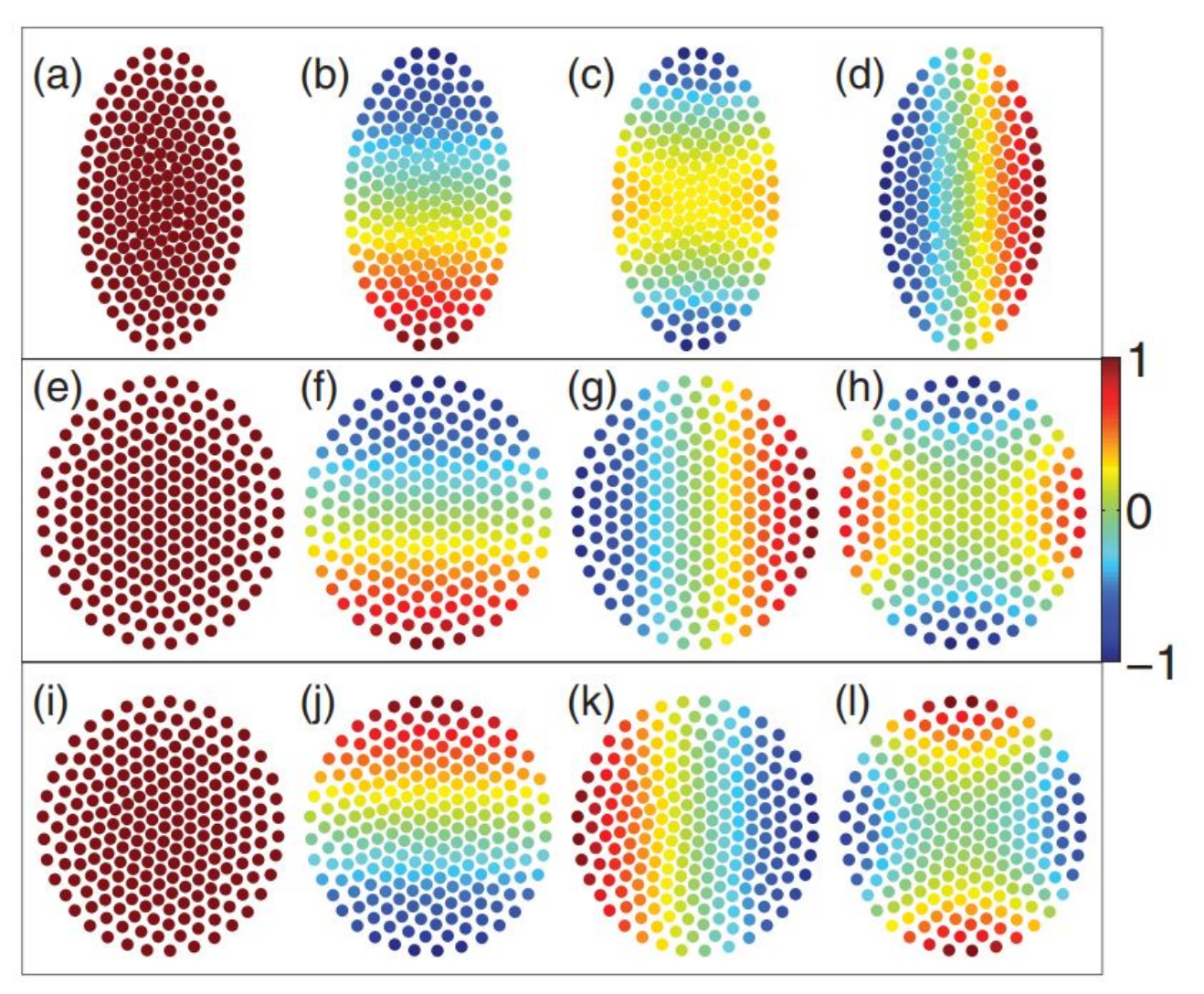} 
\caption{(Color online) The four highest frequency drumhead mode eigenvectors calculated for a single-plane ion crystal consisting of $N=217$ ions. The three rows correspond to three different in-plane confinement strengths, respectively. The color scale illustrates the eigenvector amplitude with the maximal value set to one. From~\onlinecite{Wang:2013}. \label{fig:drumhead_modes}}
\end{figure}

Figure~\ref{fig:drumhead_modes} illustrates the oscillation amplitudes of the four highest-frequency drumhead modes for a single-plane ion crystal. 
From larger to lower frequency: The highest frequency mode is the center-of-mass (or bulk) mode where all ions in the crystal oscillate as a bulk at frequency $\omega_z$. The next drumhead modes are tilt modes. Like the out-of-plane oscillations ($\hat{z}$ polarization in Fig.~\ref{fig:infinite_2D_dispersion}) in an infinite crystal, as the normal mode frequency decreases, so does the effective wavelength. In the short wave-length limit, the wavelength is of the order of the lattice spacing in the interior of the crystal and only ions in the crystal interior participate in the mode. In this spatial region the crystal approaches a regular triangular lattice and the short-wavelength drumhead modes essentially coincide with the transverse modes of the infinite crystal. See Refs.~\cite{Sawyer:2012,Kiesenhofer:2023} for examples of experimental spectroscopy of drumhead modes in single plane crystals in rf and Penning traps.

\begin{figure}
\includegraphics[width=0.45\textwidth]{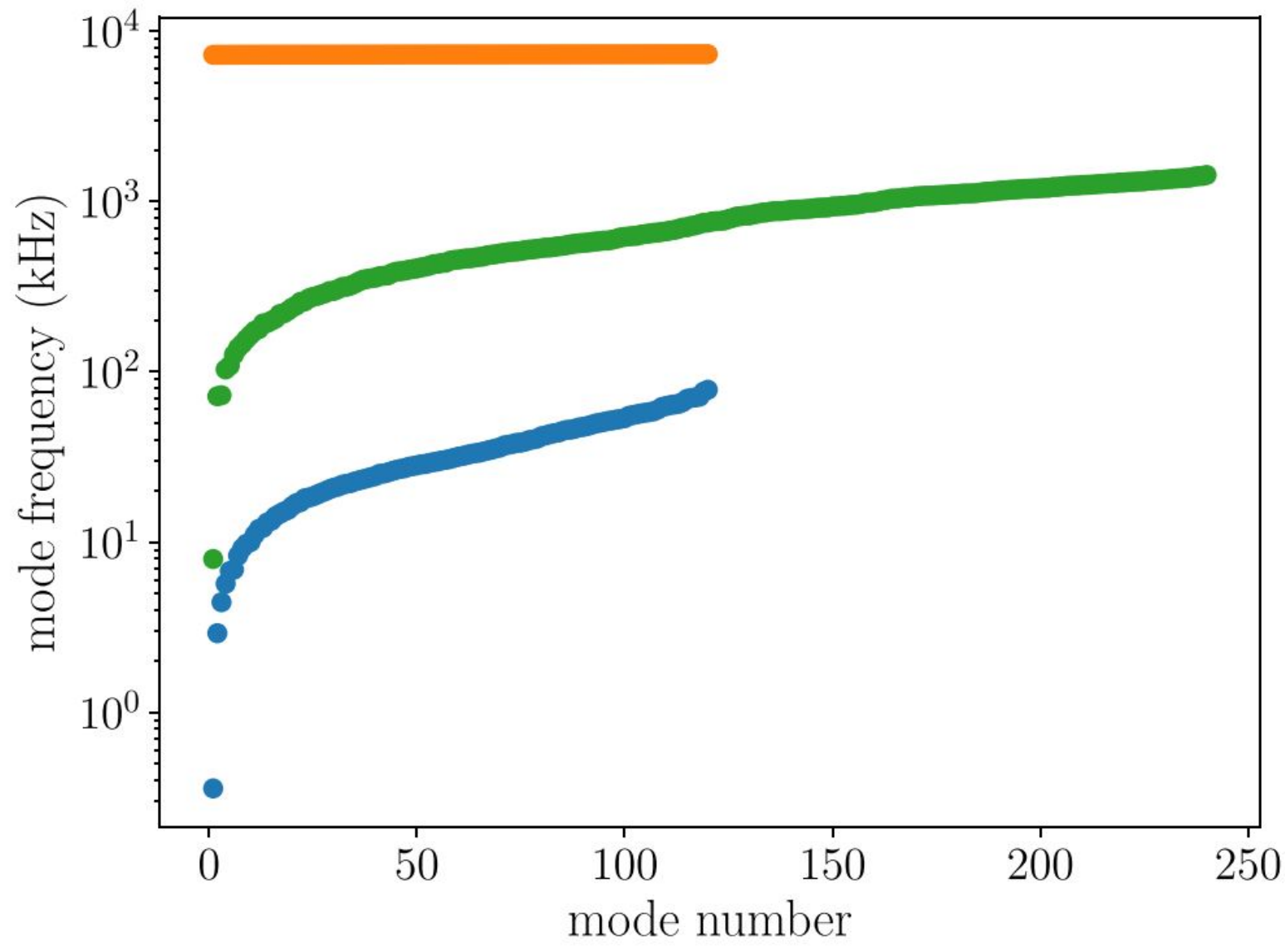} 
\caption{Example calculation of the in-plane mode frequencies for a single-plane crystal with $N=120$ ions in the presence of no magnetic field (green central branch) and in the presence of a strong magnetic field that splits the single central branch into low-frequency (blue) and high frequency (orange) branches.  The central (zero-field) branch is the in-plane mode spectrum in the pseudopotential approximation for a single-plane ion crystal in an rf trap with $\omega_z/(2\pi) = 1.59$ MHz and effective confinements $\sqrt{\beta_x} \omega_z/(2\pi) = 276$ kHz and $\sqrt{\beta_y} \omega_z/(2\pi) = 259$ kHz in the orthogonal directions. The blue and orange branches are, respectively, the low-frequency ${\mathbf E} \times {\mathbf B}$ mode and high-frequency cyclotron mode spectra for the same single-plane crystal confined in a Penning trap by an identical effective potential in the rotating frame of the crystal. 
For the Penning trap, a cyclotron frequency $\Omega_c/(2\pi) = 7.60$ MHz, a rotation frequency $\omega_r/(2\pi) = 180$ kHz and a weak quadrupole rotating wall potential characterized by  $\omega_W = 68$ kHz (see Eq.~(\ref{eq:quad_wall2})) are assumed. The mode frequencies are referenced to the rotating frame of the ion crystal. From~\onlinecite{Shankar:2020}. \label{fig:in_plane_modes}}
\end{figure}

For single-plane crystals in an rf trap the magnetic field is nominally zero and the in-plane modes exhibit acoustic-like dispersion, similar to the in-plane modes of an infinite crystal (the $\hat{x}$ and $\hat{y}$ polarizations of Fig.~\ref{fig:infinite_2D_dispersion}). The green branch in Fig.~\ref{fig:in_plane_modes} shows the results of a calculation of the in-plane mode spectrum for a single-plane crystal consisting of 120 ions in zero magnetic field~\cite{Shankar:2020}.  The calculation is performed for an rf trap in the pseudopotential approximation with an axial frequency $\omega_z/(2\pi) = 1.59$ MHz and confinement frequencies $\omega_x/(2\pi) = 0.276$ MHz and $\omega_y/(2\pi) = 0.259$ MHz in the orthogonal directions.
For $N=120$ ions, there exist 240 in-plane modes, which are ordered in the figure according to increasing frequency.  The lowest frequency modes are associated with the weak asymmetry that provides azimuthal confinement, without which there would be a zero-frequency mode corresponding to rigid body rotation about the $\hat{z}$-axis.  The higher frequency modes correspond to effective shorter wavelength oscillations where ion motion becomes anti-correlated over a distance of a few inter-ion spacings.

Adding a strong magnetic field normal to the single-plane crystal (parallel to the $z$-axis) gives rise to velocity dependent forces. This is the case of the Penning trap. Here, the mode analysis becomes more involved for the in-plane modes, requiring the solution of a generalized eigenvalue problem~\cite{Wang:2013,Dubin:2020,Shankar:2020}.  
Figure~\ref{fig:in_plane_modes} illustrates that an important consequence of the strong magnetic field is to split the single band of $2N$ in-plane modes for $B=0$ (green central branch) into a band of $N$ high frequency cyclotron modes and a band of $N$ low frequency ${\mathbf E} \times {\mathbf B}$ modes with frequency on the order of the single ion magnetron frequency $\omega_m$ (see Sec.~\ref{sec:Penning trap}).  

In a frame that co-rotates with the ion crystal, individual ions in the cyclotron modes undergo an approximate clockwise circular motion about their equilibrium positions.  In contrast, the ${\mathbf E} \times {\mathbf B}$ modes on average exhibit counterclockwise motion~\cite{Shankar:2020}. Movies that show the eigenvectors of the planar normal modes of a single-plane crystal in a Penning trap can be found in the supplementary material of Ref.~\cite{Wang:2013}. In the rotating frame, both the cyclotron modes and ${\mathbf E} \times {\mathbf B}$ modes are positive-energy modes. They exhibit the conventional property that an increase in mode amplitude corresponds to an increase in mode energy.  However, the ${\mathbf E} \times {\mathbf B}$ and cyclotron modes are more complicated than the drumhead modes, which, in the linear approximation, behave like $N$ independent simple harmonic oscillators.  In contrast to a simple harmonic oscillator,  the time-averaged kinetic and potential energy are not equal for the cyclotron or the ${\mathbf E} \times {\mathbf B}$ modes.  The cyclotron modes are dominated by kinetic energy and the ${\mathbf E} \times {\mathbf B}$ modes by potential energy.  For the normal mode analysis of Fig.~\ref{fig:in_plane_modes}, which employs conditions relevant for recent experimental work~\cite{Shankar:2020} with $^9\textrm{Be}^+$ and parameters
 $(\Omega_c, \omega_z, \omega_r)/(2\pi)=(7.60, 1.59, 0.180)\; \textrm{MHz}$, the ratio of potential to kinetic energy for a typical ${\mathbf E} \times {\mathbf B}$ mode is of order 200 and while the ratio of potential to kinetic energy for a typical cyclotron mode is 0.005.  

The large magnetic field of a Penning trap therefore produces a separation of thermal fluctuations into potential energy fluctuations that are mainly captured by the ${\mathbf E} \times {\mathbf B}$ branch and kinetic energy fluctuations that are predominantly captured by the cyclotron branch.  This separation has consequences for Doppler laser cooling. In fact, Doppler laser cooling exploits the dependence of the scattering rate on the velocity via the Doppler shift, and therefore efficiently cools cyclotron motion.  On the other hand, the ${\mathbf E} \times {\mathbf B}$ modes are dominated by the potential energy associated with positional fluctuations of the ions from their equilibrium positions and are therefore inefficiently laser-cooled. 
In addition, recent simulations show that the equilibration rate between the cyclotron branch and the ${\mathbf E} \times {\mathbf B}$ branch is exponentially weak in the ratio of the frequency of the cyclotron branch to the ${\mathbf E} \times {\mathbf B}$ branch ($\sim\Omega_c/\omega_m$)~\cite{Tang:2021}.

Evidence for elevated temperatures of the ${\mathbf E} \times {\mathbf B}$ mode has been obtained from simulations~\cite{Johnson:2024} and from the measurement of the broadening of the spectrum of the drumhead modes, which is sensitive to fluctuations in the ion positions~\cite{Sawyer:2012, Sawyer:2014,Shankar:2020}.  This motivates developing techniques for improved measurements of the temperature of the in-plane modes, in particular, for the ${\mathbf E} \times {\mathbf B}$ modes, and generalizing techniques like ``axialization”\cite{Phillips:2008}, which use time-dependent electric field gradients to couple and equilibrate the cyclotron and magnetron motions of single or a few ions stored in Penning traps.  Doppler cooling simulations of single-plane ion crystals in a Penning trap have indicated that efficient cooling of the ${\mathbf E} \times {\mathbf B}$ modes is possible by overlapping the bandwidths of the ${\mathbf E} \times {\mathbf B}$ and drumhead modes through appropriate tuning of the ion crystal parameters~\cite{Johnson:2024}.

\subsubsection{Structural instability of the planar geometry} 
\label{sec:1_3_instability}

\begin{figure}
\includegraphics[width=0.5\textwidth]{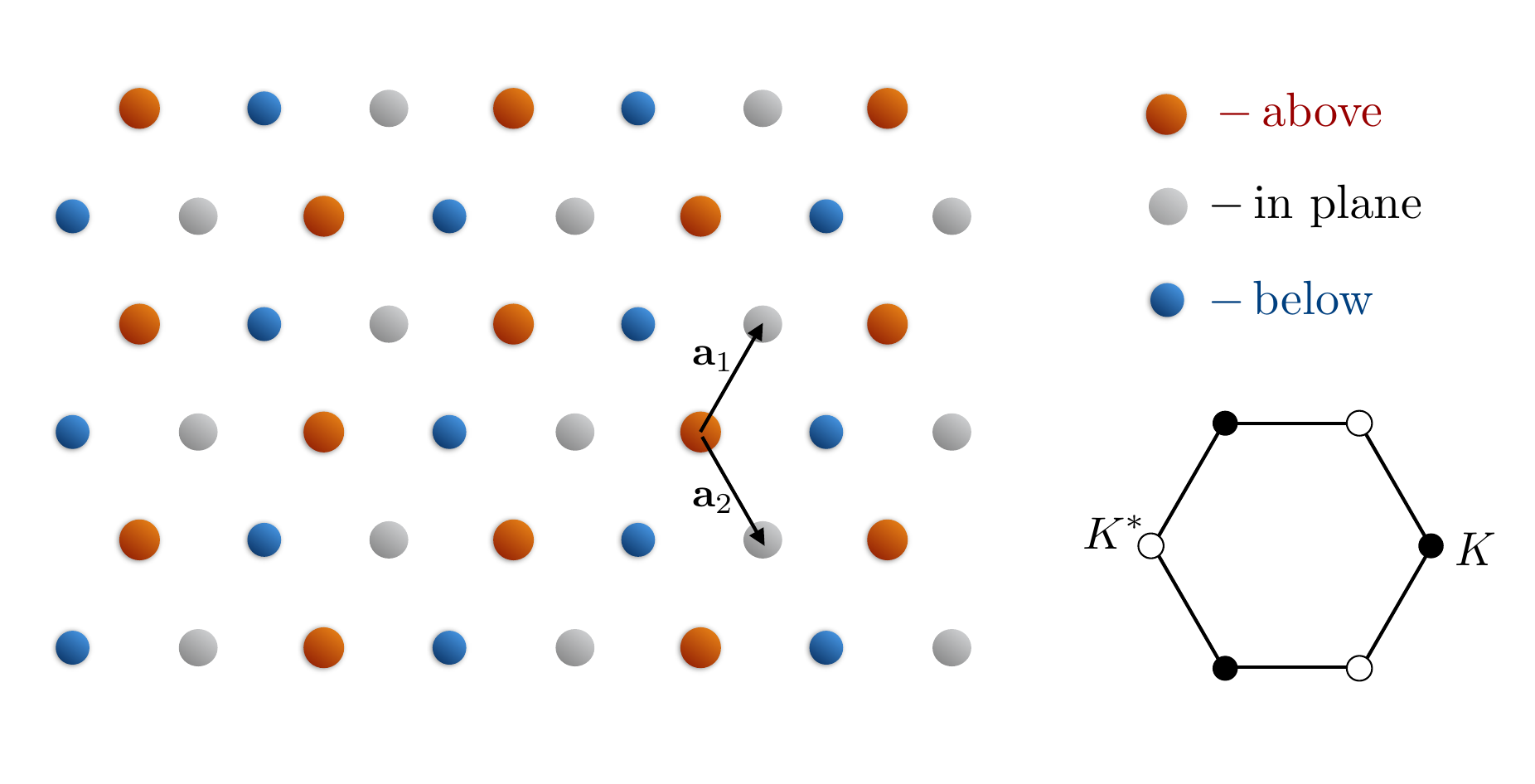} 
\caption{(Color online) A single-plane, triangular crystal undergoes a so-called buckling transition to three planes, each with triangular structure. Left panel: Height pattern in the ordered phase. The triangular lattice is split into 3 sublattices, one in the $z=0$ plane (gray medium circles), one above it (large red circles), and one below it (small blue circles). There are $3!=6$ inequivalent configurations, corresponding to the possible height assignments to the sublattices.  Right panel: The first Brillouin zone of the single-plane crystal is shown, together with the wave vectors ${\bf K}$ and ${\bf K}^*=-{\bf K}$ of the height pattern. From~\onlinecite{Podolsky:2016}}
\label{fig:1:Podolsky2016}
\end{figure}

Single-plane crystals are strictly two-dimensional geometries that are stable provided that the transverse confinement is sufficiently strong. By decreasing the transverse trap strength below a critical value, the plane becomes unstable and the crystal undergoes a transition to a three-dimensional structure \cite{Dubin:1993}. This instability is driven by the drumhead modes at the lowest frequency and manifests itself in the formation of a three-plane structure, illustrated in Fig.~\ref{fig:1:Podolsky2016}.  We call this structure the {\em buckled phase}, and the transition into this phase the {\em buckling transition}. 

The theoretical prediction of the buckling transition of Ref.~\cite{Dubin:1993} is based on minimization of the potential energy. The condition for stability of the single-plane structure can be written as $\sigma a_0^2\lesssim 1.1$, where $a_0$ is the characteristic length emerging from the competition of the Coulomb repulsion and the transverse harmonic confinement. Equivalently, stability requires the trap frequency $\omega_z$ to exceed the value $\omega_z^{\rm MF}$, with
\begin{equation}
\omega_z^{\rm MF}\approx \left(\frac{13.36\,Q^2/(4\pi\epsilon_0)}{Ma_0^3}\right)^{1/2}\,.\label{eq:omega_MF}
\end{equation}
This analysis predicts a direct phase transition from the planar to the buckled phase, which is of second order.  

Due to their large mass, ions typically have negligible kinetic energy compared to their potential energy. However, at the buckling transition, where the Coulomb and confining potential energies balance each other, kinetic energy becomes disproportionately significant. This leads to thermal and quantum fluctuations that alter the nature of the buckling instability. The role of ion kinetic energy in the buckling transition has been examined in a field theory analysis presented in Ref.~\cite{Podolsky:2016}. This theory predicts that the transition from one to three planes occurs via a two-stage process, where the planar (disordered) and buckled (ordered) phases are separated by an intermediate, partially ordered phase. This behavior is captured by mapping the field theory onto a six-state clock model, a well-studied framework in statistical physics \cite{Jose:1977,Oshikawa:2000,Lou:2007}. Below we review the basic steps leading to the mapping and its predictions.

The first step of the field theory analysis is to identify the order parameter of the buckled phase. The pattern shown in Fig.~\ref{fig:1:Podolsky2016}, for example, is encoded by the expression 
\begin{eqnarray}
z_i={\rm Re}\left[\psi e^{i {\bf K}\cdot \bm{\rho}_i}\right]\,,
\label{eq:heights}
\end{eqnarray}
where  $z_i$ is the transverse displacement of ion $i$ from the plane,  $\bm{\rho}_i$ is its in-plane lattice vector, and ${\bf K}=(4\pi/3a_{\mathrm{lat}},0)$ is the wave-vector at the corner of the Brillouin zone. Here, $\psi=|\psi |e^{i\theta}$ is a complex-valued field that serves as an order parameter. The amplitude $|\psi|$ determines the height of the buckling modulation, and the phase $\theta$ dictates the in-plane position of the maximal height modulation.
The requirement that one of the three sublattices remains level at $z=0$ (see Fig.~\ref{fig:1:Podolsky2016}) implies that there is an energetic preference for $\theta$ to take one of the discrete values $\theta=\pi(2n+1)/6$, where $n\in\{1,\ldots,6\}$. 

This naturally suggests a mapping to the six-state clock model \cite{Jose:1977,Oshikawa:2000,Lou:2007}. The mapping is performed for transverse frequencies $\omega_z$ near the mean-field critical value $\omega_{z}^{\rm MF}$ (Eq.~\eqref{eq:omega_MF}). In this regime, the displacements $z_i$, and consequently $|\psi|$, remain much smaller than the in-plane interparticle distance, which allows corrections to the planar Coulomb interaction to be incorporated via a multipole expansion.  By assuming that the complex-valued order parameter $\psi(\bm{\rho},t)$ varies slowly in space and time, a gradient expansion can be employed by treating the discrete planar distribution as a continuum. This leads to a Ginzburg-Landau (GL) theory expressed in terms of $\psi(\bm{\rho},t)$, whose form is constrained by the symmetries of the triangular lattice. The explicit formulation of the GL model can be found in Ref.~\cite{Podolsky:2016}.  Here, we present its form when the transverse trap frequency $\omega_z$ falls below the mean-field threshold $\omega_z^{\rm MF}$. In this regime, the order parameter $\psi$ acquires a finite amplitude, $|\psi| = |\psi_0|$, which is effectively described by the mean-field model. Meanwhile, the phase dynamics of $\theta(\bm{\rho},t)$ are governed by the Lagrangian density:  
\begin{equation}
{\mathcal{L}}_{\rm GL} = \frac{\rho_s^0}{2} \left(\frac{1}{c^2} |\partial_t\theta|^2 - |\nabla \theta|^2 \right) - h_6 \cos (6\theta)\,,
\label{eq:lagrangian}
\end{equation}
where the parameters $\rho_s^0$, $c$, and $h_6$ depend on $\omega_z$ and the characteristics of the single-plane crystal. The Lagrangian density in Eq.\ \eqref{eq:lagrangian} determines both the thermal and quantum phases of the six-state clock model.  

\begin{figure}
\begin{overpic}[width=0.38\textwidth]{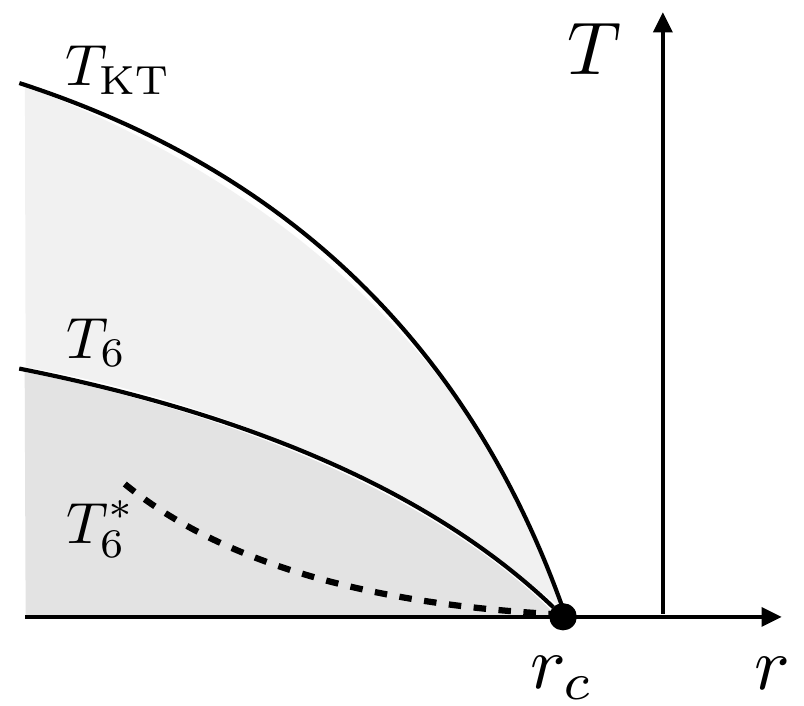}
  \put(157, 6){\Large $0$}
 \end{overpic}

\caption{Phase diagram of the structural instability of a planar ion crystal in the parameter space determined by the temperature $T$ and $r=\omega_z-\omega_z^{\rm MF}$. For $r>0$ the ions form a stable, single-plane crystal. At $r<0$, in the white region, the transverse structure is disordered because of either thermal (for $T>0$) or quantum ($T=0$) fluctuations.  The external line at $T_\mathrm{KT}$ separates the disordered phase from a critical phase, where the ions are distributed along three layers and exhibit quasi long-range order in the transverse direction. The second line at $T_{6}$ indicates the onset of the long-range ordered buckled phase, where the ions arrange in three static planes as illustrated in Fig.~\ref{fig:1:Podolsky2016}. The value $r_c<0$ indicates the critical point at $T=0$. The dashed line at $T_{6}^*$ is a crossover separating quantum and classical behaviors within the ordered phase. From~\onlinecite{Podolsky:2016}. }
\label{fig:pd}
\end{figure}

Figure~\ref{fig:pd} illustrates the phase diagram as a function of two control parameters: temperature $T$ and the transverse trap frequency, here expressed as the difference $r=\omega_z-\omega_z^{\rm MF}$ from the mean-field value $\omega_z^{\rm MF}$. At finite temperatures, the model predicts a two-stage thermal transition. At high temperatures, for $T > T_{\mathrm{KT}}$, the system remains in a fully disordered phase, where ions undergo thermal fluctuations around the single-plane structure.  In this phase, correlation functions of the atoms' heights decay exponentially with distance and time. At low temperatures, below a critical temperature denoted by $T_6$, the system enters a fully ordered phase, where the ions’ equilibrium positions form a static, three-plane buckled structure. Both transitions are of the Kosterlitz-Thouless (KT) type, with critical temperatures determined by the conditions  
\begin{equation}
\label{T:KT_T6}
T_{\mathrm{KT}} = \frac{\pi\rho_s(T_\mathrm{KT})}{2}\,,\quad T_6 = \frac{2\pi\rho_s(T_\mathrm{KT})}{9}\,,
\end{equation}  
where $\rho_s$ denotes the renormalized stiffness. In the intermediate regime $T_6<T<T_\mathrm{KT}$, a critical phase emerges where the height correlations decay as a power law (on top of oscillations at the wave-vector ${\bf K}$ of the buckling structure):
\begin{equation} 
\label{eq:powerlaw_corr} 
\langle z(\bm{\rho})z(0)\rangle\sim |\bm{\rho}|^{-\eta}\,\cos(\mathbf{K}\cdot\bm{\rho})\,,\quad \eta=\frac{T}{2\pi\rho_s}\,.
\end{equation}
Noting that $0<\eta<1/4$, this phase exhibits quasi-long-range order of the buckled structure.

As the temperature $T$ decreases, quantum fluctuations become increasingly dominant, ultimately driving a phase transition at a single quantum critical point $(T=0,\, r=r_c)$ \cite{Oshikawa:2000,Lou:2007}. For temperatures above $T_6(r)$, the clock term $\cos(6\theta)$ in Eq.~(\ref{eq:lagrangian}) is effectively suppressed by fluctuations. Consequently, the model describing the transition corresponds to the quantum XY model in two space dimensions, implying that the stiffness $\rho_s$ scales as $r$ approaches $r_c$ according to  
\begin{equation}
\rho_s (r,T) = \Delta \Phi(T/\Delta)\,,\quad \Delta \sim |r - r_c|^{\nu}
\label{eq:rho_s_vs _r}
\end{equation}  
where $\nu \approx 0.671$, and $\Phi(x)$ is a universal scaling function of a single variable $x$, which remains nearly constant for $x \leq 1$. As a result, the values of $T_\mathrm{KT}$ and $T_6$ — and thus the extent of the critical intermediate phase —
shrink as $r$ approaches $r_c$, ultimately vanishing together at the quantum critical point $r = r_c$ at $T=0$ (see Fig.~\ref{fig:pd}).  

For finite but sufficiently
small $T$, a crossover occurs within the ordered phase from
classical to quantum behavior at a temperature $T_6^*<T_6$. This
signifies the emergence of an additional characteristic length scale, $\xi_6\sim 1/T_6^*$, beyond which the long-range clock order becomes apparent 
\cite{Oshikawa:2000,Lou:2007}.
The clock term influences the system's dynamics by introducing a gap, due to breaking of the continuous phase-displacement symmetry of $\theta$. 
This gap, of order
$T_6^*$, is strongly suppressed as the system approaches the quantum critical point.

The experimental signatures of the predicted phases are the measurement of the layer separation and the structure form factor. The layer separation $h$, associated with the formation of the three planes, serves as a measurable probe of the order parameter. In particular, tracking its behavior as a function of $T$ and $r$ can distinguish the three phases. In addition, it is suggestive of the classical-quantum crossover at $T_6^*$: for a given $r<r_c$, the value $T_6^*$ can be identified with the temperature scale at which $h(T)$ saturates to a constant value. The structure form factor provides information on the enlargement of the unit cell, signaled by additional Bragg peaks, and on the critical, KT phase. The latter shall become visible in the height and lineshape of the peaks as a function of the temperature, which are expected to exhibit a power-law dependence \cite{Podolsky:2016}. Small layer separations can be detected by means of interferometric measurements, based on spin-motion entanglement \cite{Baltrusch:2011,Baltrusch:2012,Gilmore:2021}. The structure form factor is accessed by Bragg scattering measurements \cite{Itano:1998,Dantan:2010}. 

In 2D ion crystals available to date, experiments so far have not unambiguously revealed the occurrence of the buckled phase, but its verification, while challenging, is possibly within reach. 
We now provide some estimates of the required experimental conditions. Sufficiently far from the quantum critical point $r_c$, the critical temperatures $T_{\mathrm{KT}}$, $T_6$ (which are of the same order of magnitude - see Eq. \eqref{T:KT_T6}) are found to be two orders of magnitude smaller than the Coulomb energy, yielding $\sim 10$~mK for typical interparticle distances of order $15~\mu$m. Tracking of their shrinking as $r\rightarrow r_c$, and accessing the lower temperature scale $T_6^*$, would require cooling down to temperatures of the
order of hundreds of microkelvin. This regime is accessible using available sub-Doppler cooling techniques \cite{Eschner:2003,Lechner:2016}. A more serious limitation arises due to the finite size of the crystal.
Present experiments can realize
planes with $N\sim 1000$ ions   corresponding to linear sizes of  $L\sim 30$ ions (although typically an ordered triangular lattice is formed on a smaller scale). In this regime, finite size effects will strongly affect the observations. However, as argued in 
\cite{Podolsky:2016}, a systematic finite-size scaling analysis can be utilized to perform a meaningful comparison with the theory.

\subsection{Three-dimensional crystals}
\label{sec:three-dimensional crystals}

Three-dimensional (3D) trapped ion crystals have been prepared in Penning and rf ion traps since the early days of laser cooling~\cite{Diedrich:1987,Wineland:1987,Gilbert:1988}.  While 1D and 2D ion crystals have been limited to $\sim500$ ions, 3D crystals can reach sizes greater than $10^5$ ions~\cite{Tan:1995}. Large, 3D crystals provide intriguing opportunities for developing new control and characterization techniques because the current techniques employed with 3D crystals is not as advanced as that developed for 1D and 2D crystals. To date, only Doppler cooling has been applied with 3D crystals. We review here the types of 3D crystal configurations that have been prepared in the lab, discuss the importance of finite size effects, and summarize characteristic properties of the vibrational motion.

\subsubsection{Ground state configuration} 
\label{Sec:ground state config}

Before discussing the detailed structure of three-dimensional ion Coulomb crystals, it is useful to consider the cold fluid plasma model of a single species ion crystal, in which the ions are approximated by a continuous charge distribution in space. This is also called the zero-temperature, mean-field plasma~\cite{Dubin:1999}. For simplicity, consider an effective harmonic confinement potential in all three dimensions, which is routinely achieved in Penning and quadrupole rf traps.  At low temperatures this gives rise to ion crystals with constant density and sharp boundaries.  In the cold fluid model, the continuous charge distribution is uniform in space with approximately the same boundary as the ion crystal. The energy required to assemble the continuous charge distribution does not scale extensively with (that is, it is not proportional to) the number of trapped charges.

The cold fluid model is useful for determining the overall crystal shape and the number of trapped ions through the volume
enclosed by the crystal shape, but the specific lattice structure is dictated by the discreteness of the charges and, more precisely, by the short distance contributions to the energy. The latter is the so-called correlation energy and scales extensively with the lattice size.  Three-dimensional strongly coupled OCPs are peculiar systems in that three different lattice configurations have nearly the same correlation energy per ion in the infinite (bulk) lattice and at zero temperatures.  These configurations are body centered cubic (bcc), face centered cubic (fcc) and hexagonal close packed (hcp).  The bcc lattice is the zero temperature ground state~\cite{Brush:1966, Ichimaru:1982}, but the bulk energy per ion in the fcc and hcp are less than $1\times10^{-4}$ higher than the one in the bcc configuration~\cite{Dubin:1989,Baiko2001}.

The near degeneracy of the bcc, fcc, and hcp lattices means that the actual ground state configuration of finite, 3D trapped ion crystals can be strongly impacted by boundary effects.  In general, large system sizes are required  for the bcc lattice to be the thermodynamic ground state. Below we review the structures that have been studied and realized in the laboratory for different confinement geometries.

\begin{figure}
\includegraphics[width=0.40\textwidth]{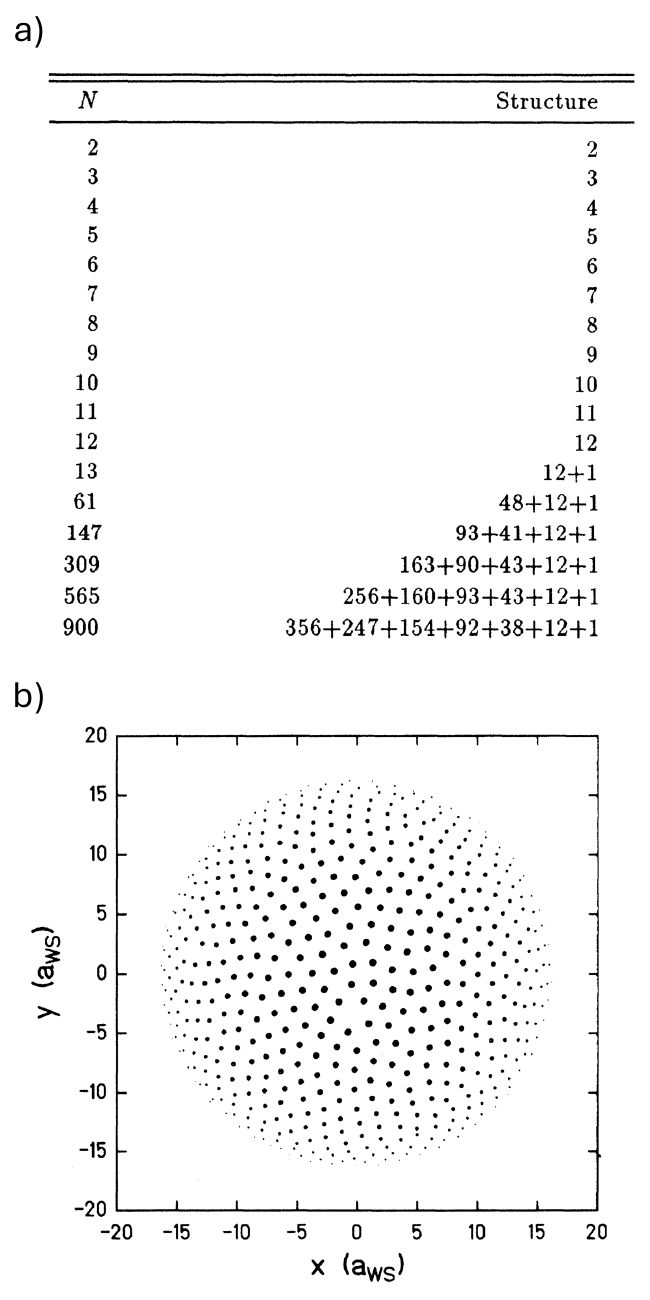}
\caption{Molecular dynamics simulations of spherical symmetric confined ion Coulomb crystals. a) Shell structure as a function of the number of ions $N$. When a sign $+$ appears in the column $Structure$, it means a new shell starts forming. In this case, $N$ is the corresponding "magic" number."  In all simulations the final Coulomb coupling parameter $\Gamma$ is larger than 10$^5$. b) Illustration of the position of the ions in the outer shell of a 5000 ion system. Note the general hexagonal structure with a few point defects, such as the one below the center. The length scale is the Wigner-Seitz radius $a_{\rm WS}$. From\ \onlinecite{Hasse:1991}.} 
\label{fig:magic numbers}
\end{figure}

\emph{Spherical geometry---} Consider the most symmetric case of isotropic confinement, where the confining force can be written in terms of a single force constant $\kappa$,

\begin{equation}
\label{Eq:F_trap}
\bm{F}_{\text{trap}}(\bm{r})=-\kappa\bm{r}\,, 
\end{equation}
with $\bm{r}$ the spherical radius vector. Due to the spherical symmetry of the confining potential, the global structure of finite Coulomb crystals is expected to be spherical as well. Crystals containing up to a few thousand ions minimize their potential energy as the temperature goes to zero by organizing themselves in concentric spherical shells with a specific maximum “magic” number of ions in each shell~\cite{Dubin:1988,Hasse:1991,Schiffer:2003,Calvo:2012}; see Fig.~\ref{fig:magic numbers}. Within each shell the ions form a two-dimensional triangular lattice (like single-plane crystals; see Sec.~\ref{Sec:GroundState2D}), but with a number of defects to account for the curvature of the surface (see Fig.~\ref{fig:magic numbers}). The ground state (zero temperature) configurations are traditionally found by Monte Carlo methods or molecular dynamics simulations combined with careful annealing procedures. It is challenging, however, to determine the absolute ground state configurations already for a modest number of ions ($\sim$100), since, as is well-known from cluster physics, the number of slightly excited metastable configurations increases exponentially with the particle number. Combined with the small deviations from a perfect harmonic confining potential in real experiments, predictions of the exact structures obtainable experimentally become extremely challenging. 

The uncertainty in the predicted magic numbers of ions within each shell is, though, so small that when observing a given spherical shell structure for crystals consisting of up to thousand ions in the laboratory, one can typically obtain an estimate of the number of ions at an uncertainty level much better than the square-root of the number of ions. Hence, measurements of the crystal structure provide a reliable way to obtain information on the number of ions in a real crystal.

\begin{figure}
\includegraphics[width=0.40\textwidth]{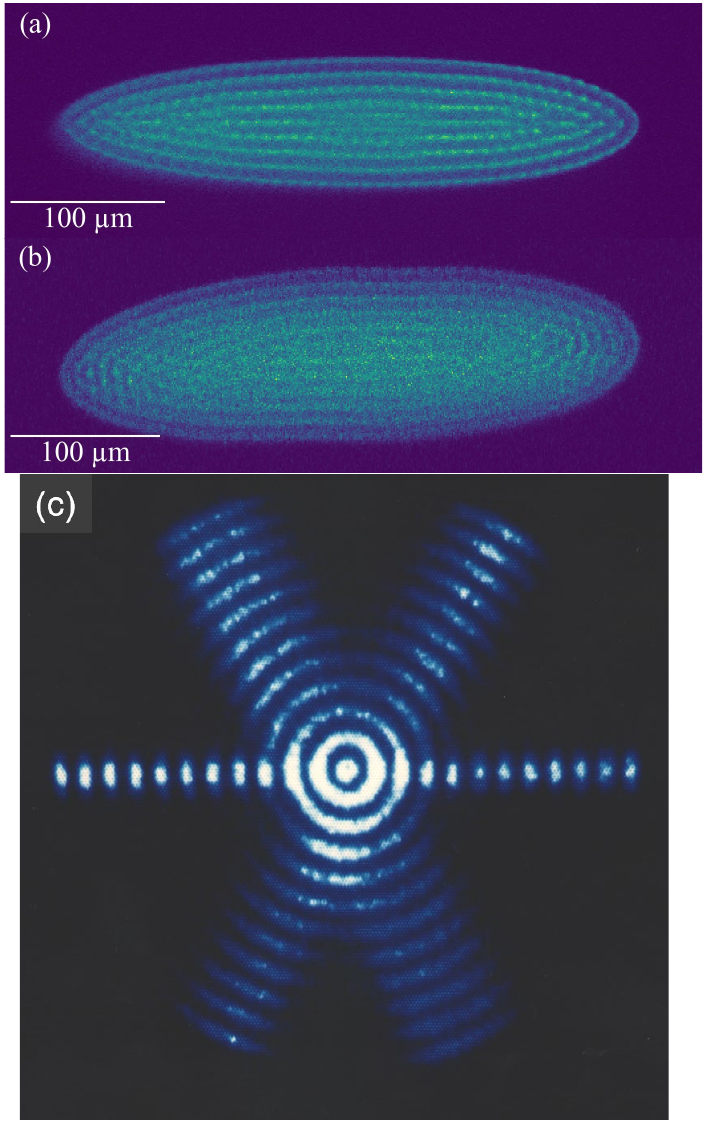}
\caption{Images of the ion fluorescence from trapped-ion crystals exhibiting shell structure.  (a) Horizontal and (b) vertical view of an ion crystal consisting of about 2000 Be$^+$ ions stored in a linear rf trap. 
From\ \onlinecite{Schmid:2022}. (c) Image of the ion fluorescence from three laser beams directed through the center of a $N\approx 15000$ Be$^+$ ion crystal that exhibits shell structure.  The ions are stored in a Penning trap and the image, averaged over many rotations of the ion crystal, is obtained along the magnetic field or rotation axis of the ion crystal. The crystal diameter is 0.6 mm. From\ \onlinecite{Gilbert:1988}.} \label{fig:shell_structures}
\end{figure}

\emph{Ellipsoidal and spheroidal geometries---} A similar shell structure is obtained for the more general ellipsoidal ion crystal shapes (or spheroidal if the confinement in two orthogonal directions is equal \footnote{A spheroid is the shape obtained by rotating an ellipse about one of its principal axes}). The Coulomb potential energy of the crystal is minimized by forming a shell on the boundary of the crystal that conforms with the overall crystal shape determined by the confining potential.  This shell structure propagates into the interior of the ion crystal through the formation of commensurately shaped shells of smaller size.  Experimentally, shell structures are routinely observed in trapped 3D ion crystals~\cite{Drewsen:1998,Gilbert:1988,Hornekaer:2002,Schmid:2022}.  See Fig.~\ref{fig:shell_structures} for experimental images.

An interesting question is at what size will it become energetically favorable for an ion crystal to have a bcc lattice as its ground state configuration. Molecular dynamics simulations with spherical OCPs indicate that the spherical bcc crystal has lower energy than shell structure for $N > 10^4$. The simulations were performed independently in \cite{Totsuji:2002,Hasse:2003} with ion numbers ranging from a few thousand to greater than $10^5$ ions. Interestingly, if the molecular dynamics simulation started with randomly distributed ions and was slowly annealed, the formation of spherical shells was always observed that advanced from the boundary of the crystal to the center. On the other
hand, starting with a spherical cutout of a bcc lattice and then annealing resulted in a bcc crystal in the interior surrounded by a few shells at the boundary.

Experimentally, stable bcc crystals have been observed in the interior of large ion crystals formed in Penning traps through Bragg scattering of the laser-induced resonance fluorescence as well as through direct imaging of the ion resonance fluorescence~\cite{Bollinger:2000}. Long range 3D periodic order was observed in approximately spherical crystals with as few as $N \approx 5\times 10^4$~\cite{Tan:1995}. The crystal diameter was $\sim70 a_\mathrm{WS}$ where $a_\mathrm{WS}$ is the Wigner-Seitz radius defined in Eq.~(\ref{eq:WS_radius}).  With $N > 2\times10^5$ ions (crystal diameter $> 130 a_\mathrm{WS}$), bcc crystals were the only type of 3D periodic crystal that was reported~\cite{Itano:1998}. This work could only place a crude lower limit on the observed bcc crystal size of $20 a_\mathrm{WS}$, which is a small fraction of the entire ion crystal size. In linear rf traps, stable fcc and bcc crystalline structures have been also observed in Coulomb crystals  with more than $10^4$ ions through fluorescence imaging~\cite{Mortensen:2006}.   

\begin{figure}
\includegraphics[width=0.48\textwidth]{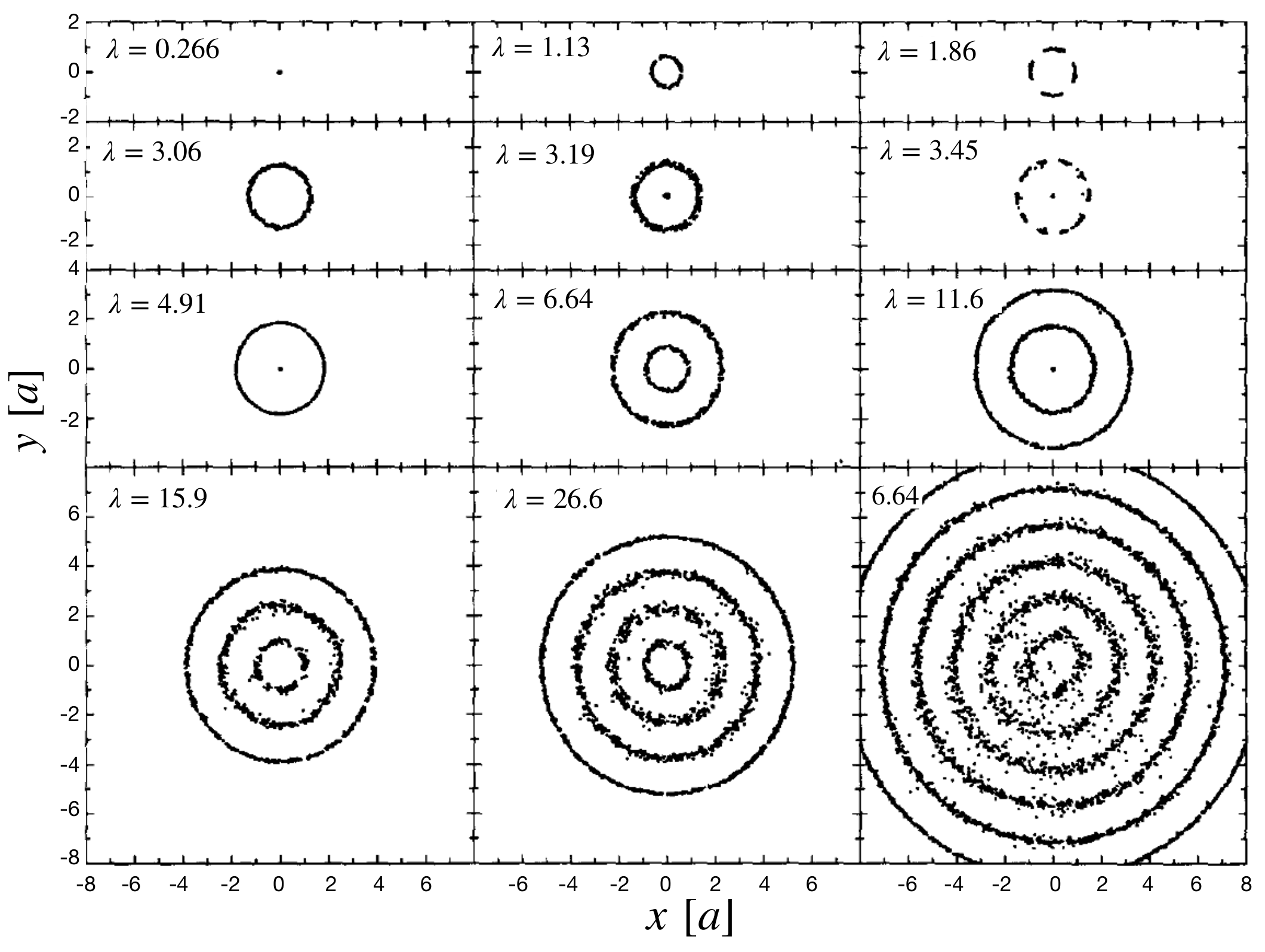}
\caption{Molecular dynamics simulation of cross sections of cylindrically confined Coulomb crystals for various normalized linear charge densities $\lambda$. For the lowest charge density, $\lambda$=0.266, the charged particles form a one-dimensional string (a single dot in the cross section view), whereas, as the charge density increases, an increasing number of concentric shells, sometimes with a central string, form. The figure is adapted from \onlinecite{Hasse:1990}, where more details on the simulations can be found.}
\label{fig:Hasse cylindrical confinement}
\end{figure}

\emph{Cylindrical geometry---} A different trapping situation for three-dimensional crystals is realized when the ions are confined by an isotropic harmonic potential in two dimensions while being free to move along the third dimension. Here, the parameter governing the crystal structure is the linear density of charged particles along the unconfined axis. For low linear densities the minimum energy states are concentric cylindrical shell structures with varying surface structures within the shells. As the linear density increases the core forms lattice structures that are translationally invariant along the axis (See Fig.~\ref{fig:Hasse cylindrical confinement}).

\begin{figure}
\includegraphics[width=0.5\textwidth]{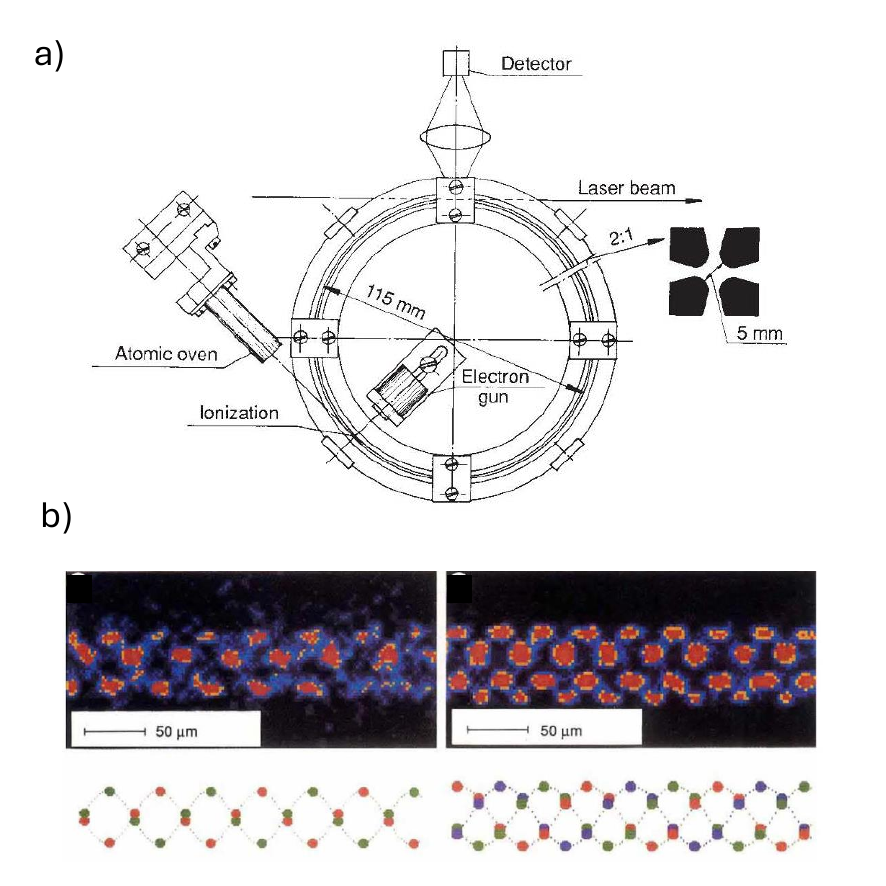}
\caption{Sketch of the rf ring trap experiment used to study ion Coulomb crystals under effective cylindrical confinement (a), together with some examples of fluorescence images of ion crystalline structures (here helical structures) obtained with different linear densities (b). The helical structures are sketched below the images. See also Fig.~\ref{Fig:LinZigzag}. From \onlinecite{Birkl:1992}.} 
\label{fig:Walther ring exp}
\end{figure}

In the laboratory this idealized situation cannot be fully implemented. It can effectively be realized by choosing a ring shaped two-dimensional rf quadrupole trap, as has been pioneered by Herbert Walther and collaborators\, \cite{Birkl:1992,Waki:1992} (see Fig.~\ref{fig:Walther ring exp}). As long as the ring's circumference is much larger than the typical ion-ion distance and the radial extent of the crystal, only very small deviations from the ideal infinite situation are expected.

More surprisingly, when two different ion species are simultaneously confined in a linear rf quadrupole trap, cylindrical crystal structures can be realized that mimick those of infinitely long systems~\cite{Hornekaer:2001}. This should also be observable in Penning traps~\cite{Larson:1986,Imajo:1997}. See the paragraph on \emph{Two-species Coulomb crystals} which concludes this subsection.

\emph{Slab geometry---} The slab geometry consists of a uniform density OCP that is harmonically confined in one direction, $q\phi_T(\rho,z) = m\omega_z^2 z^2/2$, but infinite in the orthogonal directions. This geometry allows for a careful theoretical treatment that illustrates the importance of surface effects with 3D trapped ion crystals~\cite{Dubin:1989}. The treatment assumes that the lattice planes that form in the $T=0$ limit have the same symmetry. It is found that, although bcc has a lower bulk energy per ion than fcc, the fcc lattice has a smaller surface energy contribution than bcc, making it the lower energy lattice for thicknesses with smaller numbers of planes.  Dubin showed that when the slab thickness is larger than approximately 60 planes, bcc becomes the structure that minimizes the energy, while for smaller thickness fcc prevails.

Experimentally this geometry can be approximated in ion traps by adjusting the trap radial confinement to be much weaker than the axial confinement.  In a Penning trap this condition can be achieved with low rotation frequencies $\omega_r$ that are slightly higher than the magnetron frequency $\omega_m$ of the trap (see Eq.~\eqref{eq:magnetron} and Fig.~\ref{fig:Pening_crystal_shapes}).  For a given rotation frequency, this is the regime that gives rise to single-plane crystals (see Sec.~\ref{sec:planar_geometries}) for a sufficiently small number of ions.   With larger numbers of trapped ions, multi-plane crystals form with an overall pancake (or lenticular) shape.  The lattice planes have some curvature, but near the radial center ($\rho\approx 0$) of the trap, the planes are very flat and can be compared with the structures of infinite planar systems analyzed by \cite{Dubin:1993}. 

\begin{figure}
\includegraphics[width=0.5\textwidth]{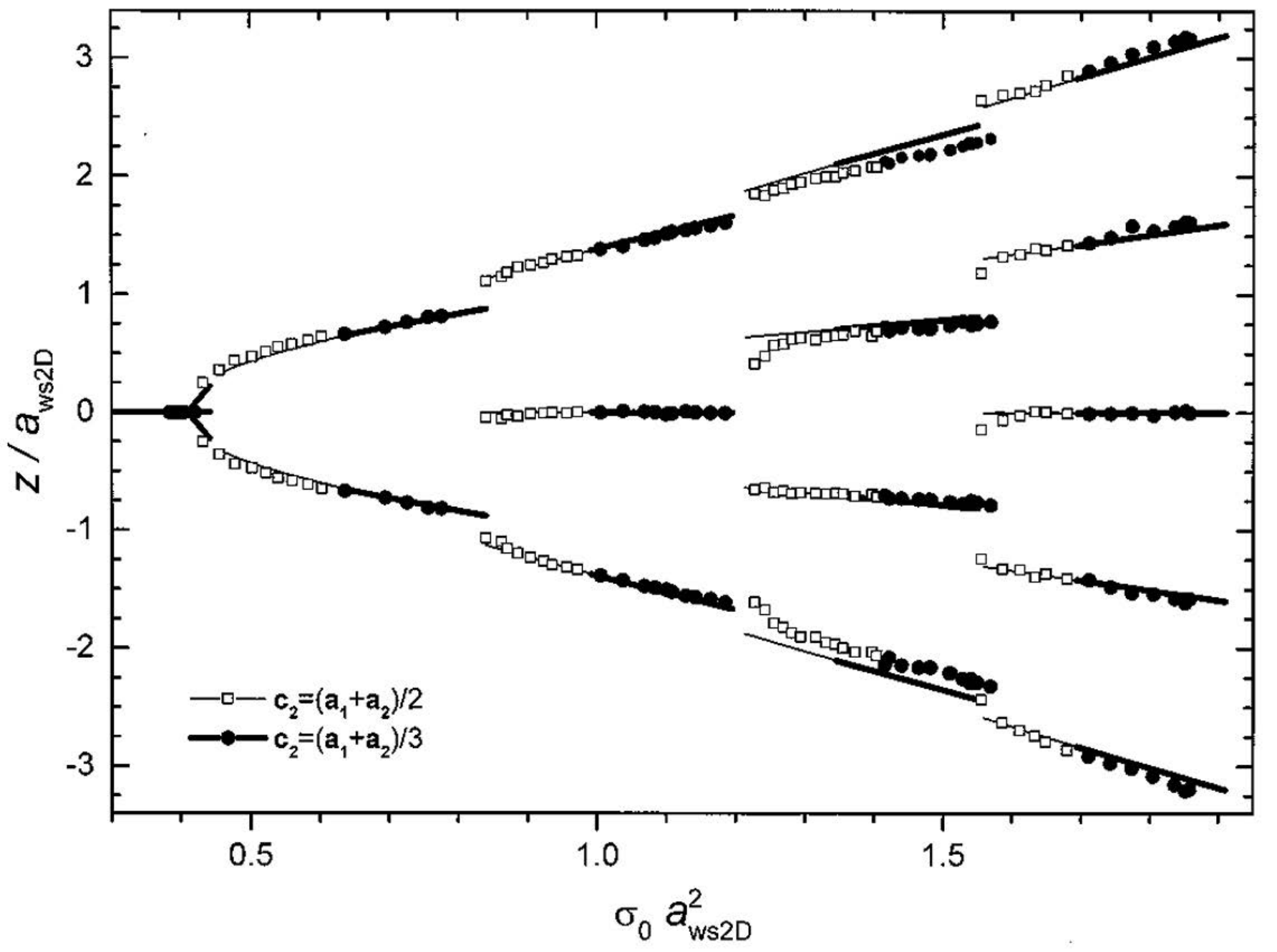}
\caption{Axial positions of the planes in the central region ($\rho \approx 0$) of small aspect ratio ion crystals as a function of the central areal number density.  The lines show predictions from theory, and the symbols show experimental measurements.  An open square (filled circle) indicates the experimental observation of a square or rhombic plane (hexagonal plane) was experimentally observed.  Similarly, a thin line (thick line) indicates that theory predicts a square or rhombic plane (hexagonal plane).  Lengths have been normalized by $a_{WS2D} = (3e^2/4\pi\epsilon_0m\omega_z^2)^{1/3}$. From\ \onlinecite{Mitchell:1998}, where the interlattice displacement vector $\bf{c}_2$ is defined.} \label{fig:Mitchell_slab_geo}
\end{figure}

Figure~\ref{fig:Mitchell_slab_geo} summarizes the ion crystal structures observed in the laboratory as a function of the 2D areal number density ($\sigma_0 a_{ws2D}^2$) obtained by integrating over the axial extent of the crystal~\cite{Mitchell:1998}.  The crystal structure was determined from direct images of the ion fluorescence from the central ($\rho \approx 0$) region of Doppler laser-cooled ion crystals with aspect ratios $\alpha\equiv Z_p/R_p < 0.1$ and ion number $N \lesssim 10^4$ ions.  Figure~\ref{fig:Mitchell_slab_geo} also compares the measured separation of the planes with the theory of \cite{Dubin:1993}. Starting with low 2D number density, the experiment reported single-plane crystals with a triangular lattice (see Sec.~ \ref{Sec:GroundState2D}). 
With increasing $\sigma_0 a_{ws2D}^2$, theory predicts a ``buckling-transition" to three closely spaced planes that is stable over a very small interval of $\sigma_0 a_{ws2D}^2$ (see Sec. \ref{sec:1_3_instability}).  This small region of three planes is followed by two, more widely separated planes.  Initially each plane has a square and then rhombic symmetry.  As the planes move further apart, they transition into planes with hexagonal symmetry.  With increasing  $\sigma_0 a_{ws2D}^2$, eventually 3 planes with square symmetry form, and the pattern repeats. As visible from the figure, the overall agreement between experimental observations with the $T=0$, infinite extent theory of \cite{Dubin:1993} is good, with some minor discrepancies~\cite{Mitchell:1998}. One disagreement is that the transition from a single plane to a small region of closely spaced three planes was not observed. This transition is particularly interesting (see Sec.~\ref{sec:1_3_instability}) and motivates a more careful study at lower temperatures with new experimental tools. 

Interestingly, the crystal planes with rhombic symmetry, observed in \cite{Mitchell:1998}, have peculiar thermodynamic properties such as negative Poisson ratios.  A negative Poisson’s ratio means that stretching the crystal in a particular direction, in this case, along the $\hat{z}$ or axial direction of the trap, causes an expansion in a lateral direction.  This behavior contradicts common experience for ordinary materials that act as incompressible, such as rubber bands.  Negative Poisson ratios were measured in ~\cite{Baughman:2000}.  

We close this section on slab geometries by pointing out that multilayer 3D crystals of trapped ions consisting of flat planes can be formed through the addition of anharmonic terms to the trapping potential.  This was numerically demonstrated for crystals in Penning traps, where the anharmonic terms are additions to the usual harmonic trapping potential (see Eq.~\ref{eq:quadratic}) and are readily implementable in the lab~\cite{Hawaldar:2024}.  The formation and control of ion crystals in traps with significant anharmonic components to the trapping potential is a subject of ongoing experimental investigation.  

\begin{figure}
\includegraphics[width=0.5\textwidth]{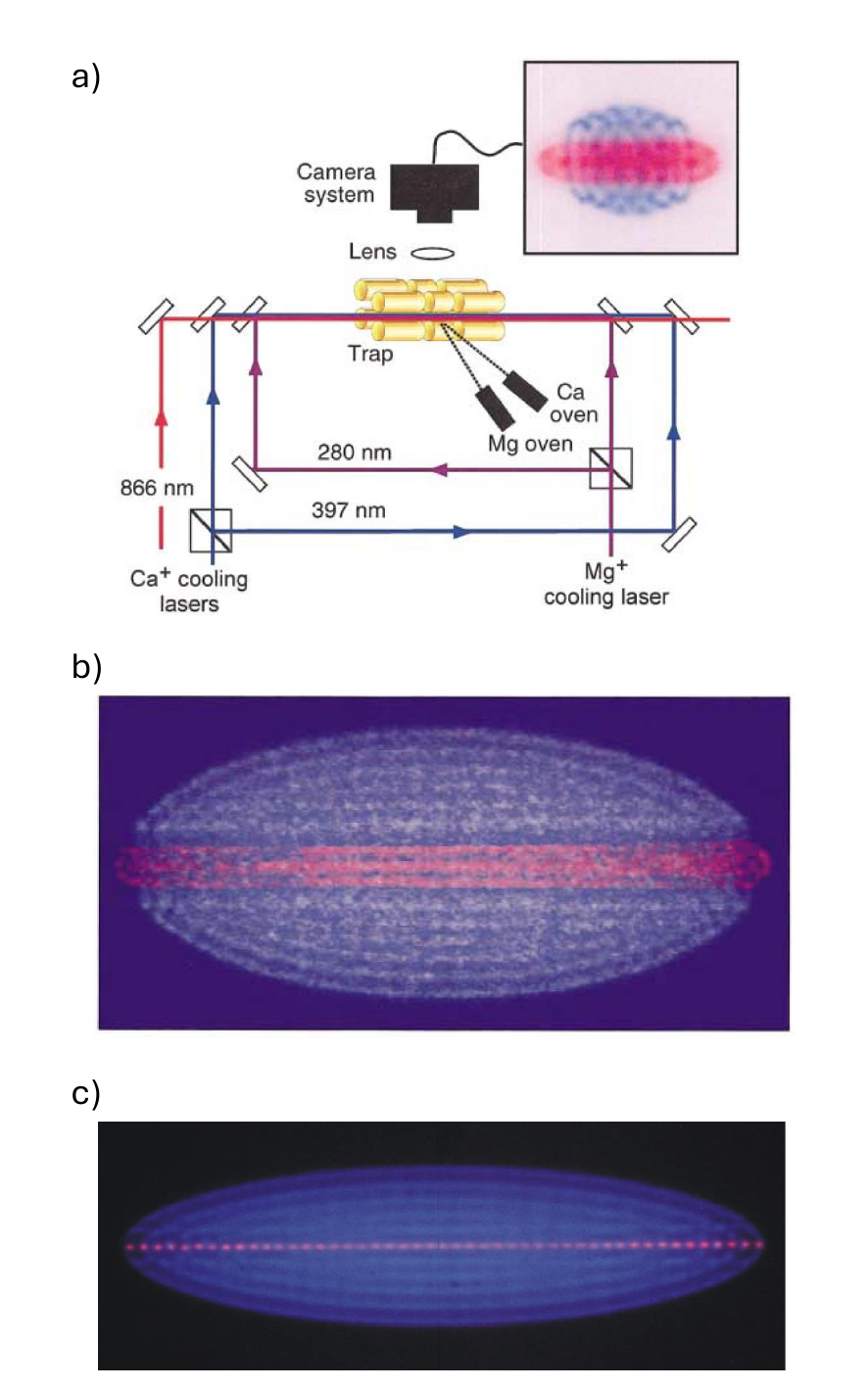}
\caption{(Color online) (a) Sketch of a linear rf trap experiment to study two-species ion Coulomb crystals consisting of Mg$^{+}$ and Ca$^{+}$ ions (a). The lower plots show fluorescence images for (b) 300 Mg$^{+}$ ions (red central region) and 3000 Ca$^{+}$ ions and (c) 47 Mg$^{+}$ ions and 1300 Ca$^{+}$ ions. Both crystals have a spheroidal global shape with a cylindrical symmetry axis coinciding with the central axis of the Mg$^{+}$ component. From \onlinecite{Hornekaer:2001}.} 
\label{fig:Two species crystals}
\end{figure}

\emph{Two-species Coulomb crystals---} As indicated previously, two-species Coulomb crystals constitute a very interesting and special type of Coulomb crystal. It is instructive to draw a comparison with normal solids formed through electronic (chemical) bindings between the atomic constituencies. In this case, when more atomic species are involved, various alloys form. In contrast, in Coulomb crystals the confined ions are mainly interacting through the repulsive Coulomb forces at a distance that inhibits chemical bonding. Consequently, the structural organization of two simultaneously trapped ion species is the one that minimizes the potential energy of all ions. Since in both linear Paul (rf) traps and Penning traps, the effective confining potential in the plane perpendicular to the static confining direction depends on both the charge and mass of a specific ion species, a two-species Coulomb crystal will generally lead to a separation of the two species with respect to this plane. This type of separation consists of the strongest bound species forming a cylinder-like central structure within an embedding spheroidal, globally shaped crystal and has been reported in\,\cite{Hornekaer:2001,Mortensen:2007}. Examples are shown in Fig.~\ref{fig:Two species crystals} and Fig.~\ref{fig:Two isotope crystals}. When the charge-to-mass ratio of the two species is the same, the two species mix. This has been observed in experiments where the two trapped species were $^{26}$Mg$^{+}$ and $^{24}$MgD$^{+}$ ions \cite{Molhave:2000}. It was also reported in molecular dynamics simulations where one species has both double the mass and charge of the other \cite{Matthey:2003}. In the latter case, the ground state of an infinite system is expected to be a simple cubic lattice with the unit cell consisting of one species at a corner and the other displaced half way the $\langle 111\rangle$ diagonal, similar to chemically bound CsCl crystals in solid state.

\subsubsection{Metastable and perturbation induced structures} 
\label{sec:Metastable:3D}

The theoretical analysis of the experimental results discussed so far have assumed that the trapping potential is conservative and purely harmonic. However, this is only an approximation of the experimental situations. Specifically, in rf traps the time dependence of the trapping potential can produce measureable deviations from this idealized situation, thus impact the formation of ion crystal structures and even the resulting structures.

For instance, in linear rf traps operated with $a_\mathrm{dc}=0$ the pseudopotential is symmetric under rotation about the $z$-axis in the geometry of Fig.\ \ref{fig:linear_rf_traps}. However, this continuous symmetry is only an approximation of the exact discrete symmetry, imposed by the four electrode configuration of the trap. The effect of this discrete symmetry is already evident in clusters composed of two ions, which are forced by a strong confining force to lie in the $x - y$ plane: molecular dynamics simulations (as well as expansions of the pseudo potential beyond the leading order \cite{Broener:2000}) show that at equilibrium the two ions always align along one of the (1,1) or (1,-1) directions in the plane, where the orientation of the micromotion is perpendicular to the line connecting the two ions. Similarly, for weak confinement along the $z$-axis, small zigzag structures always align in the same directions. 

\begin{figure}
\includegraphics[width=0.5\textwidth]{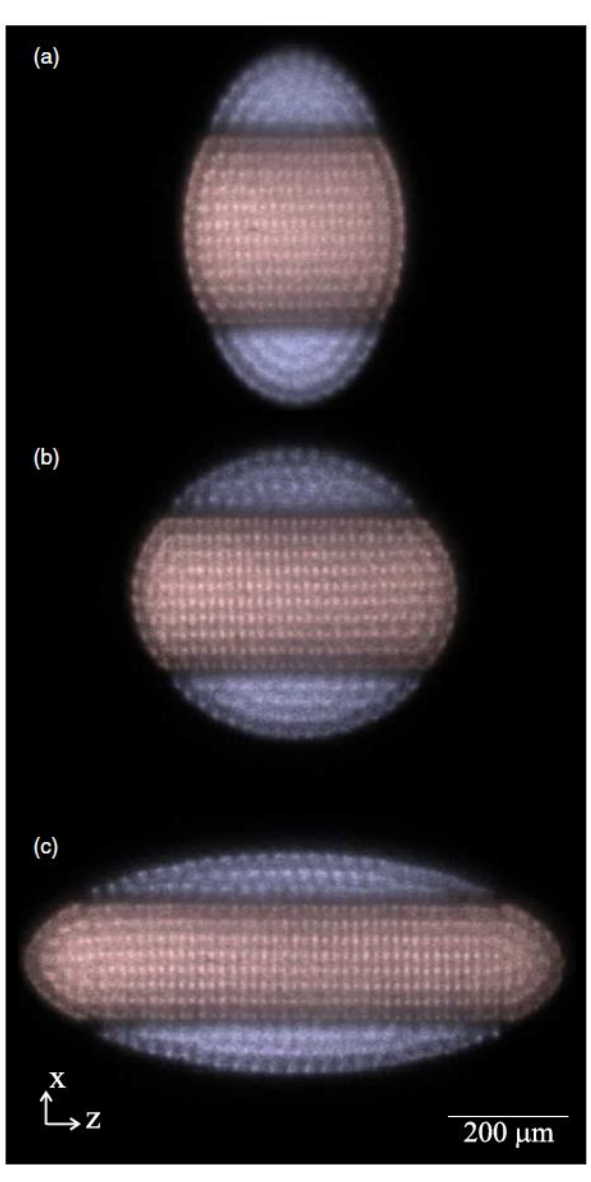}
\caption{(Color online) Fluorescence images of two-species ion Coulomb crystals containing $\sim$1500 $^{40}$Ca$^{+}$ ions in the central region (red)
and $\sim$2000 $^{44}$Ca$^{+}$ ions in the surrounding regions (blue) for the same ion densities, but different aspect ratios of the trapping potential. From~\onlinecite{Mortensen:2007}.} 
\label{fig:Two isotope crystals}
\end{figure}

In three-dimensional ion crystals the micromotion occurs along several directions and the resulting order is therefore more challenging to predict. There are however some general properties extracted from experimental evidence. 
Three-dimensional crystals that are laser-cooled by beams propagating along the axis of the linear trap are never observed to rotate around the same axis. This occurs for single as well as for multi-species crystals and indicates that the trap potential tends to pin the crystals. Moreover, translationally periodic crystal structures of single species crystals seem to have a propensity to occur with, e.g., bcc or fcc configurations at specific orientations with respect to the trap electrodes~\cite{Mortensen:2006,Mortensen:2007}. More pronounced effects have been observed in two-species crystals consisting of two isotopes of Ca$^+$ ($^{40}$Ca$^{+}$ and $^{44}$Ca$^{+}$), as shown in Fig.~\ref{fig:Two isotope crystals} \cite{Mortensen:2007}. Here, the inner $^{40}$Ca$^{+}$ component, in addition to having a global cylindrical shape, had a very persistent internal structure (presumably orthorhombic) with fixed relation to the trap electrodes and with life times longer than 10 seconds.
Similar results were found by simulations, which included the full rf trap potential~\cite{Mortensen:2007}.

\begin{figure}
\includegraphics[width=0.5\textwidth]{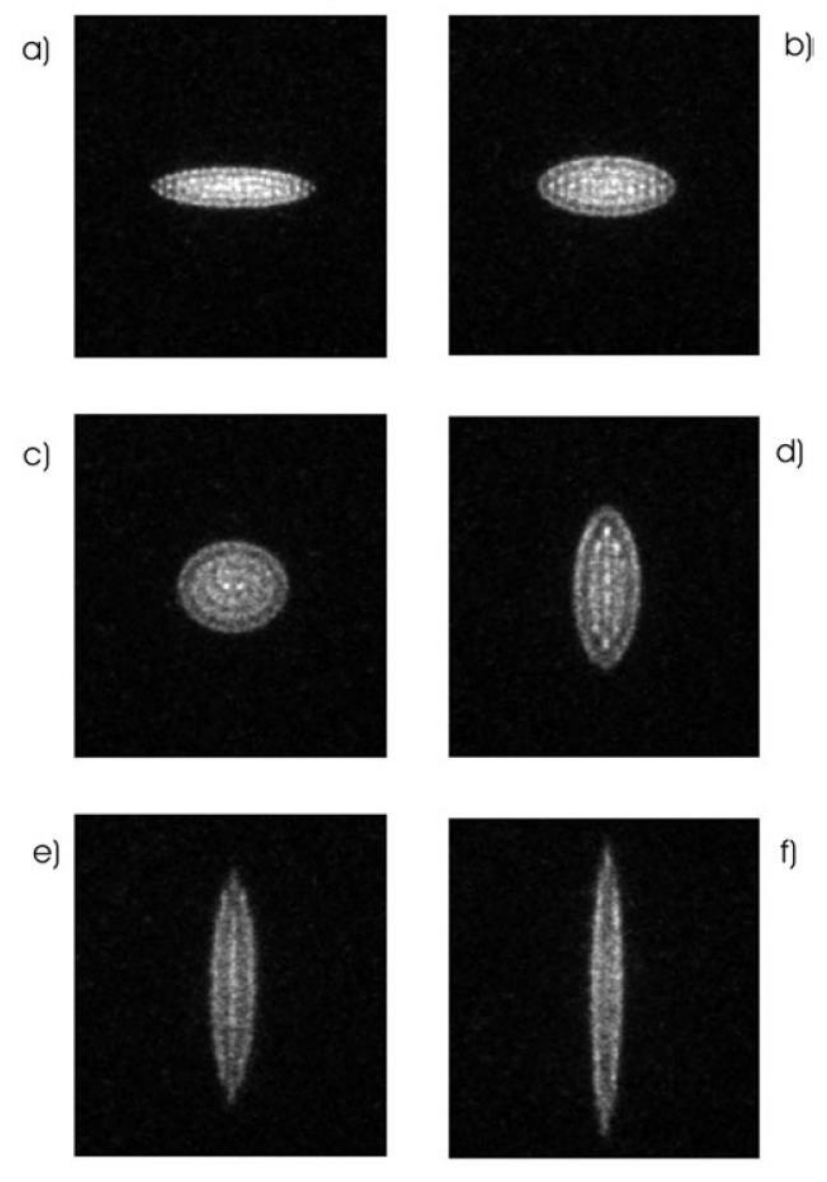}
\caption{Images of an ion crystal consisting of $\sim$200 $^{24}$Mg$^+$ ions illuminated by two traveling laser beams. One laser beam propagates along the the $z-$axis and laser cool the crystal. The other beam  propagates within the $x-y$ plane and exerts a torque on the crystal forcing a rotation, whose axis is left-right in the figure. The image in (a) corresponds to a stationary crystal. The other images show rotating crystals with increasing rotational frequency $\Omega_{\rm rot}$, namely: (b) $\Omega_{\rm rot}=0.4\omega_{\rm r,trap}$, (c) $\Omega_{\rm rot}=0.7\omega_{\rm r,trap}$, (d) $\Omega_{\rm rot}=0.9\omega_{\rm r,trap}$, (e) $\Omega_{\rm rot}=0.96\omega_{\rm r,trap}$ and (f) $\Omega_{\rm rot}=0.98\omega_{\rm r,trap}$, where $\omega_{\rm r,trap} = \omega_y=\omega_x$ is the radial trap frequency. In all experiments, $\omega_{\rm r,trap} =2\pi\times$180 kHz and $\omega_{z} = 2\pi\times$86 kHz. 
From \onlinecite{Drewsen:2003}.} 
\label{fig:Rotating crystals}
\end{figure}

The application of optically-induced dipole forces ~\cite{Grimm:2000} can generate spatially localized perturbations to the ions' trapping potential. A standing-wave light field can force the ions to localize on gratings of potential minima separated by half the laser wavelength, $\lambda/2$~\cite{Cormick:2012,Enderlein:2012,Linnet:2012,Laupretre:2019,Wipfli:2023}. In general, the competition with the Coulomb repulsion gives rise to frustration (see Sec.~\ref{Sec:Frustration}). However, for crystalline structures with bcc, fcc, or hcp symmetry, one could tune these gratings so that the periodicity is commensurate with the distance between the lattice planes. Molecular dynamics simulations showed that in this way one can control the crystal structures and orientations as well as adiabatically change the crystal structure from bcc to fcc structures, and vice versa~\cite{Horak:2012}. In a cylindrically symmetric pseudopotential, the radiation pressure of a single traveling laser beam that does not propagate across the trap center exerts a torque and can set the crystal into rotation.
In~\cite{Drewsen:2003} this was observed by illuminating a crystal in a rf-trap by a beam propagating within the $x - y$ plane, but off-set with respect to the center point. The measured rotational frequencies varied from zero all the way up to 98\% of the radial trap frequency dependent on the specific position and detuning of the laser beam~\cite{Drewsen:2003} (see Fig.~\ref{fig:Rotating crystals} as well as Sec.\ \ref{Sec:Frustration} for an analogous experiment performed in Penning traps).

\begin{figure}
\includegraphics[width=0.5\textwidth]{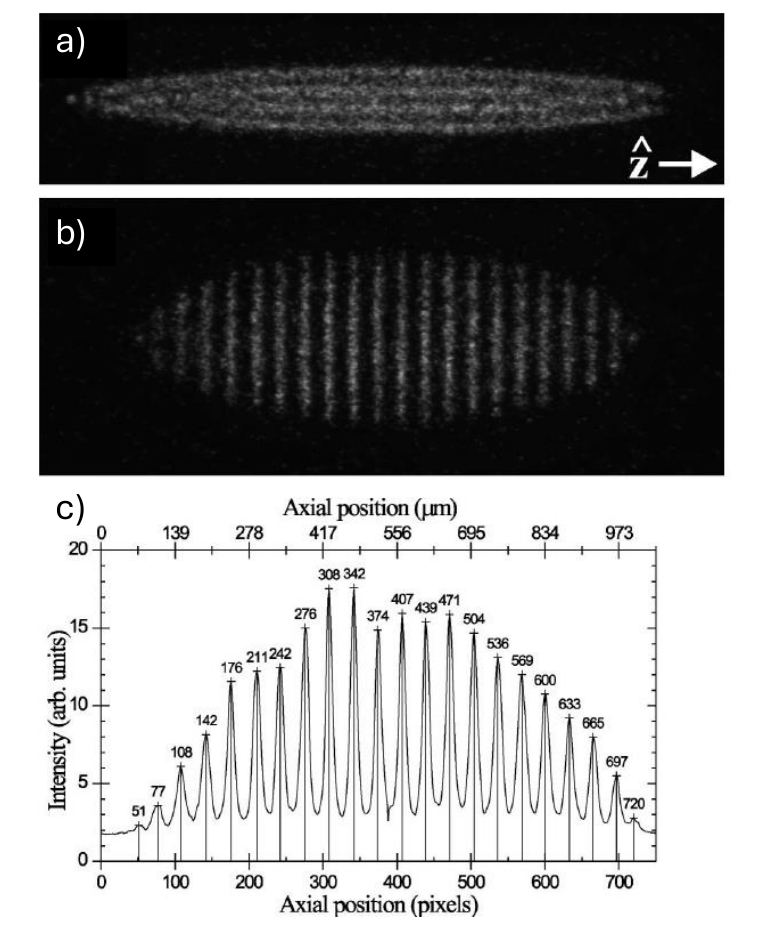}
\caption{Laser cooled ion ensemble containing approximately 370 $^{24}$Mg$^+$ ions in Coulomb crystal configurations just below (a) and above (b) the 2$\omega_{p}$=$\omega_\mathrm{rf}$ dynamical instability. (c) The integrated axial intensity distribution of the string-of-disks
configuration (b), which shows the disks are near-equidistantly spaced. From \onlinecite{Kjaergaard:2003}.} 
\label{fig:parametric instability}
\end{figure}

In rf traps, a more dramatic kind of perturbation can be induced when the density of a crystal is increased to the point, where the plasma frequency of the crystal $\omega_p$ becomes close to the half that of the rf drive frequency, $\Omega_{\rm rf}$, namely, when $\omega_p\sim \Omega_{\rm rf}/2$. 
In such situations, the parametric resonances lead to heating of the crystal~\cite{Bluemel:1989}, and in many instances even to melting. However, if the crystal is set into rotation, its density and hence its plasma frequency will be lowered, and the crystal will eventually get out of parametric resonance with the rf frequency.  By appropriate laser cooling and laser torque conditions, one can stabilize structures where the ions form rotating discs~\cite{Kjaergaard:2003}, see Fig.~\ref{fig:parametric instability}. A full understanding of these types of dynamics is still not available. Nevertheless these configurations seem to satisfy a detailed balance condition between the laser-cooling and the rf heating processes.
  
\begin{figure}
\includegraphics[width=0.5\textwidth]{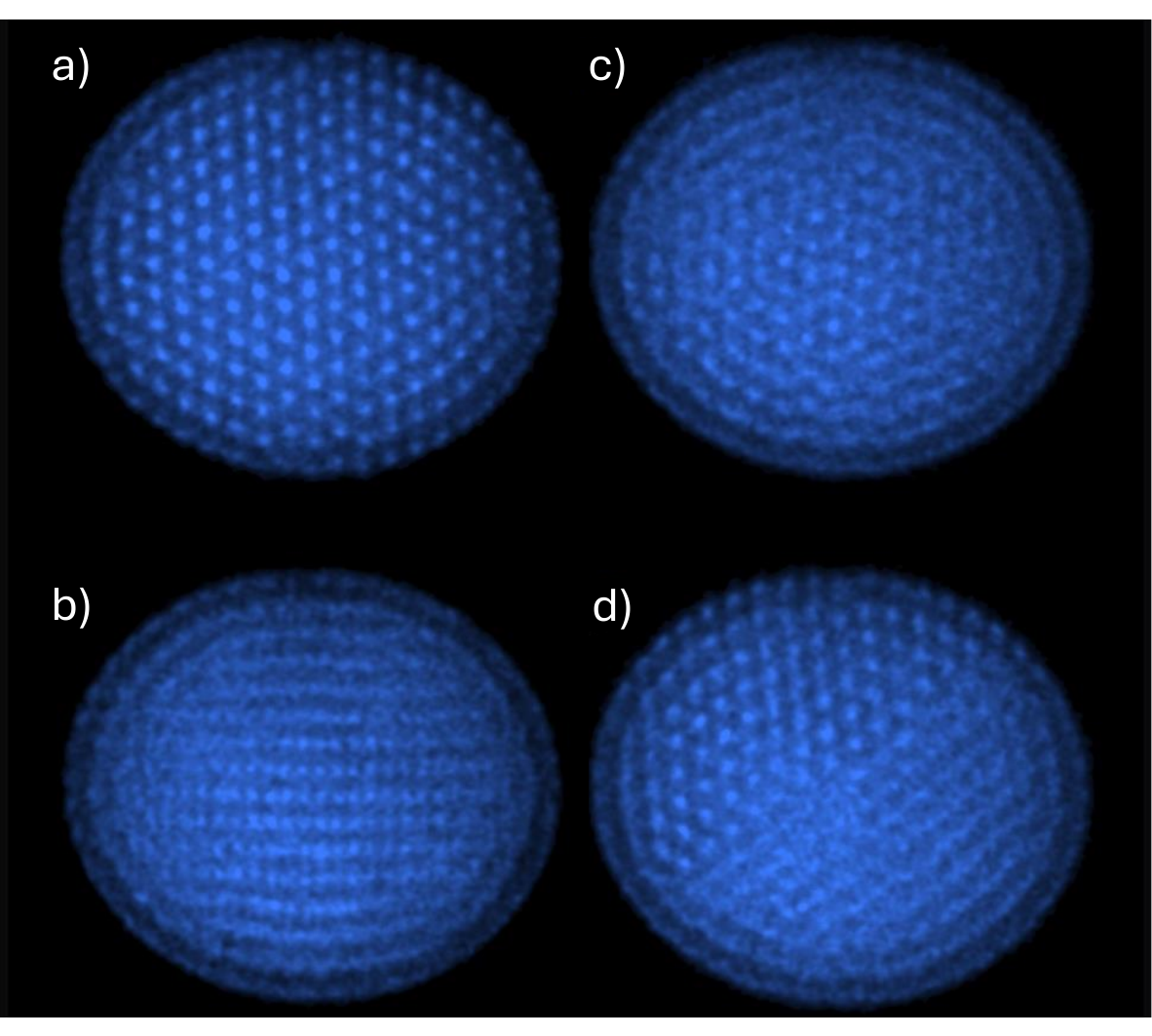}
\caption{Fluorescence images of metastable spheroidal Coulomb crystals containing $\sim$2700 $^{40}$Ca$^+$ ions: a) Dominating bcc structure viewed from the $\textless$111$\textgreater$ direction. b) Crystal structure containing a central fcc structure viewed from the $\textless$211$\textgreater$ direction. c) Crystal structures with indications of nucleations of defects at the center. d) Crystal indicating the coexistence of a bcc and a fcc structure. All images have been obtained with a 0.1 s integration time. From ~\onlinecite{Drewsen:2012}.} 
\label{fig:Metastable crystals}
\end{figure}

As mentioned in Sec.~\ref{Sec:ground state config}, the number of ordered metastable configurations within a given energy interval typically increases exponentially with the number of ions in a Coulomb crystal. Hence, in experiments where the ions are cooled to finite temperatures, one will expect to observe thermally excited metastable configurations. 
Figure~\ref{fig:Metastable crystals} shows four images of the same cold ion ensemble of approximately 2700 ions stored in a linear rf trap. The images were recorded seconds apart, showing various 3D periodic lattices structures. These are all excited metastable configurations, e.g., they correspond to energies higher than the ground state energy.  We note that Coulomb crystals as small as $\sim$1000 ions have been observed to possess a central, metastable bcc ordered core~\cite{Mortensen:2006}.  
In Penning traps, examples of metastable structures are the fcc lattices observed in ion crystals with more than 50000 ions~\cite{Tan:1995,Itano:1998}, see the previous section.

\subsubsection{Normal modes of three-dimensional crystals} 

Normal mode phonon spectra of bulk, crystalline strongly coupled OCPs have been calculated with varying degrees of accuracy over the past 6 decades both in the absence ~\cite{Clark:1958,Chabrier:1992,Baiko2001} and presence of a strong magnetic field~\cite{Usov:1980,Nagai:1982,Baiko:2009}. Here, the interest has frequently been in predicting the thermodynamic properties of dense astrophysical OCPs. As discussed in previous sections, trapped-ion crystals formed and studied in the laboratory typically are not of sufficient size to form a body-centered cubic lattice in a large fraction of the interior of the crystal. In addition, the normal modes that are most straightforwardly excited in the laboratory typically have wavelengths that are long compared to the inter-ion spacing. The dispersion relation of long wavelength modes is insensitive to the details of the lattice structure but can be influenced by the boundary or shape of the ion crystal.   

A trap geometry that is frequently employed in experimental work provides quadratic (frequently called harmonic) confinement in the axial and radial directions.  See Eqs. (\ref{eq:pseudopot})
and (\ref{eq:rotating_frame}).  As discussed in Secs.~\ref{sec:rf trap} and \ref{sec:Penning trap}, at low laser cooling temperatures a quadratic confining potential results in uniform density crystals with a spheroidal shape. Interestingly, a simple exact analytic solution exists for the normal modes of a uniformly charged spheroid, both in the absence as well as in the presence of a magnetic field. The treatment assumes that the normal mode wavelength is long compared to the inter-ion spacing so that the mode is well described by cold fluid equations.  See Refs.~\cite{Dubin:1991,Bollinger:1993} for a detailed description of the cold-fluid equations.  

The modes can be catalogued by two integers $l$ and $m$ where $0 \leq m < l$, following the conventions of \onlinecite{Dubin:1993}.  Here, $l$ and $m$ characterize the functional dependence of the mode  potential $\psi$ in terms of associated Legendre functions.  For example, the solution for the mode potential outside the spheroidal crystal is given by
\begin{equation} 
    \psi\propto Q^{m}_{l}(\xi_1/d) P^{m}_{l}(\xi_2) e^{i(m\phi-\omega t)}.
    \label{eq:3Dmode}
\end{equation}
Here, $Q^{m}_{l}$ and $P^{m}_{l}$ are associated Legendre functions, the coordinate $\xi_1$ is a generalized distance coordinate (similar to spherical radius), $\xi_2$ is a generalized latitude (similar to the polar angle $\theta$), $\phi$ is the usual azimuthal angle, $\omega$ is the mode frequency, and $d^2\equiv Z_p^2 - R_p^2$, where $2Z_p$ is the axial extent and $2R_p$ is the diameter of the spheroid.  The specification of the mode potential in the interior of the ion crystal is more complicated.  

\begin{figure}
\includegraphics[width=0.45\textwidth]{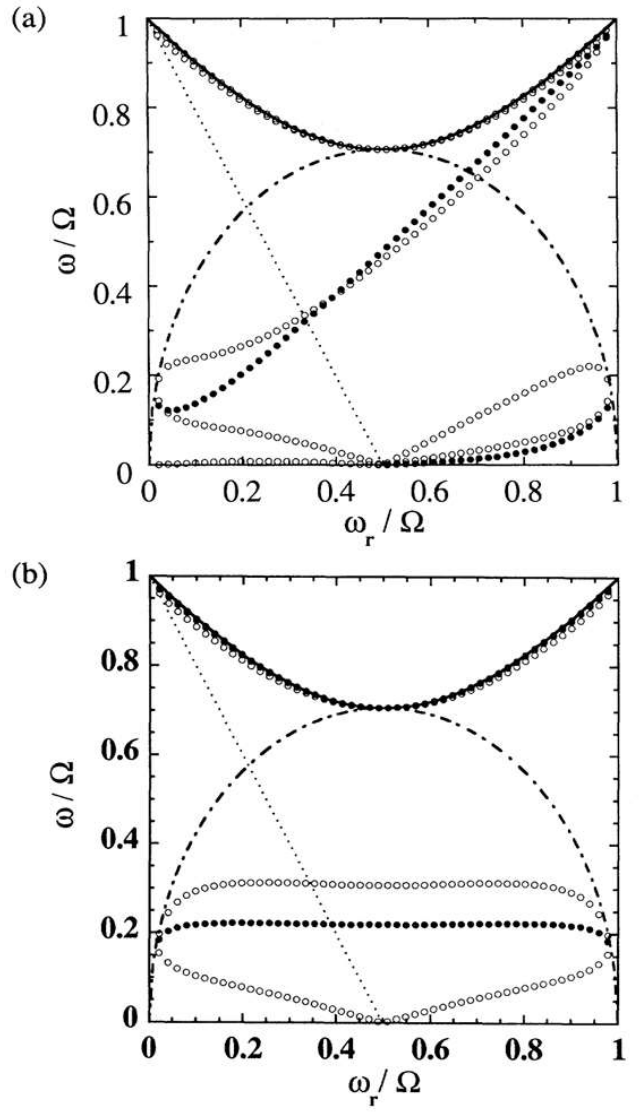}
\caption{Normal mode frequencies $\omega$ in the rotating frame of the ion crystal vs rotation frequency $\omega_r$ for $\omega_z/\Omega=0.151$ and different values of $(l,m)$, where $\Omega\equiv\Omega_c$ is the cyclotron frequency.  Frequencies $\omega < 0$ are allowed and obtained in the diagrams by reflecting through the point $(\omega_r,\omega)=(\Omega/2,0)$.  For $m\neq0$ the sign of $\omega$ determines the direction of the azimuthal motion of the mode potential.  Also shown for comparison are the vortex frequency $\Omega_v(\omega_r)=\Omega-2\omega_r$ (dotted line), the plasma frequency $\omega_\mathrm{p}(\omega_r)$ (see Eq. \eqref{eq:ocp_melt}) (dot-dashed curve), and the upper hybrid frequency $\Omega_u(\omega_r)=(\omega_\mathrm{p}^2 + \Omega_v^2)^{1/2}$ (solid curve). (a) $(l,m)=(2,1)$, filled circles; $(l,m)=(5,1)$, open circles.  (b) $(l,m)=(2,0)$, filled circles; $(l,m)=(4,0)$, open circles. From\ \onlinecite{Bollinger:1993}.} \label{fig:93mode_calc}
\end{figure}

We discuss some example calculations of $m=1$ and $m=0$ modes in Figs.~\ref{fig:93mode_calc}(a) and (b), respectively.  Both figures show the calculated mode frequency $\omega$ normalized to the ion cyclotron frequency $\Omega$ ($\Omega\equiv\Omega_c$ in Secs.~\ref{sec:Penning trap} and \ref{sec:planar_geometries}) as a function of the ion crystal rotation $\omega_r$ in a Penning trap with axial frequency $\omega_z/\Omega = 0.151$. The mode frequencies are plotted in the rotating frame of the ion crystal.  For $\omega_r \sim 0$ (or $\omega_r \sim \Omega$) the ion crystal has a lenticular shape (see Fig.~\ref{fig:Pening_crystal_shapes}) and the mode frequencies separate into three different bands. These bands are the 3D analogues of the ${\mathbf E} \times {\mathbf B}$, drumhead, and cyclotron modes discussed in the planar geometries section~\ref{sec:planar_geometry_mode}. A difference with respect to single-plane modes is that the ${\mathbf E} \times {\mathbf B}$ (and to a much lesser extent the cyclotron modes) now acquire a component of motion that is parallel to the magnetic field~\cite{Dubin:2020,Hawaldar:2024}.

As $\omega_r/\Omega$ increases from zero, the separation of the modes into three bands becomes less distinct.  At $\omega_r/\Omega = 0.5$ the effective magnetic field in the rotating frame of the ion crystal goes to zero.  At this special point the equations of motion of the normal modes are equivalent to the ones of a spheroidal ion crystal with the same aspect ratio in an rf trap in the psuedopotential approximation, and the modes can be classified with the integers $(l,m)$ of Eq.\ \eqref{eq:3Dmode}~\cite{Dantan:2010}. For a discussion about the limits of validity of this equivalence and a mathematical analysis of modes that includes the time-dependent dynamics of rf Paul trap potentials, see Refs.~\cite{Landa:2012,Landa:2012a}.   
  
Experimentally, the normal modes of 3D trapped ion crystals have been excited by applying time-dependent potentials to the trap electrodes~\cite{Mitchell:1998b,Dantan:2010} or by means of radiation pressure~\cite{Kriesel:2002}. Typically, the ion motion produced by exciting the mode induces Doppler shifts that can be used to detect the mode. In Ref~\cite{Dantan:2010} several $(l,m=0)$ modes of spheroidal ion crystals stored in a rf trap were excited through rf excitation and the resulting Doppler shifts detected through modification of the collective coupling of the ion crystal to an optical cavity.  Excellent agreement (better than 1\%) was observed with the mode frequency predictions of the cold, spheroidal theory of Dubin~\cite{Dubin:1991}.  In Penning traps, detection of the Doppler shifts produced by an excited mode through changes in the ion fluorescence (so-called Doppler velocimetry) was used to measure the frequencies of modes $(l,m)$ with $l$ as large as 9 and $m$ as large as 2~\cite{Heinzen:1991,Mitchell:1998}.  

\begin{figure}
\includegraphics[width=0.40\textwidth]{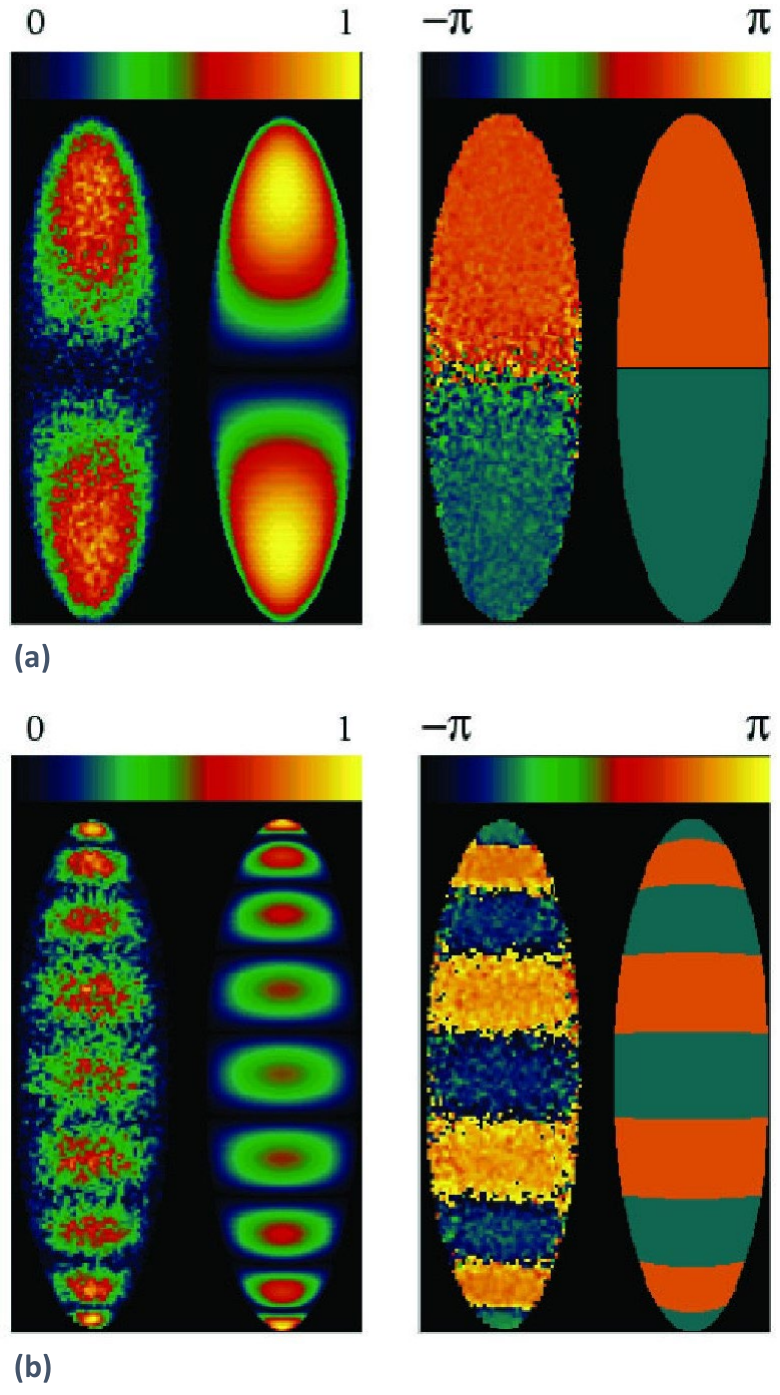}
\caption{Sideview image of the velocity profile of an $(l,m=0)$ mode excited in a spheroidal, laser-cooled cloud of $^9$Be$^+$ ions in a Penning trap.  Here, $\omega_r/(2\pi)=1.00$ MHz, $\omega_z/(2\pi)= 1.13$ MHz, and  $\Omega/(2\pi)= 7.608$ MHz, where here $\Omega\equiv\Omega_c$ is the cyclotron frequency.  The magnetic field and axial laser beam point up.  The ion cloud spheroid dimensions are $2Z_p=0.76$ mm and $2R_p=0.24$ mm, with the density $n_0=2.70\times10^9$ cm$^{-3}$.  At each point of the image, the measured ion resonance fluorescence is fitted to an amplitude and phase.  (a) The left panel compares the measured amplitude of a (2,0) mode excited at $\omega_{2,0}/(2\pi)=1.656$ MHz with the predictions of linear theory. The right panel compares the measured phase with theory.  (b) The left panel compares the measured amplitude of a (9,0) mode excited at $\omega_{9,0}/(2\pi)=2.952$ MHz with the predictions of linear theory.  The right panel compares the measured phase with theory. From\ \onlinecite{Mitchell:1998b}.} \label{fig:20_90_mode}
\end{figure}

Images of the excited mode eigenfunction of a trapped ion crystal can be obtained through the laser-induced ion fluorescence.  Figure~\ref{fig:20_90_mode} shows experimentally measured images of $(2,0)$ and $(9,0)$ modes excited in cold, spheroidal clouds of $^9$Be$^+$ ions in a Penning trap.  Here, a laser beam is directed parallel to the magnetic field axis (the vertical direction in Fig.~\ref{fig:20_90_mode}) with a frequency tuned below resonance with a cycling transition of  $^9$Be$^+$, see Fig.~\ref{Fig:1:Mitchell} for the experimental schematic.  Ions moving with a velocity in the direction opposite to the laser wave vector \textbf{k} fluoresce more strongly than ions moving in the same direction as \textbf{k}.  By recording side-view images of the ion crystal with a photon-counting imaging camera that measures the spatial and temporal coordinates of each detected photon, it is possible to construct side-view images of the excited mode velocity profile that are phase synchronous with the external radio frequency drive used to excite the mode. Comparison of the experimental images with the theoretical predictions in Fig.~\ref{fig:20_90_mode} shows that experiment and theory are in good agreement.  
\section{Out-of-equilibrium dynamics of ion Coulomb crystals}
\label{Sec:3}

Understanding how strongly-correlated many-body systems reach equilibrium is a central objective of statistical physics and is essential for applications to quantum technologies. By offering a controllable environment, ion Coulomb crystals are a prominent platform for simulating models that allow verification of equilibration and thermalization hypotheses. The intrinsic physical properties make the system peculiar by itself. In fact, the unscreened Coulomb interactions make ion Coulomb crystals a unique laboratory for studying the out-of-equilibrium dynamics of long-range interacting systems, which can clarify hypotheses and conjectures formulated for cosmological and plasma systems  \cite{Campa:2009,Defenu:2023}.  
In this section we review experimental and theoretical work on out-of-equilibrium dynamics in ion Coulomb crystals. We focus in particular on transport and relaxation of ion Coulomb crystals after a perturbation from equilibrium, such as a slow or sudden change (quench) of the trapping parameters or of the temperature. One relevant aspect is the formation, meta-stability and decay of defects and dislocations, which in ion crystals can be imaged both in real as well as in Fourier space. As the concept of out-of-equilibrium is central to this section, we start by defining what can be understood as equilibrium in a trapping environment, and when the laser-cooled crystal can be said to be in a thermal state.

\subsection{Equilibrium and temperature}
\label{sec:Equilibrium and temperature}

In the Penning trap (see Sec.~\ref{sec:I.B}), the trapping forces break time reversal symmetry and confinement is achieved by setting the ions in rotation. Collisions between the ions drive the system into a state where the ion ensemble rotates rigidly (without shear) about the symmetry axis of the trap, which is parallel to the direction of the magnetic field. Rigid body rotation has been observed in the laboratory~\cite{Brewer:1988,Driscoll:1988}. In the frame co-rotating with the ions, the system can evolve into a thermal equilibrium state characterized by a global temperature~\cite{Dubin:1999}.

 The Paul trap, instead, achieves confinement by means of fast-oscillating forces. Often the confinement can be characterized by a conservative potential in the pseudopotential approximation, enabling a conventional statistical mechanics description that can potentially be described by a global temperature.  However, strictly speaking the dynamics is a many-body Coulomb problem in the presence of time-dependent, periodic forces.  A theoretical study accounting for the full temporal dependence of the trap  unveiled the existence of ordered structures that are periodic in time and that are otherwise not captured by the pseudo-potential description \cite{Landa:2012,Landa:2012a}. 

The concept of temperature for characterizing laser cooled ensembles is often convenient, even though not strictly appropriate. Temperature is often employed as a unit of measure for the energy width of the single-particle distribution in the pseudopotential approximation (or, for Penning traps, in the rotating frame of the crystal), assuming that the crystal's degrees of freedom have equilibrated. However, the crystal is not in thermal contact with a heat bath and the stationary state is reached through laser cooling, namely, by means of the mechanical forces associated with scattering of laser photons. This leads to two key issues: One issue lies at the core of laser cooling, and is that the single-particle distribution is strictly speaking not a Maxwell-Boltzmann  distribution. The second issue is intrinsic to the Coulomb crystal: 
even under ideal conditions, the steady state vibrational occupation of the individual vibrational modes are cooled to different temperatures, depending on the ratio between the effective linewidth and the vibrational frequency of the mode~\cite{Eschner:2003}. The possibility of non-thermal motional distributions resulting from laser sideband cooling has been discussed in the context of optical atomic clocks, where a detailed evaluation of the motional distribution is important for characterizing time-dilation shifts~\cite{Chen:2017}.

On longer time scales, anharmonicities can redistribute the energy between the modes leading to a thermal state~\cite{Morigi:2001b,Morigi:2004b}.  At first glance, because radio frequency (rf) and Penning traps can routinely confine ensembles of charges of the same sign for many hours to days, the achievement of global thermal equilibrium states appears feasible.  However, depending on the details of the trap parameters, the time to equilibrate to a global thermal state can exceed even these time scales. Consider, for example, a long linear chain of ions. Here, the confinement in the transverse direction, characterized by frequency $\nu_t$ (see Eq.~\eqref{Eq:potential}) is much stronger than the confinement in the longitudinal direction, characterized by frequency $\nu$. This results in normal mode frequencies for motion transverse to the trap axis that are much higher than the normal mode frequencies of motion parallel to the trap axis.  In this case it has been shown theoretically that the rate of equilibration between the transverse and parallel motional degrees of freedom can be exponentially small in the ratio $\nu_t/\nu$, leading to equilibration time periods that exceed many months for cold ion chains with $N > 10$~\cite{Chen:1993}. Physically, this result can be understood through energy conservation arguments. For large ratios $\nu_t/\nu$ the coupling between the transverse (optical) and the longitudinal (acoustic) bands is a weak perturbation. In terms of vibrational quanta an interband transition requires that several low-frequency acoustic phonons are annihilated to create a single high-frequency optical phonon.  Such multi-phonon processes arise as high-order terms in the Coulomb interaction and have very small effective rates.  

Long equilibration rates are also anticipated with single-plane crystals with large ion number $N$, which require a trap asymmetry parameter $\beta\ll 1$ (see Eq.~\eqref{eq:azimuth_symmetric}). Lastly, in Penning traps, the strong magnetic field splits the normal modes that describe in-plane motion into high frequency cyclotron modes and low frequency ${\mathbf E} \times {\mathbf B}$ modes (see Fig.~\ref{fig:in_plane_modes}).  Numerical studies show that the equilibration of the cyclotron and ${\mathbf E} \times {\mathbf B}$ branches is exponentially weak in the ratio of the frequencies of the two branches~\cite{Tang:2021}.

\begin{figure}
\includegraphics[width=0.45\textwidth]{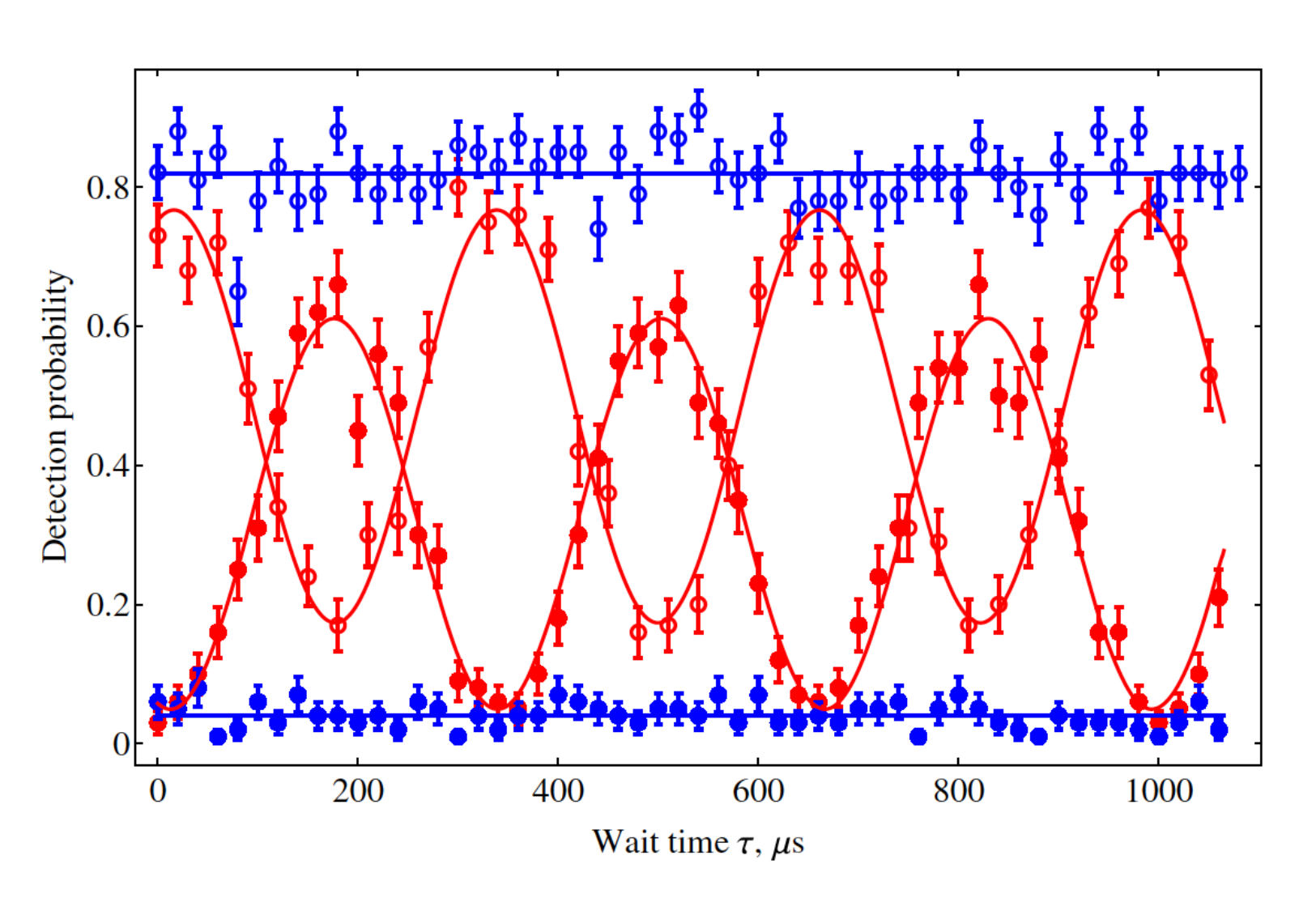}
\caption{(Color online) Parametric coupling between an axial ($a$) and a radial ($r$) normal mode of a two-ion system. The mode frequencies are tuned such that $\omega_a=2\omega_r$. The data points indicate the time evolution of the excitation of the axial (solid dots) and the radial (open circles) mode when the system is initially prepared with either one or two quanta of excitation in the radial mode. In the latter case (data points fitted by a sinusoidal red solid curve), a coherent exchange of energy is observed between the two modes. In case the radial mode is initially prepared  with only one quantum of excitation (data points fitted by solid blue horizontal lines), no coherent exchange is observed and the modes are effectively decoupled. The error bars show one standard deviation statistical uncertainty. From\ \onlinecite{Ding:2017}.} \label{Fig_Parametric_coupling}
\end{figure}

In summary, although ion Coulomb crystals can be confined for long times, this might not be sufficient to observe equilibration between different bands, especially when these are separated by relatively large energy gaps.  Equilibration processes can be experimentally accelerated by various strategies, such as fine tuning of the trap parameters.  As a simple example, for a two-ion crystal aligned along the axis of a linear rf trap, the axial stretch mode frequency can be tuned to be approximately equal to twice the frequency of the radial rocking mode. In this case the intrinsic non-linear Coulomb force between the ions leads to a strong non-linear coupling between these two modes, as illustrated in Fig.~\ref{Fig_Parametric_coupling}~\cite{Ding:2017}.  Furthermore, in a linear crystal of three ions it has been possible to tune and exploit the resonant coupling of three normal modes ~\cite{Maslennikov:2019}. In Penning traps, numerical studies show that the coupling between in-plane ${\mathbf E} \times {\mathbf B}$ modes and the out-of-plane drumhead modes for single-plane crystals can be strongly enhanced by tuning trap parameters so that their respective bandwidths overlap~\cite{Johnson:2024}.  

An alternative strategy accelerates the process of equilibration by inducing effective mode-mode couplings with time-dependent electric fields \cite{Gorman:2014,Hou:2024}. This has been successfully applied to the equilibration between two modes of small ion crystals.  Its extension to large ion crystals with widely separated bandwidths of many modes is an area of current investigation.

We conclude by discussing temperature measurement in rf-traps. Temperature is determined by extracting the kinetic energy associated with the rf-driven coherent micromotion, which is typically several orders of magnitude larger than the thermal motion. This can be done, e.g., by recording in molecular dynamics simulations the change in the individual ion velocities at the same phase of the rf fields over time \cite{Schiffer:2000}. In this way, one can also gain knowledge of the heating of the Coulomb crystals due to the rf drive \cite{Bluemel:1989,Schiffer:2000,Tarnas:2013,Nam:2014,Poindron:2023}. In experiments with 3D Coulomb crystals, where the kinetic energies of the individual ions (or equivalently the various normal mode excitations) are difficult to measure, comparison between the experimentally measured fluorescence images and the ones generated from molecular dynamics simulations are often a practical way to estimate the temperature~\cite{Ostendorf:2006,Okada:2010,Kiethe:2021}.

\subsection{Transport in trapped ion chains}

The possibility to perform normal mode spectroscopy of an ion chain and to address and monitor a single ion within the chain make these systems a powerful laboratory for studying thermodynamics of and heat transport in many-body systems at the quantum level. This has led to investigations of coherent transport of excitations along ion chains varying from few to dozens of ions, and with temperatures ranging from the Doppler cooling limit to near zero point energies. 

\begin{figure}
\includegraphics[width=0.45\textwidth]{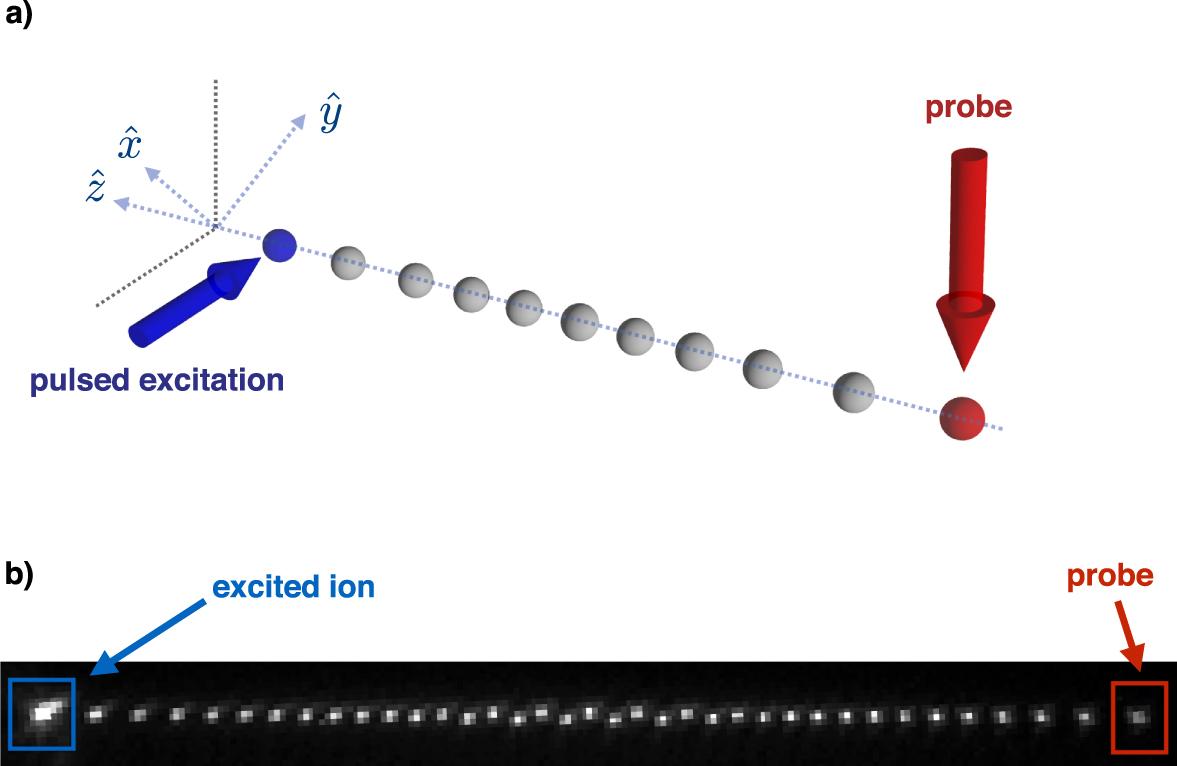}
\caption{(Color online) Schematic of the energy transport experiment of \onlinecite{Ramm:2014}. (a) An ion at the chain edge is excited by a pulsed beam (horizontal blue arrow).  Following free evolution, the energy is read out along the chain with a probe beam (vertical red arrow). (b) CCD image of the ion chain
with 37 ions. The ions are $^{40}$Ca$^+$ and initially Doppler cooled. The pseudopotential frequencies of the Paul trap are $(\omega_x,\omega_y,\omega_z)/(2\pi)=(2.25,2.0,0.153)$ MHz. The vibrations are locally excited along the $x$ direction, corresponding to the steepest confining potential. From\ \onlinecite{Ramm:2014}. }\label{Ramm2014_1}
\end{figure}

\begin{figure}
\includegraphics[width=0.45\textwidth]{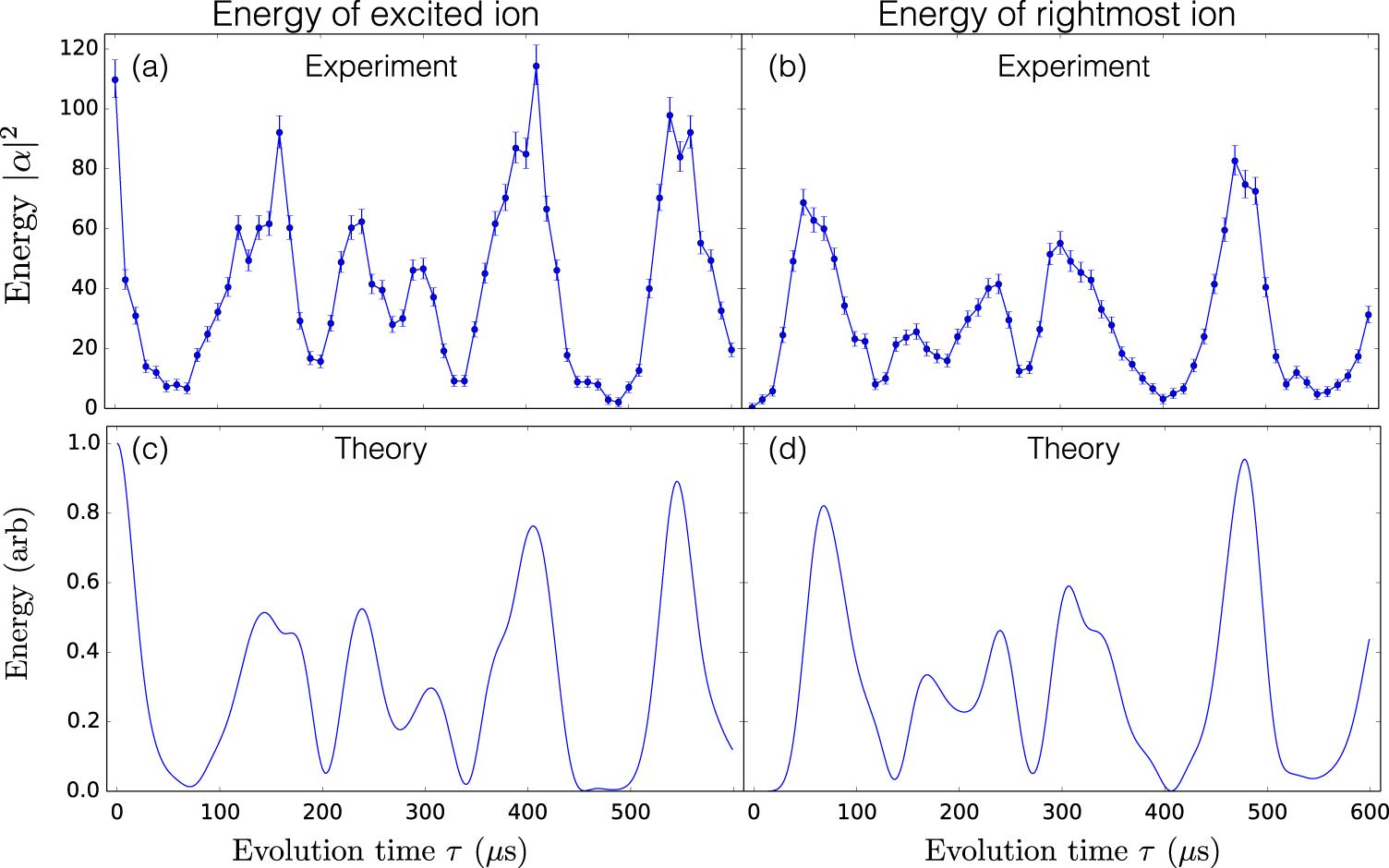}
\caption{Energy transport in a chain of five ions. The leftmost ion is given a kick and
its energy is measured after a subsequent evolution time $\tau$ as shown in (a). Energy revivals approaching the initial energy are observed. Panel (b) shows the energy of the rightmost ion: initially unexcited, the energy from the kick is rapidly transferred across the chain.  Panels (c) and (d) show the energies of the ions obtained from molecular dynamics simulations.  From~\onlinecite{Ramm:2014}.} \label{Ramm2014_3}
\end{figure}

Coherent transport of vibrational excitations was studied in ion chains in linear Paul traps by \cite{Haze:2012,Ramm:2014}. We discuss in more details the experiment of \cite{Ramm:2014}, as this is representative of the state of the art reached on the observation of these dynamics.
Here, transport of a transverse vibrational excitation was measured along chains of $^{40}$Ca$^+$ in a linear Paul trap with the number of ions ranging from 5 to 37, as illustrated in Fig.~\ref{Ramm2014_1}. 
After initializing the chain motion to Doppler cooling temperatures, an edge ion was excited by a laser pulse with a finite duration. The intensity of the pulse was modulated at the trap radial frequency, thereby locally exciting a transverse vibration. The dynamics of energy propagation was monitored by resolving individual ions along the chain using focused, near resonant laser pulses to convert motional excitation to electronic excitation, which is routinely measured. The measured data is consistent with a model of coupled oscillators (or equivalently, with a harmonic crystal). In a quantized form, the model for the vibrational dynamics along the transverse $x$-direction reads:
\begin{equation}
\label{Eq:H:Transport}
    H=\sum_i\hbar \omega_{x,i}a_i^\dagger a_i+\sum_{i\neq j}\hbar t_{ij}(a_i^\dagger a_j+a_j^\dagger a_i)\,,
\end{equation}
where $i$ labels the ions along the chain ($i=1,\ldots,N$), and $a_i$ is the boson annihilation operator of a quantum of transverse vibration at site $i$. In this representation, an excitation hops between sites with a non-local hopping amplitude $t_{ij}$. The frequency of the local oscillators is the transverse trap frequency $\omega_x$, here along the $x$-direction, and includes a position-dependent frequency shift due to the Coulomb interactions: $\omega_{x,i}=\omega_x-\sum_{i\neq j} t_{ij}$. In the experiment, the hopping element between the leftmost ion and its neighbor ranges in the interval $t_{ij}\in 2\pi \times (2.21,6.7)$ kHz, where the smallest (largest) value is for a chain of 5 ions (25 ions). The measurements show that the time required for the excitation to propagate across the entire chain is comparable with the inverse of the coupling rate between nearest neighbors. This behavior was interpreted as a manifestation of the long-range hopping, which permits the excitation to directly reach to a distant ion. In this configuration, long-range hopping favors direct transfer between the edge ions, since they possess the same local frequency. These features also explain why adjacent ions could get very different energy excitations. 

Revivals of the excitation energy are visible in Fig.~\ref{Ramm2014_3}. They were reported even for long chains. The revivals persisted for times of the order of 40 ms with measured time scales which seem weakly dependent on the number of ions. Their decay was attributed to the change of the radial trap frequency over this time.  These revivals can be understood as due to the rephasing of multiple phonon normal modes and signal the coherent dynamics of the propagating excitation, showing that the chain had not thermalized. In these settings, in Ref.~\cite{Abdelrahman:2017} the authors measured the spreading and refocussing of a vibrational excitation  on a chain composed of 42 ions, by means of a Ramsey-type of measurement as in~\cite{DeChiara:2008}. They could access the autocorrelation functions of the phonons as well as the quantum discord. The experiments so far described were performed with cold, yet thermal, chains. Deep in the quantum regime, transport of a single quantum of vibration was first realized in Ref.\ \cite{Tamura:2020} on a chain composed of four $^{40}$Ca$^+$. The experiment reported coherence time exceeding 10 ms. The experimental measurements were in excellent agreement with the predictions of Eq.\ \eqref{Eq:H:Transport}. 

Heat transport in ion chains is the subject of several theoretical studies. In Ref.~\cite{Lin:2011} the energy distribution along the chain was determined when two ions in the chain realize point contacts to heat baths at different temperatures. This configuration can be implemented by continuously laser cooling the two individual ions to 
different stationary temperatures. A rich behavior was predicted as a function of the cooling rate and of the position of the laser-cooled ions within the chain. Heat transport through an ion chain that undergoes the linear to zigzag transition was experimentally measured in Ref. \cite{Ramm:2014}. A theoretical study  reported that the onset of zigzag order gives rise to a qualitative change in transport, from ballistic to diffusive  \cite{Ruiz:2014}. The origin of the diffusive behavior was attributed to the nonlinearity induced by coupling between the longitudinal and transverse phonons to the zigzag modes. Moreover, the heat conductivity was found to be maximimal near the critical point, where the transverse phonon spectrum becomes gapless. Transport in a chain embedding two mass impurity defects was theoretically analyzed in Ref.\ \cite{Fogarty:2013,Taketani:2014}. Here, it is argued that for certain mass ratios, the motion of the mass defects can become entangled via the resonant exchange of vibrational excitations with the rest of the chain. Heat transport in the presence of a topological defect was investigated in \cite{Timm:2023} and is reviewed in Sec. \ref{Sec:Defects}. The properties of transport using the radial modes of a linear chain are at the basis of proposals for quantum information with trapped ion chains \cite{Serafini:2009}.

In microtraps, where the ion structure is designed by the external potential, transport experiments realized the coherent exchange of phononic excitations \cite{Brown:2011,Harlander:2011,Hakelberg:2019}. This experimental progress demonstrates the capability of controlling and tailoring phonon hopping between the ions and sets the basis for simulating many-body dynamics of strongly-correlated bosons \cite{Porras:2004} as well as for revealing quantum features of transport \cite{Bermudez:2013}. Further work has been devoted to  
transport controlled by the coupling with the internal degrees of freedom of the ions. We here mention the realization of vibrationally-assisted transfer of energy \cite{Gorman:2018} and the simulation of an effective Jaynes-Cummings type of dynamics \cite{Schneider:2012} in order to achieve effects such as phonon-blockade and designed phonon-phonon interactions \cite{Debnath:2018,Ohira:2021}.

\subsection{Topological defects and metastability}
\label{Sec:Defects}

\begin{figure}
\includegraphics[width=0.48\textwidth]{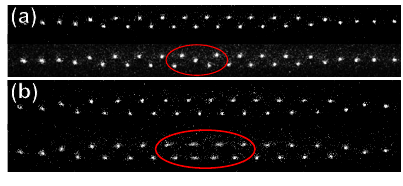}
\caption{\label{Fig1:Partner2013}(Color online) Experimentally observed kinks in an ion chain  of about 30 ions. The kinks (red ellipses) are created during a radial quench of a crystal and imaged with exposure times of 40 ms. (a) Uniform zigzag and localized kink for trap aspect ratio $\nu/\nu_t\approx 8$. (b) Zigzag and extended kink for $\nu/\nu_t\approx 5.5$. 
Adapted from~\onlinecite{Partner:2013}.} 
\end{figure}

As mentioned in Sec.\ \ref{sec:Metastable:3D}, manifestations of prethermalization in ion Coulomb crystals such as defects and dislocations can be imaged on a CCD camera. They are typically created by means of temporal transformations of the trap potential, or by rapidly quenching the temperature while laser cooling the ion cloud to a crystal. In low dimensional crystals, and especially in the zigzag chain, it is possible to experimentally monitor the formation and the dynamics of individual defects. This has led to a series of insightful studies, some of which are reviewed in the rest of this Section.

Figure \ref{Fig1:Partner2013} reports some examples of defects (kinks) measured in a zigzag chain: the upper panel displays a so-called localized kink. The lower panel shows an example of extended kink, within which the ions form two parallel ion chains. Other kinds of kinks are composed by dislocations in all three dimensions \cite{Mielenz:2013,Nigmatullin:2016}. In \cite{Mielenz:2013} the formation of kinks was observed by laser cooling a cloud of $^{24}$Mg$^+$ ions in a linear Paul trap into a zigzag configuration. By imaging the ion positions, kinks were measured at the center of the zigzag chain with an occurrence rate that depended on the number of ions. The kinks decayed by propagating from the center to the edges of the zigzag chain on a timescale of order 10~s, exceeding the oscillation period of the axial trap by five orders of magnitude.
 
Defects such as those shown in Fig.\ \ref{Fig1:Partner2013} were also observed in experiments where the trap aspect ratio was slowly varied across the linear-zigzag transition while the chain was continuously laser cooled \cite{Ejtemaee:2013,Partner:2013,Pyka:2013,Ulm:2013}. They were reproduced by molecular dynamics simulations in the pseudopotential approximation, showing that micromotion is not essential in determining their onset and metastability (see Sec.\ \ref{sec:Metastable:3D} for comparison). Their metastable nature identifies them as topological defects. Interestingly, distant localized kinks interact, as verified by a comparison between experimental data and molecular dynamics simulations \cite{Landa:2013}. 

\begin{figure}
 \begin{center}
 \begin{overpic}[width = 0.48\textwidth, unit=1pt]{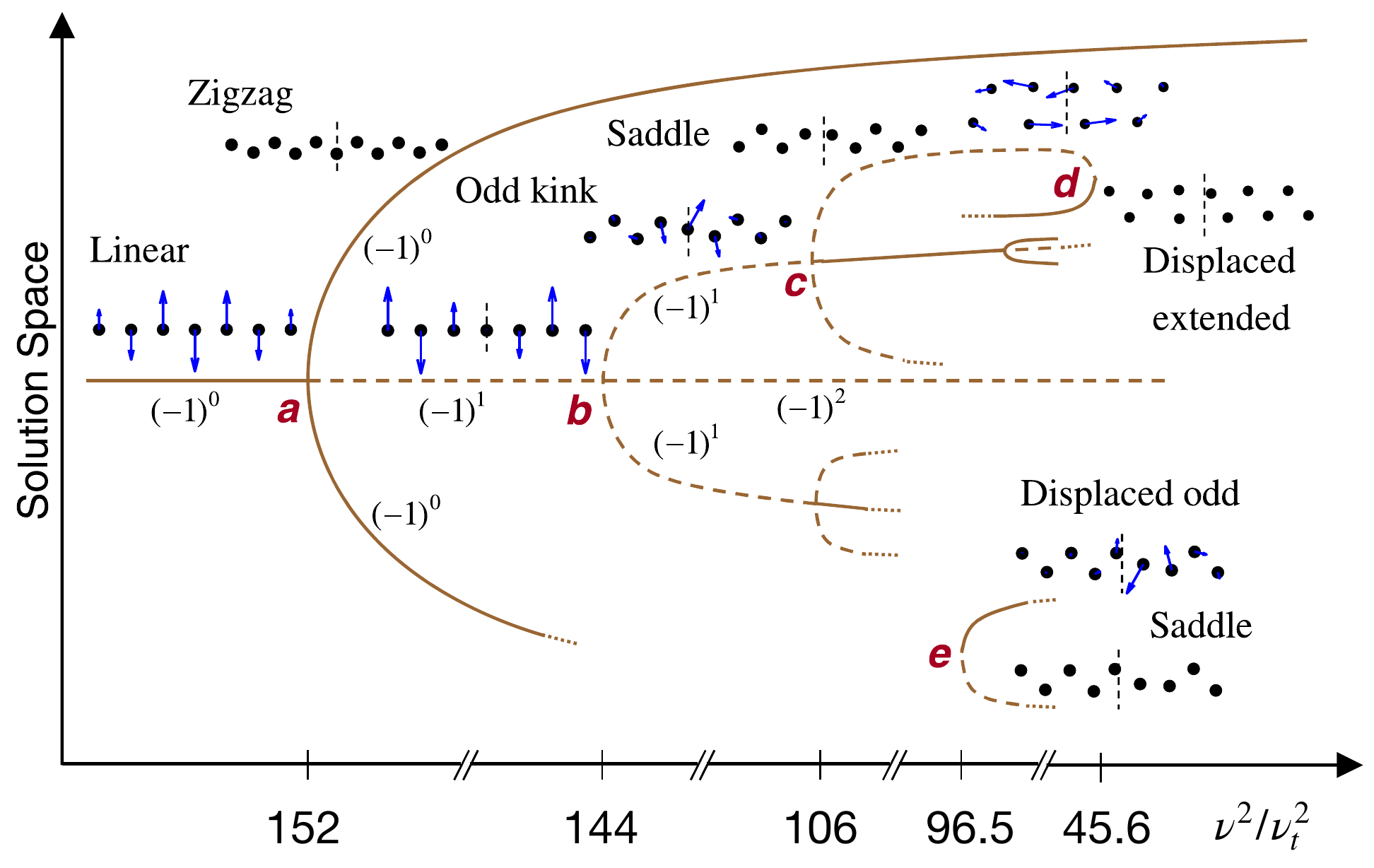}
 \end{overpic}
 \end{center}
\caption{\label{Fig1:Landa2013} Schematic illustration of bifurcations and stability of one-kink configurations in a planar crystal of 31 ions. The possible ion configurations are depicted as a function of the ratio of confinement in the axial and one of the transverse directions, where confinement in the second transverse direction is assumed to be very strong. Solid lines indicate stable configurations (local minima of the potential), dashed lines are unstable solutions. The arrows illustrate the spatial oscillations of the normal mode which crosses zero energy at the bifurcation and generates the bifurcating solutions. From~\onlinecite{Landa:2013}.} 
\end{figure}

A classification of kinks was carried out in \cite{Landa:2013} using tools of topological degree theory. The stability of a given defect is determined by the analysis of the eigenvalues of the corresponding Hessian matrix; a bifurcation occurs every time an eigenvalue of the Hessian changes sign.  Figure \ref{Fig1:Landa2013} illustrates the resulting metastable and unstable equilibrium configurations of one planar kink as a function of the trap aspect ratio, for a fixed number of ions \cite{Landa:2013}. 

The dynamics of a kink is characterized by localized modes, which include a translational mode as well as internal modes. Theoretical analysis of the internal modes of both localized and extended kinks predict well-resolved spectroscopic features and long coherence times, even at Doppler-cooling temperatures \cite{Landa:2010}. The translational energy of the topological defects was studied by calculating the Peierls-Nabarro potential, namely, the potential energy of the soliton as a function of the kink center position \cite{Landa:2013,Partner:2013}. Numerical simulations of energy transport in ion chains with a topological defect reported energy localization at the kink \cite{Timm:2020}. 

The spectroscopic features of the localized defects were measured in \cite{Brox:2017} by parametric heating. Experimental observations show features that suggest directed transport as a function of the topological charge \cite{Brox:2017}. The stability of the extended kink as a function of the trap ratio and its dynamics were extensively analyzed in \cite{Kiethe:2018}, unveiling typical features of static friction, see Sec.~\ref{Sec:Frustration}. 

The prospect of controlling the formation and dynamics of topological defects in ion chains has inspired several theoretical proposals. In Ref.~\cite{Landa:2014}, it is argued that discrete solitons in a chain confined to a ring geometry can generate entanglement between ions. In a linear Paul trap, when the edge ions are heat point contacts at two different temperatures, the presence of a kink can substantially modify the heat flow, giving rise to a pronounced temperature gradient at its location \cite{Timm:2021}. Together, these insights suggest that the interactions between kinks and normal modes of the chain can modify and even tailor the transport properties (see also Sec.\,\ref{Sec:Frustration}). 

Mass impurities embedded in the chain give rise to an additional class of topological defects \cite{Landa:2013}. It has been argued that the combination of mass impurities and static electric fields can allow for deterministic manipulation and creation of kinks \cite{Landa:2013}. Theoretical studies further predict that interactions between mass impurities and the surrounding chain can generate entanglement between the localized modes associated with these defects \cite{Fogarty:2013}. Chains containing mass or charge impurities are routinely realized in the ion-trap laboratories \cite{Feldker:2013,Keller:2024,Hausser:2025}. Structural phase transitions in the presence of mass impurities have been characterised in \cite{Rueffert:2024}. Dual species ion crystals can be deterministically prepared \cite{Zawierucha:2024}. By positioning impurities at selected locations within the chain, the dispersion spectrum can be designed \cite{Sosnova:2021}, with applications in quantum logic \cite{Morigi:2001b,Home:2013} and metrology \cite{Schmidt:2005,Home:2013}. 

We conclude by mentioning structural defects realized by combining spin-dependent forces and individual ion excitations in ion Coulomb crystals \cite{DeChiara:2008,Baltrusch:2011,Li:2012,Feldker:2013,Mallweger:2025}. Such dynamics can give rise to entanglement between vibrational and internal degrees of freedom \cite{DeChiara:2008,Baltrusch:2011,Gilmore:2021} and employed for quantum technological applications  \cite{Gilmore:2021,Martins:2026}.    

\subsection{Slow quenches and Kibble-Zurek mechanism}
\label{sec:KZ}

Experimental imaging of kinks in ion Coulomb chains enables a statistical analysis of their formation as a function of the experimental parameters \cite{Ejtemaee:2013,Mielenz:2013,Partner:2013,Pyka:2013,Ulm:2013}. On this basis, the authors of \cite{DeChiara:2010,DelCampo:2013} argued that one can test the so-called Kibble-Zurek (KZ) hypothesis by performing slow ramps across the linear-zigzag structural transition. 

The KZ hypothesis was first formulated in an attempt to describe inhomogeneities at the cosmological scale as defects formed during the slow expansion of the universe \cite{Kibble:1980} and then proposed as a paradigm for adiabatic expansion across a continuous, symmetry breaking phase transition \cite{Zurek:1985}. It predicts the scaling of the excess heat produced by a slow ramp of a control field across a second-order phase transition, thereby connecting the equilibrium critical exponents with features of the out-of-equilibrium dynamics.

In the experiments, the statistics of defect formation was measured after slowly ramping the trap aspect ratio across the symmetry-breaking linear-zigzag transition \cite{Ejtemaee:2013,Pyka:2013,Ulm:2013}.  The behavior of the excess heat as a function of the ramp velocity appeared to follow power-law behavior \cite{Ejtemaee:2013,Pyka:2013,Ulm:2013}, compatible with the KZ hypothesis. In this Section we review the theoretical model and the experimental results. 

In \cite{DeChiara:2010} the dynamics of defect formation was modeled starting from the Landau model of Sec.\ \ref{sec:linear-zigzag} for the transverse displacement field $\psi$ and included the dissipative dynamics of laser cooling by means of a damping and a Langevin force (see \cite{Puebla:2017} for a related semiclassical model). For a chain aligned along the $x$ axis and assuming the motion along $z$ is frozen out, the displacement field is along the $y$ direction and its dynamics is governed by the equation
\begin{equation}
\label{Eq:GL:KZ}
\frac{\partial^2}{\partial t^2}\psi-v(x)^2\frac{\partial^2}{\partial x^2}\psi+\delta(x,t)\psi+2{\mathcal A}(x)\psi^3=-\eta \partial_t\psi+\varepsilon(t)\,.
\end{equation}
On the right-hand side, $\eta$ is the damping rate and $\varepsilon(t)$ is the Langevin force (recall: $\langle \varepsilon(t)\rangle=0$ and $\langle \varepsilon(t)\varepsilon(t')\rangle=(2\eta k_\mathrm{B}T/m)\delta(t-t')$ where $T$ is the stationary temperature of Doppler cooling \cite{Stenholm:1986,Morigi:2001a}). The equation of motion on the left-hand side was obtained by taking the continuous limit of Eq.~\eqref{eq:GinzburgLandau:0} with $V_0$ given by Eq.\ \eqref{eq:V0}. Here, the first two terms describe a wave-like propagation with velocity $v(x)$, which is position dependent when the chain's charge distribution is inhomogeneous (see e.g.\ Eq.\ \eqref{Jacobi}).  The parameter $\delta(x,t)\approx \nu_t(t)^2-\nu_t^{(c)}(x)^2$ gives the amplitude of the local force in the harmonic approximation, where $\nu_t^{(c)}(x)$ is a position-dependent critical trap frequency, at which the chain becomes locally unstable \footnote{For a chain in a linear Paul trap, at the center the functional $\nu_t^{(c)}(x)$ is maximum and takes the value $\nu_t^{(c)}(0)=\nu_t^{(c)}$ (see Sec.\ \ref{sec:linear-zigzag}), while it decreases with the distance $|x|$ from the chain's center.}. The term $2\mathcal A\psi^3$ is the force in next order, see Eq.\ \eqref{eq:V0}.  

\begin{figure}
\includegraphics[width=0.48\textwidth]{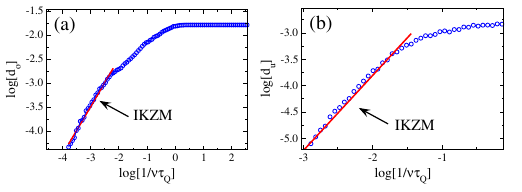}
\caption{\label{Fig:DelCampo2010}(Color online) Density of defects $d$ for an ion chain in a linear trap as a function of the inverse of the sweeping rate (a) in the overdamped and (b) in the underdamped regime. 
 The circles are obtained from a molecular dynamics simulation with $50$ ions in a linear Paul trap, the defects are only
counted in the central 30 ions. The red solid line is the fitting function $d\sim \tau_Q^{-\alpha}$ giving (a) $\alpha =0.995$ and (b) $\alpha=1.427$. Both results are consistent with the predictions of the inhomogeneous KZ mechanism. The curves saturate mostly because of finite size effects. From~\onlinecite{DelCampo:2010}. 
} 
\end{figure}

Let the transverse trap frequency be ramped from above to below the critical point with rate $1/\tau_Q$, such that $\delta(t)=-\delta_0 t/\tau_Q$ with $\delta_0$ a constant while the time is varied in the interval $t\in[-t_0,t_0]$, so that at $t=-t_0$ the linear chain is the stable equilibrium configuration of Eq.\ \eqref{Eq:GL:KZ} while at $t=t_0$ the stationary configuration is the zigzag. According to the KZ hypothesis, the linear and zigzag ground states can be connected by an adiabatic transformation when $\tau_Q$ is much larger than the relaxation time scale $\tau_R$ of the time-independent problem: In this case, the chain will be able to instantaneously adapt to the temporal variation of $\delta(t)$, implying that $\tau_R$ will depend on the instantaneous value of $\delta(t)$, $\tau_R=\tau_R(\delta(t))$. However, close to the critical point $\tau_R$ diverges and the adiabatic condition $\tau_Q\gg\tau_R$ will become invalid. The KZ mechanism identifies the so-called freeze-out time $\hat t$, at which  $\hat \delta=\delta (\hat t)$ so that $\tau_Q=\tau_R(\hat \delta)$. At the freeze-out time scale the KZ theory postulates that the dynamics become frozen out until after the ramp has crossed the critical region. Effectively, it is as if $\delta$ were suddenly quenched from $\hat\delta$ to $-\hat\delta$. According to this conjecture, domains formed in the linear phase at $\hat\delta$ become then topological defects in the symmetry-broken phase. This simple hypothesis permits one to determine the scaling of the density of defects $d$ with the ramp time, since $d$ is inversely proportional to the size $\xi(\hat \delta)$ of the domains.

In the homogeneous case the coefficients of Eq.\ \eqref{Eq:GL:KZ} are independent of $x$ and the critical value is independent of the position along the chain. Close to the critical point, $\tau_R\sim \delta^{-\nu z}$ where $\nu=1/2$ is the mean-field exponent of the correlation length $\xi$ and $z$ the dynamical critical exponent. The scaling of defects is determined by the size of the domain walls at the freeze-out time, with $d\sim 1/\xi(\hat \delta)$ and $\xi(\hat\delta)\sim |\hat\delta|^{-\nu}$, giving $d\sim \tau_Q^{-\alpha}$ with $\alpha=\nu/(1+z\nu)$. Depending on whether the dynamics is overdamped ($\eta\gg \sqrt{\hat\delta}$, corresponding to $z=2$) or underdamped ($\eta\ll \sqrt{\hat\delta}$, corresponding to $z=1$), then $\alpha={1/4}$ or $\alpha={1/3}$, respectively \cite{DeChiara:2010}.

In a linear Paul trap the ion density along $x$ is non-uniform. For long chains there is a propagating front, from the center to the edges, with which $\nu_t(x)$ crosses the local critical value. This front tends to give rise to independent spatial regions where the transition occurs, and thus to nucleation of defects. The speed of the propagating front should be compared with the velocity $v(x)$, at which a perturbation propagates from the center to the edges of the chain causally connecting the chain. Both quantities depend on the local density but with different functional behaviors. This leads to different scaling exponents, namely, $\alpha=1$ for the overdamped regime and $\alpha=4/3$ for the underdamped dynamics \cite{DeChiara:2010,DelCampo:2010}. These scalings are confirmed by molecular dynamics simulations, as visible in Fig.\ \ref{Fig:DelCampo2010}.

\begin{figure}
\includegraphics[width=0.48\textwidth]{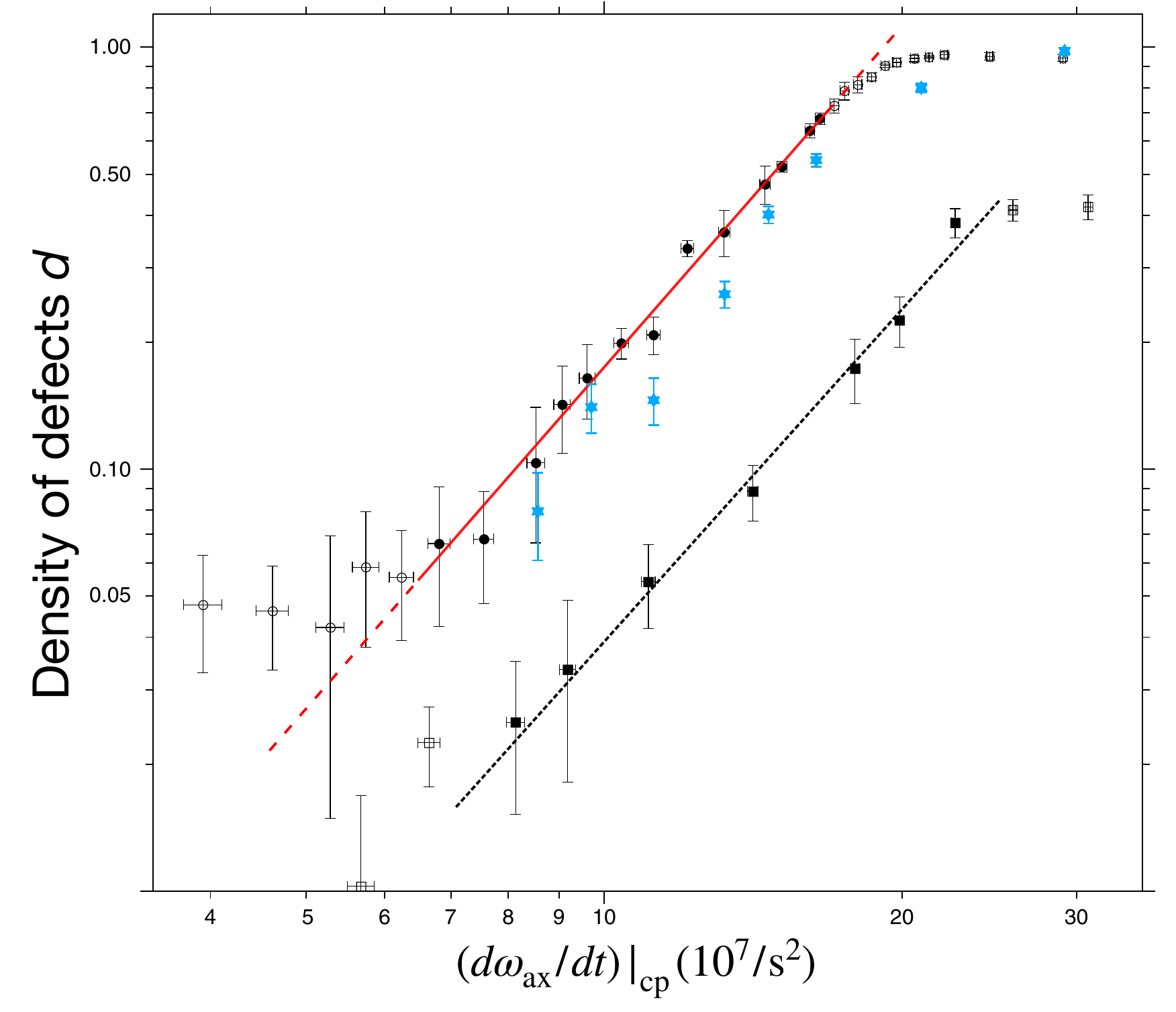}
\caption{\label{Fig4:Pyka2013} 
(Color online) Defect formation rates in a chain of 16 $^{40}$Ca$^+$ ions by changing the axial trap frequency across the linear-zigzag transition. Double-logarithmic plot of the measured density of defects $d$ versus the rate of change of $\omega_{\rm ax}$ at the critical point, where $\omega_{\rm ax}$ is the trap confinement frequency along the direction of the ion chain. All circles correspond to 60000 measurements at a radial trap anisotropy of $1.03$. The fitted function of the form $d\propto 1/\tau_Q^\alpha$ (red solid line) gives an exponent $\alpha = 2.68\pm 0.06$. The constant offset visible at lower ramping rates stems from background gas collisions. The saturation is due to the maximum number of defects being limited by the system size. For comparison, the rate
of defects measured with a higher radial trap anisotropy of 1.05 is plotted (squares), showing a loss of defects but a similar fitted exponent of $\alpha= 2.62 \pm 0.15$ (dotted line). Solid data points are used for the fits. The blue stars depict the result of molecular dynamics simulations. Adapted from \onlinecite{Ulm:2013}}.
\end{figure}
Experimentally the slow quench was performed in three different platforms, with different ion numbers and species. The trap aspect ratio was varied across the linear-zigzag transition by either sweeping the transverse \cite{Pyka:2013} or the axial frequency \cite{Ulm:2013}, or both simultaneously \cite{Ejtemaee:2013}. In all experiments the ions were initially cooled to the Doppler limit and the cooling laser was on during the whole duration of the quench. The density of defects was measured after the quench using a CCD camera per sweep and averaging over several sweeps, yielding consistent results. Figure \ref{Fig4:Pyka2013} shows the resulting statistics of kinks \cite{Ulm:2013}. The solid and dashed lines show the function $d\sim \tau_Q^{-8/3}$, which fits the interval in which the sweep rate exceeds the rate of background collisions but is still sufficiently small so that defect generation does not saturate. This scaling is compatible with the measurements reported in \cite{Ejtemaee:2013} and was also reported in the defect statistics measured in \cite{Pyka:2013} on a larger parameter interval. Molecular dynamics simulations accounting for the experimental setup reproduced the results of Ref.\ \cite{DeChiara:2010,DelCampo:2010} for other ramps than the experimental ones and reported the exponent $\alpha=8/3$ for the experimental regime \cite{Ejtemaee:2013,Pyka:2013}. This power-law scaling has been attributed to a form of KZ mechanism, even though a theoretical model justifying this scaling has not been yet identified, see \cite{DelCampo:2013}. 
 
The agreement between experiments and molecular dynamics simulations shows that the observed scalings emerge from classical dynamics. Indeed, the experiments are performed with laser-cooled ions at Doppler temperatures at which quantum effects are not expected to be visible (see Sec.\ \ref{Sec:2}A and Ref.\ \cite{Shimshoni:2011a}). Quantum effects could be measured for slower quenches and purely coherent dynamics. In this case, it should be possible to observe the scaling of defects determined by the critical exponents of the Ising model \cite{Silvi:2013}, namely, the quantum KZ mechanism. In Ref. \cite{Silvi:2016} a DMRG simulation was used to determine the defect statistics generated by slow quenches across the quantum critical model of Sec.\ \ref{Qlinear_zigzag}.  A crossover from classical to quantum KZ scaling was reported as a function of the quench rate, identifying the requirements on the ramp speed in order to unveil quantum effects on defect formation.

\subsection{Stick-slip motion and nano-friction}
\label{Sec:Frustration}

\begin{figure}
\includegraphics[width=0.44\textwidth]{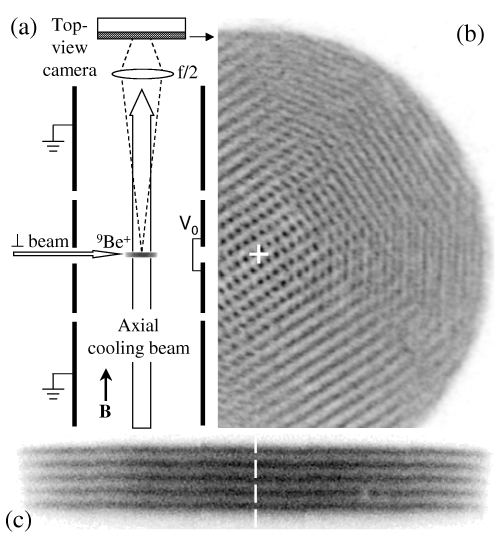}
\caption{\label{Fig:1:Mitchell}
Setup used to study stick-slip motion on a Coulomb crystal.  (a) Schematic of the cylindrical Penning trap and the
top-view imaging system. (b) Strobed top-view image of a five-axial plane $^9$Be$^+$ 
ion crystal with a bcc structure.
(c) Side-view image (unstrobed) of the same ion crystal. A rotating electric field creates a torque on the crystal. From \cite{Mitchell:2001}.
}
\end{figure}

\begin{figure}
\includegraphics[width=0.45\textwidth]{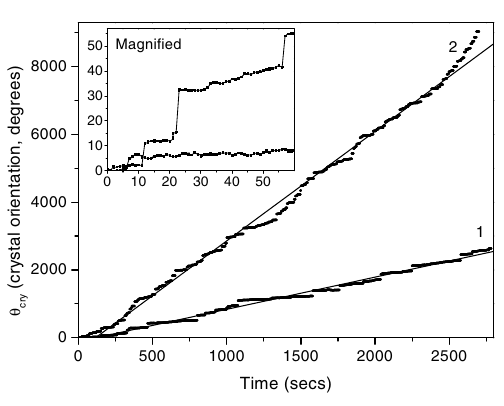}
\caption{\label{Fig:2:Mitchell}
Crystal orientation in the frame of the rotating
perturbation for measurements with two different perpendicular beam torques.  The torque is greater in measurement 2. The lines are from a linear regression fit. The inset shows a magnified plot of the first 60 s of data. From \onlinecite{Mitchell:2001}. 
} 
\end{figure}

The possibility to image the ions within a Coulomb crystal and to control the forces at play make these systems ideal platforms for investigating static friction \cite{Vanossi:2013}. Static friction refers to the finite work needed to put in motion an object in contact with a surface. It is associated with dynamical phenomena such as stick-slip motion, which consists of intermittent sliding and stuck phases and can characterize the motion of a mesoscopic object dragged over a surface. Stick-slip motion has been observed in laser--cooled ion crystals rotating in Penning traps in the setup shown in Fig.~\ref{Fig:1:Mitchell}. Stress was created by two opposing torques generated by a rotating electric field that controls the ion crystal rotation through forces applied to the crystal boundary, and radiation pressure forces from a focused laser beam, implementing Doppler cooling \cite{Mitchell:2001}. Stick-slip dynamics was manifested in sudden angular jumps (“slips”) of the crystal orientation alternating with intervals when the crystal orientation was phase locked (“stuck”) relative to the rotating wall perturbation, as seen in Fig.\ \ref{Fig:2:Mitchell}. Interestingly, the slip amplitudes were shown to exhibit a power-law distribution reminiscent of self-organized critical phenomena \cite{Mitchell:2001}.

\begin{figure}
\includegraphics[width=0.45\textwidth]{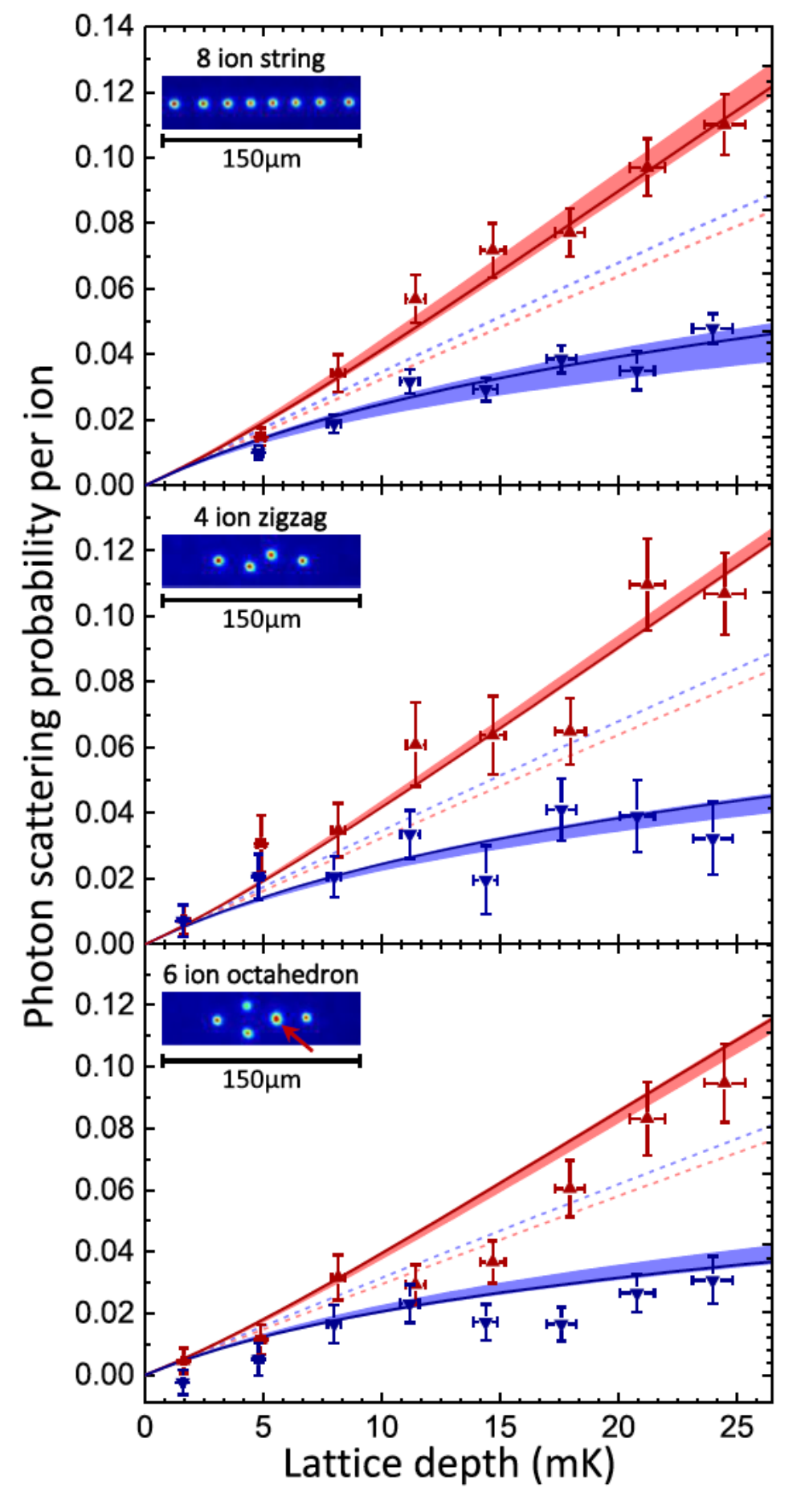}
\caption{\label{Fig:Pinning}
(Color online) The probability of scattering a photon from a one-dimensional optical lattice as a function of the optical lattice depth for a small one-
(top), two- (center), and a three-dimensional Coulomb crystal (bottom). The red (up-pointing) and blue (down-pointing) triangles are experimental data points for red- and blue-detuned lattices, respectively. As expected for pinned ions, the scattering probability is higher for pinning at intensity maxima (red detuning) than at intensity minima (blue detuning). For comparison, the red (lower) and blue (upper) dashed lines show the theoretical scattering probabilities expected for delocalized ions. From \onlinecite{Laupretre:2019}, which includes a discussion of the error bars and uncertainties. 
} 
\end{figure}

\begin{figure}
\includegraphics[width=0.48\textwidth]{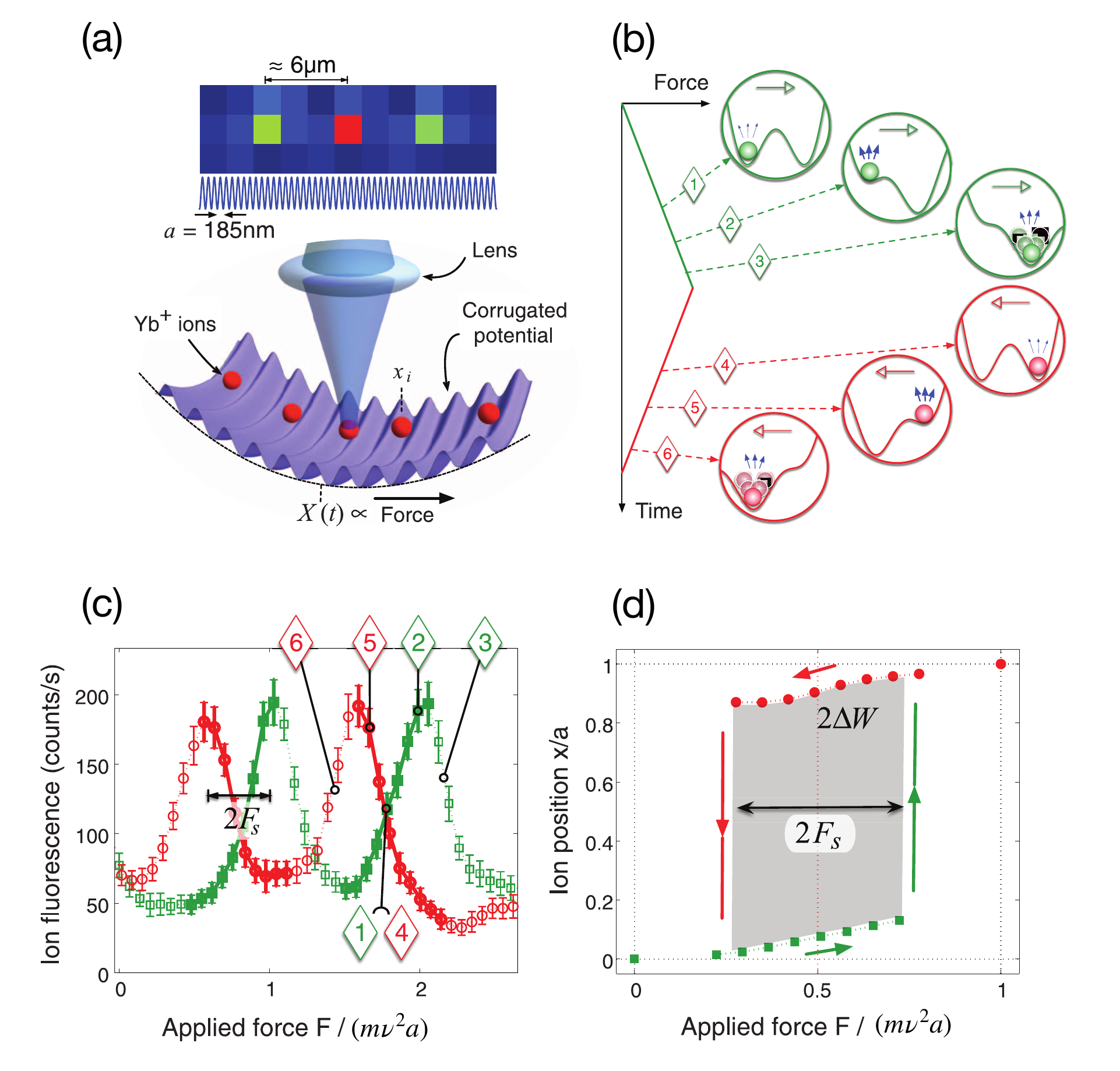}
\caption{\label{Fig:1:Bylinskii} (Color online) Ion-crystal emulator of stick-slip friction. (a) A Coulomb crystal of $^{174}$Yb$^+$ ions subjected to an optical lattice potential.  Ions are imaged with a single-ion-resolving microscope. The typical ion spacing is 6 $\mu m$, and the lattice period is $a$=185 nm. In the bottom illustration of the corrugated potential, the lattice period and the corrugation are strongly exaggerated. (b) Stick-slip results from bistability, illustrated here for a single ion. A shear force is linearly ramped, causing the ion to jump between minima.  The stages of the stick-slip process are labeled by numbered diamonds and illustrate the dynamics of the ion during the ramp.
The ion's position is extracted from its fluorescence, which is proportional to the lattice potential energy.
(c) Fluorescence versus applied force during the forward transport (green squares) and reverse transport (red circles), showing hysteresis that is used to measure the maximum static friction force $F_s$. The bold data points indicate the ion’s position before a slip, and only those data are used to reconstruct the force-displacement curve. (d) The force-displacement hysteresis loop encloses an area equal to twice the dissipated energy per slip $\Delta W$. The unit $m\nu^2 a$ of the applied force corresponds to $2.8 \times 10^{-19}$ N; here, $\nu = 2\pi \times 364$ kHz. Adapted from \onlinecite{Bylinskii:2015}.
}
\end{figure}

Ion chains offer a controlled environment for studying static friction. Pinning at the sub-wavelength scale can be observed using resonance fluorescence and has been measured in small ion clusters confined by optical lattices \cite{Enderlein:2012,Linnet:2012,Laupretre:2019,Bylinskii:2016,Gangloff:2020}. Figure \ref{Fig:Pinning} shows the resonance fluorescence of trapped ions as a function of lattice depth for various geometries, allowing the identification of the ions' positions within the lattice. The dragging of an object over a surface can be implemented by setting the chain in motion relative to the optical lattice, as illustrated in Fig.~\ref{Fig:1:Bylinskii}: Stick-slip motion results in bistability in the potential energy and manifests itself as hysteresis in the ion position. In Ref.~\cite{Bylinskii:2015,Gangloff:2015} the maximal friction force was extracted from measurements of the hysteresis of the ion position. 

A further advantage of studying nano-friction with ion chains is that these platforms are amenable to microscopic modeling and mapping to paradigmatic models \cite{Garcia-Mata:2007,Benassi:2011,Pruttivarasin:2011,Cormick:2013,Vanossi:2013}. In particular, when the ion chain can be described by an elastic crystal, the dynamics is captured by a paradigmatic model of static friction, the so-called Frenkel-Kontorova (FK) model \cite{Braun:2004,Pruttivarasin:2011,Cormick:2013}:
\begin{equation}
\label{Eq:FK}
    V_{\rm FK}=\frac{1}{2}\sum_{i,j}{\mathcal K}_{ij}(q_i-q_j)^2+\frac{U}{2}\sum_j(1-\cos(2\pi x_j/a))\,,
\end{equation}
where the second term is the optical lattice with depth $U$ and period $a$ and $q_j=x_j-x_j^{(0)}$ is the displacement of ions with respect to the equilibrium positions $x_j^{(0)}$ at $U=0$. The harmonic crystal approximation is valid when the average distance between ions, $d_0$, satisfies $d_0\gg a$ and the Coulomb force varies smoothly over $a$. This is typically fulfilled in ion chains, where $d_0\gtrsim 5 \mu$m and is thus at least one order of magnitude larger than the optical lattice periodicity $a$. 

For uniform chains and nearest-neighbor couplings, when $x_j^{(0)}=jd_0$ and ${\mathcal K}_{ij}=K_0\delta_{j,i+1}$, the ground state and dynamical properties of the FK model are fully determined by two dimensionless parameters, the ratio $U/K_0$ between the substrate potential and the elastic force, and the ratio $d_0/a$ between the two characteristic length scales. At fixed, incommensurable values of the ratio $d_0/a$, the critical value $[U/K_0]_c$ signals the Aubry transition, separating the sliding phase, where forces giving rise to sticking cancel, from the pinned phase, where the motion alternates stick and slip phases and the minimal force needed to initiate sliding is finite \cite{Braun:2004}. In the pinned phase, continuous translational invariance is broken and pinning is signaled by a nonzero frequency of the lowest phonon mode. At fixed lattice depth, a given commensurate (periodic) phase at a commensurable ratio $d_0/a$ remains stable for small variations $d/a=d_0/a\pm\delta$, where the mismatch $\delta$ quantifies the deviation from the commensurable ratio. The system undergoes a discontinuous transition to an incommensurate (sliding) phase at a critical value $\delta_c>0$ where formation of solitons is energetically favorable and the phase becomes sliding. 

\begin{figure}
\includegraphics[width=0.45\textwidth]{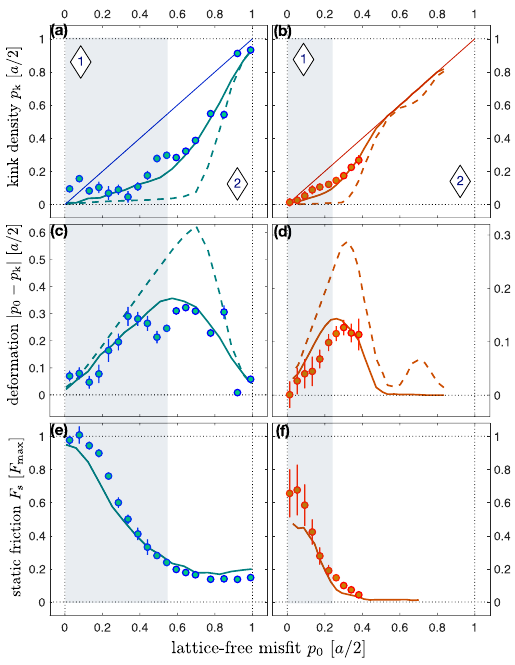}
\caption{\label{Fig:3:Gangloff} 
Commensurate-incommensurate phase transition in an extension of the setup of \cite{Bylinskii:2016}, for two (left) and five ions (right).  The plots are the measured kink density $p_k$ (top), deformation $|p_0-p_k|$ (middle), and static friction force $F_\mathrm{s}$ (bottom), vs misfit $p_0$.  Lines in (a) and (b) are the kink density $p_k=p_0$ for $U=0$.  Solid curves in (a)-(f) are numerical simulations for parameters matching those of the data shown, using $k_\mathrm{B} T/U=0.2$ [(a), (c), and (e)] and 0.5 [(b), (d), and (f)].  Dashed curves are $T=0$ simulations.  $F_\mathrm{max}$ is the maximal friction in the single-slip regime.  From \onlinecite{Gangloff:2020}.}
\end{figure}

Implementing these dynamics with ions requires one to take into account the long-range Coulomb interactions. Moreover, the external potential of a linear Paul trap breaks the translational invariance, and the lattice-free interparticle distance $d_{j,j+1}=x_{j+1}^{(0)}-x_{j}^{(0)}$ is inhomogeneous, but symmetric under reflections about the trap center. For small ion chains, from two to five ions, the long-range interactions do not significantly modify the dynamics with respect to the short-range model. The commensurate phase is realized by adjusting the Paul trap harmonic potential, tuning the misfit to $p_0=0$, where the misfit is defined as \cite{Gangloff:2020}
\begin{eqnarray}
  p_0=\frac{2}{N-1}\sum_{j=1}^{N-1}(d_{j,j+1}/a)\bmod 1 \,. 
\end{eqnarray}
In these settings, changing $p_0$ is equivalent to tuning the mismatch $\delta$. Figure \ref{Fig:3:Gangloff} displays measurements on small chains for different misfits $p_0$ across the commensurate-incommensurate transition \cite{Gangloff:2020}. The grey region corresponds to the commensurate phase, where formation of kinks is suppressed at zero temperature, while in the white region the friction force is minimal. The discrepancy of the experimental data from the zero temperature simulation is attributed to thermal effects. This hypothesis is supported by comparison with molecular dynamics calculations that assume a finite temperature. 
Note that in the Paul trap the friction force does not vanish in the sliding phase, but becomes independent of the misfit $p_0$ or of the lattice depth $U$. In this environment, the sliding-pinning Aubry transition breaks the reflection symmetry about the trap center \cite{Braiman:1990,Benassi:2011,Fogarty:2015}. 

\begin{figure}
\includegraphics[width=0.40\textwidth]{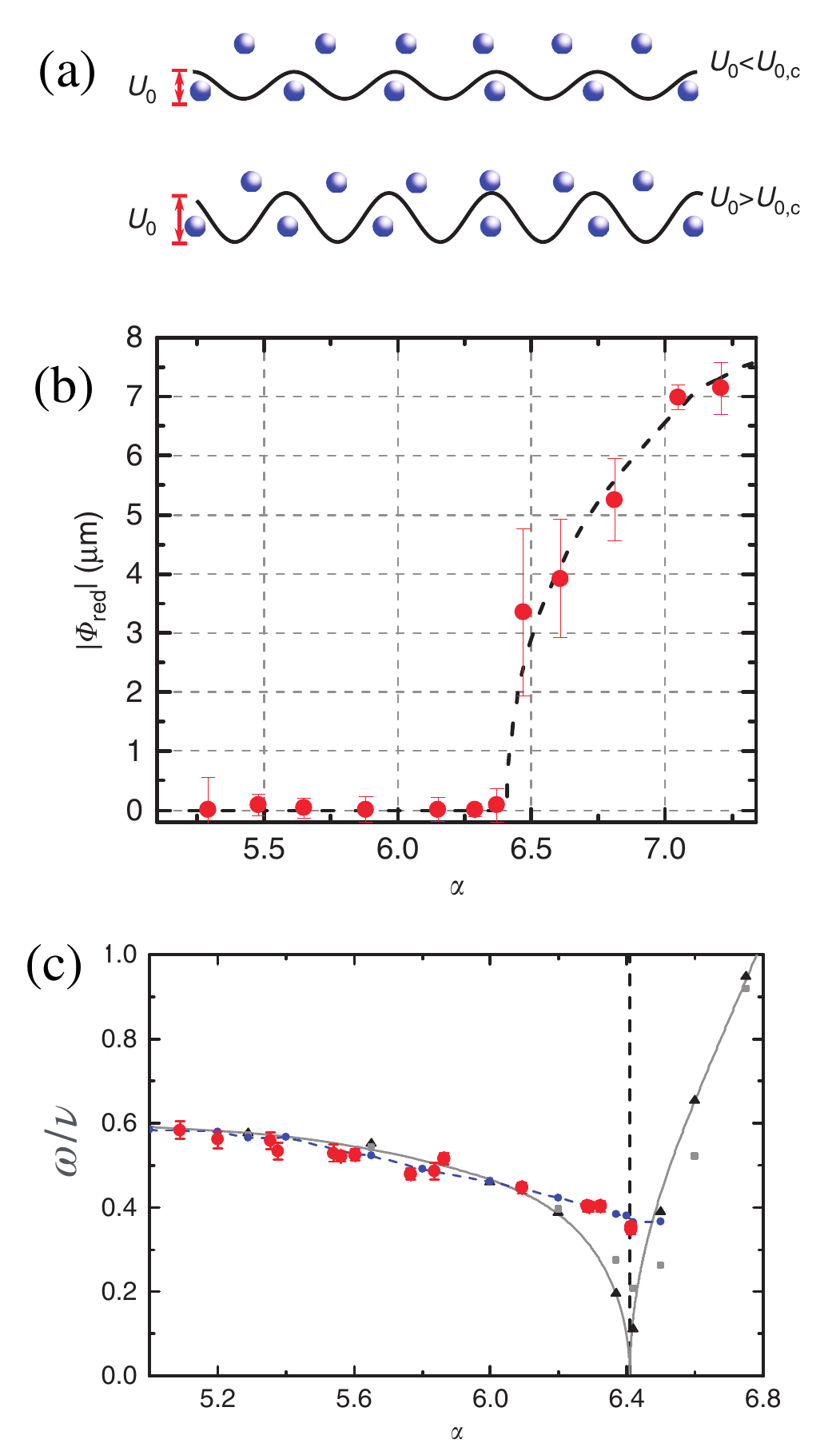}
\caption{\label{Fig:Kiethe} (Color online) 
(a) Illustration of the sliding phase (extended kink, upper) and pinned phase (zigzag chain, lower). The Coulomb potential of one row of ions acts as the corrugation potential for the other row. The depth of the corrugation $U_0$, controlled by the trap aspect ratio, determines the dynamics of the system. Below a critical corrugation depth $U_{0c}$ the system displays horizontal mirror symmetry. Above the critical value the symmetry is broken. (b) The experiment is performed with $^{172}$Yb$^+$ ions in a linear Paul trap laser cooled at about 1\,mK. The absolute value of the order parameter $|\Phi|$, which characterizes the horizontal mirror symmetry, plotted against the aspect ratio transverse to axial frequencies, $\alpha=\frac{\nu_t}{\nu}$. In the pinned phase, the symmetry is broken ($|\Phi|>0$). Experimental data
(red circles) are shown in comparison to numerically obtained values for $T=0$ K (black dashed line). Error bars are one standard deviation estimates. 
(c) Frequency of the soft mode in a 30-ion crystal. The experimental data is shown in red circles and the error bars are given by uncertainties of the measured soft mode and common mode frequencies. The solid line displays the numerically calculated
dispersion relation at $T=0$ K. Frequencies extracted via a Fourier transformation from molecular dynamics simulation are given by black triangles at
$T=5~\mu$K, grey squares at $T=50~\mu$K and blue circles at $T=1$ mK. The dashed blue line is a guide to the eye. All frequencies are plotted in units of the
axial secular frequency $\nu=2\pi\times(25.6\pm 0.2)$ kHz. The pinning transition is marked with a vertical line. Adapted from~\onlinecite{Kiethe:2017}.
} 
\end{figure}

The Aubry transition was observed in the setup of Fig.\ \ref{Fig:1:Bylinskii} by measuring the static friction force as a function of optical lattice depth \cite{Bylinskii:2016}. It was also reported in a setup \cite{Kiethe:2017}, which simulated the FK model in a chain of about 30 ions in the absence of a periodic lattice. In the realization of \cite{Kiethe:2017}, in fact, the elastic crystal and the substrate are the upper and the lower chain of an extended kink, which is metastable in the zigzag phase for a finite range of aspect ratios, see Fig.\ \ref{Fig:Kiethe}(a). In this setup, a cooling laser induces differential forces between the upper and the lower chain and can thus set one into motion with respect to the other \cite{Kiethe:2017,Kiethe:2018}. The parallel chains forming the extended kink can slide on top of one another. By decreasing the transverse trap frequency, the extended kink becomes unstable and the stable structure is a zigzag, corresponding to locking one chain with respect to the other, realizing a pinned phase. The symmetry breaking is revealed by measuring
the relative axial distance $\Phi$ between the ions of the different layers as a function of the trap aspect ratio. The results of the measurement are shown in Fig. \ref{Fig:Kiethe}(b). Panel (c) reports the measurement of the frequency of the lowest vibrational mode, which is detected by a second laser performing vibrational spectroscopy.
The smoothing of the transition is attributed to thermal fluctuations and is reproduced by both molecular dynamics simulations and a microscopic model that incorporates thermal effects \cite{Kiethe:2021}.

Despite the analogy with the FK model, the two chains in Fig.\ \ref{Fig:Kiethe} mutually interact and hence proper modeling of their dynamics must take into account the back action of the elastic crystal on the surface on which it is dragged. This is indeed closer to the typical scenario of nanofriction, where the substrate is deformable, than the idealized FK model. Another implementation of deformable substrates is realized when the ion chains interact with the standing wave field of an optical cavity \cite{Guthoehrlein:2000,Keller:2004,Linnet:2012,Cetina:2013,Casabone:2015}. For sufficiently large couplings, the cavity optical lattice depth depends nonlinearly on the ion density, such that it is constant in the sliding phase and can become steeper (or shallower, depending on the experimental parameters) in the pinned phase \cite{Cormick:2012,Cormick:2013,Fogarty:2015}. For this setup the sliding and pinned phases are separated by bistable regions where superlubric and stick-slip dynamics coexist \cite{Fogarty:2015}. Moreover, photon-scattering into the cavity mode can cool the chain vibrations even down to the zero-point motion \cite{Fogarty:2016}. Entanglement in the pinned phase has been theoretically studied in \cite{Kahan:2024}. Phase, temperatures, and correlations can be measured in the spectrum of the intensity at the cavity output \cite{Cormick:2013,Fogarty:2016}. 

\subsection{Kink dynamics and quantum sliding}
\label{Sec:quantum-Kink}
Kinks of the FK model are quasiparticles of tunable mass and velocity that can be revealed by resonance fluorescence \cite{Chelpanova:2023}.  In \cite{Gangloff:2020} a kink was created at one end of a 5-ion chain. Its dynamics as a function of the misfit $p_0$ was measured: In the sliding phase the kink travels through the chain with a velocity that increases with decreasing $p_0$, while it fades away in the commensurate phase \cite{Gangloff:2020}. This behavior is qualitatively reproduced in numerical simulations \cite{Chelpanova:2023,Chelpanova:2024}. Semiclassical simulations of sudden quenches from the commensurate into the incommensurate phase indicate that, deep in the quantum regime, the dislocation can tunnel into the chain, such that the chain is a transient superposition between a commensurate and an incommensurate (sliding) structure \cite{Chelpanova:2023,Chelpanova:2024}. 

The experimental level of control suggests that it could be possible to observe quantum dynamics of kinks. This question is at the center of studies  in material science of quantum lubrication, where sliding is initiated by quantum tunnelling  \cite{Krajewski:2004,Vanossi:2013}. Path Integral Quantum Monte Carlo simulations identified the regime of parameters  where quantum-induced sliding phases could be observed in chains of 5 ions \cite{Bonetti:2021}. Semiclassical analysis discussed the dynamics of tunnelling of the kink in the platform of \cite{Kiethe:2017}, see \cite{Timm:2021,Timm:2023}. 

A general framework for quantum lubrication in ion chains was developed in \cite{Landa:2020,Menu:2024a,Menu:2024b}, which permits one to derive a quantum field theory for the solitons as quasiparticles. These works showed that the long-range Coulomb interactions substantially modify the features of the topological kinks with respect to the short-range FK model, favoring soliton formation in long chains \cite{Menu:2024a,Menu:2024b}.

\begin{figure}
\includegraphics[width=0.44\textwidth]{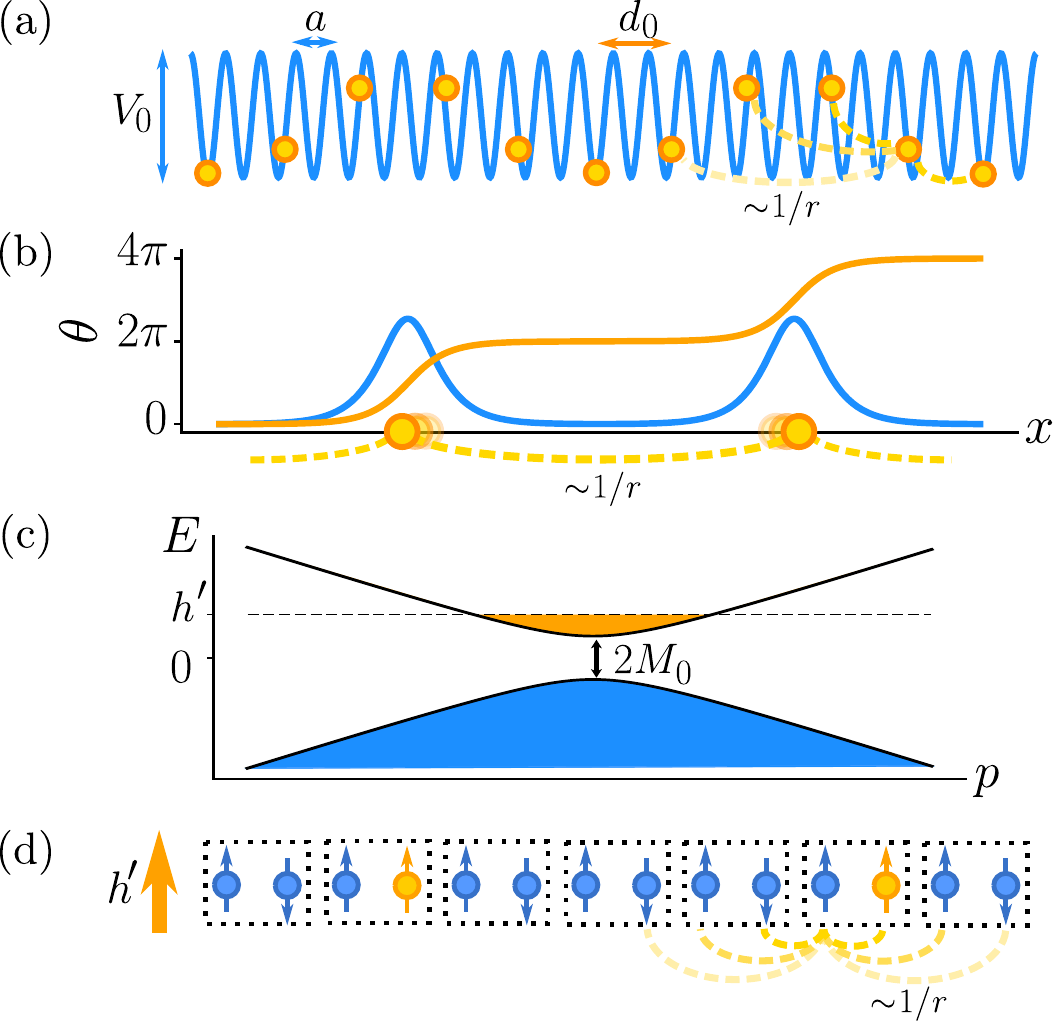}
\caption{\label{Fig:1:Menu} (Color online) 
(a) Schematic representation of the Wigner crystal in the presence of an externally applied optical lattice, mimicking a substrate potential. The substrate potential leads to dislocations of the ions from the crystal's equilibrium position. (b) The dislocations (solitons) are displayed as phase shift $\theta(x)$ along the chain axis $x$ (monotonous orange curve). The first-order derivative of $\theta(x)$ (blue) exhibits local maxima where the solitons are located. The yellow dashed curves illustrate that the solitons behave as interacting charges. (c) In the Thirring model the solitons are positive-energy excitations over a filled Dirac sea, where the gap is twice the soliton mass $M_0$. The phase is commensurate when the chemical potential $h'$ (controlled by the mismatch) falls within the gap. For $h'>M_0$, the solitons proliferate and the phase is incommensurate. (d) Interpretation of the spin-fermion equivalence in the discretized Thirring model, corresponding to a long-range interacting XXZ antiferromagnetic spin chain: Solitons are defects (co-polarized adjacent spins highlighted in orange) in the staggered ordering of a spin chain. From \onlinecite{Menu:2024a}.
}
\end{figure}

Figure \ref{Fig:1:Menu}(a) sketches a chain displaying two kinks. The kink configuration is described by the finite displacement $\delta x_j$ from the position of the commensurate configuration and can be cast in terms of a phase $\theta_j=2\pi \delta x_j/a$, such that a change by $2\pi$ corresponds to a kink, see Fig.\ \ref{Fig:1:Menu}(b) and \cite{Chelpanova:2024}. For long chains and $d_0\gg a$ the phase $\theta_j$ is captured by the field $\theta(x,t)$, which is the solution of the modified sine-Gordon equation \cite{Braun:1990,Landa:2020}:
\begin{align}
    \dfrac{1}{v_s^2} \partial^2_t \theta &=  \partial^2_x\theta - M^2 \sin \theta \notag \\
    &+ \dfrac{1}{3}\partial_x\int_{1}^{N/2}{\dfrac{\partial_x\theta(x+u) + \partial_x\theta(x-u)}{u}\mathrm{d}u}\label{Eq1}\, ,
\end{align}
and where $x$ is here rescaled by $a$.  The equation is parametrized by the velocity $v_s = \sqrt{{3K_0}/({2m})}$ and the mass term $M = \sqrt{{8\pi^2U}/({3K_0a^2})}$, with $K_0=2Q^2/(4\pi\epsilon_0 d_0^3)$. The solutions of Eq.\ \eqref{Eq1} are constrained to satisfy the mismatch $\delta=\min_{n_0\in \mathbb{N}}(d_0-n_0a)/a$. In the short-range FK model the integral term vanishes and the solution is a sine-Gordon soliton with sound velocity $v_s$ and length proportional to $d_0/M$, while the mass $M$ scales the energy of a kink. The Coulomb interactions give rise to the integral term. As a consequence, at the center the kink has the form of a sine-Gordon soliton whose width now scales with the size of the chain as $\log N$ while it asymptotically decays as $1/r$ with the distance $r$ from its center \cite{Landa:2020}. This asymptotic decay gives rise to Coulomb-like interactions between distant solitons \cite{Menu:2024a}. 

Equation \eqref{Eq1} permits to map the dynamics of a frustrated ion chain to a quantum field theoretical model, the massive Thirring model, where the commensurate phase is an effective Dirac sea and  the solitons are charged fermionic excitations, as illustrated in Fig.\ \ref{Fig:1:Menu}(c). The chemical potential is proportional to the mismatch and shows that, when this exceeds the gap, excitations (solitons) are generated. Because of the Coulomb interactions, the chemical potential also depends on the chain's size: $h' \propto \delta\ln N$. This scaling of $h'$ indicates that the Coulomb interactions dictate the order in the thermodynamic limit. Correspondingly, the critical mismatch $\delta_c$ shrinks as $1/\ln N$, giving rise to {\it interaction-induced} sliding \cite{Menu:2024a,Menu:2024b}. An equivalent spin model can be derived from the Thirring Hamiltonian, which is amenable to numerical analyses. This is a long-range XXZ Hamiltonian of spin-$\frac{1}{2}$ particles in an external field. The solitons are mapped into long-range interacting defects, as shown in Fig.\ \ref{Fig:1:Menu}(d). In these models quantum effects in the soliton's dynamics scale with the effective Planck constant $\beta^2$ \cite{Menu:2024a},
\begin{equation}
\beta^2 = \left(\frac{2\pi d_0}{a}\right)^2 \sqrt{\frac{2\hbar^2/(3m d_0^2)}{2Q^2/(4\pi\epsilon_0 d_0)}}\,,
\end{equation}
which is proportional to the square root of the ratio between kinetic and Coulomb energy. This quantity is proportional to $\tilde{\hbar}$ introduced in Sec. \ref{Qlinear_zigzag}, which determines the size of quantum effects at the linear-zigzag transition. The proportionality factor contains the ratio $(d_0/a)^2$ between the characteristic length scales of chain and lattice and suggests that size of quantum effects can be tuned by changing the trap aspect ratio and/or the wavelength of the confining lattice. Taking experimentally accessible parameters, it can range from $\beta^2\sim 10^{-4}$, deep in the mean-field regime, to $\beta^2\sim 0.1$, where quantum sliding could be measured \cite{Menu:2024a,Menu:2024b}. 

A mean-field study predicted the regions of stability of the commensurate configurations at fixed $N$ as a function of the mismatch $\delta$ and of the temperature $T$ \cite{Menu:2024b}. Using the parameters of Ref. \cite{Bylinskii:2016}, for a chain of 100 ions $^{173}$Yb$^+$ with interparticle distance $d_0 = 6 ~\mu\mathrm{m}$ and lattice periodicity $a=185~\mathrm{nm}$, commensurate phases can be observed at temperatures $T\lesssim 1 \,\mathrm{mK}$ \cite{Menu:2024b}. Quantum sliding could be revealed by means of spectroscopic measurements for relative precisions below the effective Planck's constant $\beta^2$.

\section{Concluding remarks and outlook}
\label{Sec:4}

This review reports on advances with ion Coulomb chains and crystals, starting from the state-of-the-art of more than two decades ago, when the non-neutral plasma phase was at the center of the studies~\cite{Dubin:1999}. It draws from  progress by the quantum optics and quantum information communities that, through advances in laser cooling techniques and quantum engineering, exploited the utility of ion chains and crystals for quantum sensing and information processing. 
We chose to emphasize the physical properties of ion Coulomb crystals and examined several studies of out-of-equilibrium dynamics carried out with ion Coulomb crystals. 

In contrast with electronic systems in condensed matter, ion Coulomb crystals give access to a regime that is dominated by interactions, rather than the kinetic energy of the particles. Their equilibrium and out-of-equilibrium dynamics is strongly shaped by the long-range potential, where the dominant energy contribution is non-additive, thus breaking crucial principles of textbook statistical mechanics, such as ensemble equivalence~\cite{Campa:2009}.  This makes ion Coulomb crystals a prominent laboratory for testing theoretical hypotheses formulated for astrophysical and plasma systems~\cite{Campa:2009}. Moreover, the experimental progress opens the possibility for accessing regimes so far unexplored that can shed light on the onset of equilibrium~\cite{Defenu:2024} and on quantum phases~\cite{Defenu:2023} in the presence of long-range, non-additive forces.

Exotic quantum phases and critical phenomena are, for instance, the zigzag and buckling structural phase transitions, as well as the topological features of frustrated geometries, such as the Aubry and commensurate-incommensurate transition in ion chains. These are also examples of how discrete internal degrees of freedom, such as spins, are naturally encoded in the structure of a low dimensional crystal. This paradigm could be generalized to multilayer crystals where the layer index is a discrete degree of freedom that serves as an effective spin, mapping the dynamics to models analogous to antiferromagnetism in frustrated geometries. The long-range nature of topological defects, such as dislocations and kinks, is only partially characterized in one dimension, and unexplored in higher dimensions. One question is how the coupling between lattice vibrations and topological defects (dislocations, disclinations, kinks, vortices, etc.) might give rise to metastable structures and novel phases in frustrated geometries with long-range Coulomb forces. This can be seen as an exotic realization of quantum magnetism in dynamic Wigner lattices, see \cite{Nath:2015} for a related proposal. It holds the promise for simulating dynamical features of lattice-field theory with trapped ions \cite{Bauer:2023,Bazavan:2024,Menu:2024a,Kahan:2025}. It would be interesting,  for example, to investigate how the influence of the coupling to the lattice vibrations and defects affects the nature of the magnetic phases (and thus of the layer ordering) as well as the transitions between them \cite{Shamai:2018,Samanta:2022,Abutbul:22,Sarkar:2023}.

One peculiar aspect of these systems is that exchange interaction and quantum statistics are very small and generally not measurable. Yet, an ingenious experiment measured the Aharonov-Bohm effect in the macroscopic tunneling of a chain of three ions realizing a quantum rotor \cite{Noguchi:2014}. Protocols have been proposed for revealing quantum statistics in ion Coulomb crystals \cite{Roos:2017}, indicating an avenue for unveiling these phenomena with larger crystals.

The progress reported in this review focused on fundamental aspects of ion Coulomb crystals, while intentionally leaving aside the remarkable advances in their use for engineered quantum dynamics in quantum metrology, quantum simulation \cite{Blatt:2012,Britton:2012,Monroe:2021}, quantum sensing, and quantum computing \cite{Cirac:1995,Bruzewicz:2019,Moses:2023}. Closely related developments include fundamental insights into tailoring correlations between ions with photons by means of optical elements and cavities \cite{Eschner:2001,Casabone:2015} and advanced detection schemes \cite{Moehring:2007,Slodicka:2013,Richter:2021}, which underpin powerful applications in quantum communication \cite{Duan:2010,Saha:2025}. Importantly, the advances reported here have evolved hand-in-hand with the development of these applications, leading to a strong cross-fertilization of methods and physical insights.

This interplay suggests that both the scientific reach and technological impact of ion Coulomb crystals will continue to grow. Many of these applications are rooted in the fact that the internal states of individual ions in a Coulomb crystal are only weakly perturbed by the presence of other ions, making these systems a highly interesting platform for high-resolution spectroscopy, sensing, and metrology \cite{Mehlstauebler:2018}. Representative examples in quantum sensing include extremely accurate optical atomic ion clocks \cite{Huntemann:2012,Tofful:2024,Marshall:2025}, ultra-high-resolution spectroscopy of molecular ions \cite{Chou:2020,Kortunov:2021}, and precision sensing of weak electromagnetic fields \cite{Kotler:2011,Ruster:2017,Gilmore:2021,McKay:2021}. Notably, a recent multi-ion clock based on a linear chain of 4 $^{115}$In$^+$ ions sympathetically cooled by 8 $^{172}$Yb$^{+}$ ions demonstrated a fractional frequency instability of $9.2\times 10^{-16}$ at 1 second averaging time~\cite{Hausser:2025}. 

These quantum sensing capabilities can in turn be used as powerful probes of fundamental physics, enabling tests of Lorentz invariance \cite{Megidish:2019,Driessen:2022}, searches for temporal variations of fundamental constants \cite{Rosenband:2008,Dzuba:2024}, for particles beyond the Standard Model \cite{Gerbert:2015,Counts:2020,Solaro:2020,Door:2025}, and for signatures of potential extensions to quantum mechanics \cite{Lenler-Eriksen:2024}.

We thus expect much more to come in this field, where atomic, molecular, optical, statistical, and, most recently, condensed-matter physics merge in verifying hypotheses and providing novel insights into the physics of Wigner crystallization with trapped ions. This progress opens new paths towards understanding the onset and stability of long-range interacting quantum structures and offers exciting perspectives for technological advances in the quantum control of complex systems.

\acknowledgments
The authors are indebted to our close collaborators Daniel Dubin, Shmuel Fishman, Herbert Walther, and David Wineland, with whom we had the privilege to work over the years on this fascinating subject. We further had the privilege to share many useful discussions and insights with a number of colleagues and collaborators. For reasons of space we refrain from mentioning them all, and explicitly list our collaborators on ion Coulomb crystal work in alphabetical order: Matt Affolter, Francois Anderegg,  Jens Baltrusch, Rainer Blatt, Joseph Britton, Andreas Buchleitner, Tommaso Calarco, Adolfo del Campo, Florian Cartarius, Allison Carter, Caroline Champenois, Oksana Chelpanova, Maria-Luisa Chiofalo, Cecilia Cormick, Gabriele De Chiara, J. Ignacio Cirac, Aurelien Dantan, C. Fred Driscoll, J\"urgen Eschner, Thomas Fogarty, James Freericks, Jeffrey S. Hangst, Jan-Petter Hansen, Peter Horak, Liv Hornek\ae{}r, X.-P. Huang, Wayne Itano, Endre Kajari, Martina Knoop, Jan Kiethe, Niels Kj\ae{}rgaard, Haggai Landa, Thomas Laupr\^etre, Dietrich Leibfried, Ian D. Leroux, Jamir Marino, Thierry Matthey, Tanja E.\ Mehlst\"aubler, Raphael Menu, Travis Mitchell, Simone Montangero, Thomas O'Neil, Ramil Nigmatullin, Scott Parker, Martin Plenio, Dirk Rei\ss, Alex Retzker, Ana Maria Rey, Benni Reznik, Brian Sawyer, Tobias Sch\"atz, Ferdinand Schmidt-Kaler, John P. Schiffer, Wolfgang Schleich, Athreya Shankar, Pietro Silvi, Cyrille Solaro, Bruno Taketani, Joseph Tan, Lars Timm, Vladan Vuleti\'c, Jorge Yago-Malo, Christof Wunderlich. The authors are grateful to Daniel Dubin and to Jennifer Lilieholm for the critical reading of the manuscript.

%

\end{document}